\documentclass{pas}

\usepackage{graphicx}
\usepackage{natbib}
\usepackage{longtable}
\usepackage{hyperref}

\begin{document}

% Journal metadata placeholders
\jnlPage{1}{10}
\jnlDoiYr{2025}
\doival{10.1017/pasa.xxxx.xx}

\articletitt{Research Paper}

\title{The Astro2Geo Project I. Radio astrometric offsets correlated with $\gamma$-ray brightness.}

% PASA Author format: \sn{Surname} \gn{GivenName}
\author{\sn{Hodgson} \gn{Jeffrey A.}$^{1}$,
\sn{Kr\'asn\'a} \gn{Hana}$^{2}$,
\sn{de Witt} \gn{Aletha}$^{3}$,
\sn{van Zyl} \gn{Pfesesani}$^{3}$, and \sn{Valverde} \gn{Janeth}$^{4,5}$
}

% Affiliations
\affil{$^1$Sejong University, Neungdong-ro, 209, Gwangjin Gu, Seoul, 05006, South Korea}
\affil{$^2$TU Wien, Department of Geodesy and Geoinformation, Wiedner Hauptstra{\ss}e 8, 1040 Vienna, Austria}
%\affil{$^3$South African Radio Astronomy Observatory (SARAO), South Africa}
\affil{$^3$South African Radio Astronomy Observatory (SARAO), Farm 502 JQ, Broederstroom road, Hartebeesthoek, 1740, South Africa}
\affil{$^4$Department of Physics, Marquette University, Milwaukee, WI 53201, USA}
\affil{$^5$NASA Goddard Space Flight Center, Greenbelt, MD 20771, USA}

% Correspondence
\corresp{J.A. Hodgson, Email: jhodgson@sejong.ac.kr}

% Running heads
\lefttitle{Publications of the Astronomical Society of Australia}
\righttitle{Hodgson et al.}

% Citation summary for the first page
\citeauth{Hodgson, J.A. et al., 2026. Astro2Geo: Connecting Astronomy and Geodesy I. Radio astrometric offsets correlated with $\gamma$-ray brightness. {\it Publications of the Astronomical Society of Australia}.}

\history{(Received xx xx xxxx; revised xx xx xxxx; accepted xx xx xxxx)}

\begin{abstract}

Precision geodesy relies on the stability of the International Celestial Reference Frame (ICRF), yet its reference sources, Active Galactic Nuclei (AGN), exhibit intrinsic changes in source structure that can manifest as apparent shifts in their astrometric positions. The high-precision radio measurements used to maintain the ICRF therefore provide a powerful means to investigate the astrophysical mechanisms driving these position changes. In particular, the observed astrometric variability offers a unique opportunity to link positional shifts in AGN to high-energy astrophysical processes.
We therefore investigated the relationship between the astrometric positions of ICRF AGN and their $\gamma$-ray emission. We measured the positional offsets of radio cores relative to the third realisation of the ICRF at both S/X (2.3/8.4~GHz) and K (24~GHz) bands and compared them to Fermi-\textit{LAT} (Large Area Telescope) $\gamma$-ray fluxes within $\pm$30 days of the radio observation. Out of an initial sample of 92 radio sources selected for having extensive radio astrometric observations, we identified 57 that met our selection criteria of having sufficient overlapping $\gamma$-ray data points to allow for regression analysis. We find a high incidence of statistically significant ($p<0.05$) power-law correlations, with $\sim$ 90\% of sources exhibiting this behaviour. The nature of this correlation is complex: we observe both positive and negative correlations, and the sign of the correlation can differ between the two frequency bands for the same source. To explain the correlations, we tested several scenarios, including variable $\gamma$-ray emission locations, changes in nuclear opacity, and variations in jet position angle. Our analysis reveals no single, universally applicable explanation. Instead, the results suggest that the observed correlation is driven by a complex interplay of multiple physical mechanisms, the dominance of which likely varies between sources. A search for time lags between the radio position offsets and $\gamma$-ray fluxes revealed tentative - and highly caveated - evidence for a time-delay in only five sources, with no evidence in other sources. A statistical comparison with the Optical Characteristics of Astrometric Radio Sources (OCARS) catalogue shows that, although our sample is biased towards optically brighter sources with better-constrained astrometric solutions due to their larger number of radio observations, it remains representative of the broader AGN population in terms of redshift distribution.

\end{abstract}


\maketitle

\section{Introduction} \label{sec:intro}

The interface between geodesy and astrophysics is realised through the Very Long Baseline Interferometry (VLBI) radio sources that constitute the International Celestial Reference Frame (ICRF). For geodesy, the primary requirement is stationarity; however, the reference sources that make up the ICRF are mostly radio-loud Active Galactic Nuclei (AGN), which can exhibit intrinsic changes in structure and flux over time and across frequency, manifesting as apparent shifts in their astrometric positions. The vast majority of the ICRF radio-loud AGN included in this study are blazars, AGN whose relativistic jets are aligned close to the line of sight. These sources are physically dynamic, exhibiting flares, ejecting new jet components, and showing frequency-dependent structures. Because blazars dominate the extragalactic $\gamma$-ray sky, they provide a natural class in which to investigate connections between high-energy emission and astrometric radio variability.

%Traditionally, geodesy has treated these structural changes as noise sources to be mitigated. However, the wealth of accumulated astrometric data offers a powerful tool for astrophysics. By linking the stability of the ICRF to the physics of the sources, we can probe the mechanisms of jet production. Specifically, if astrometric ``wobbles'' are driven by the ejection of bright, optically thick jet components, high-energy emission may serve as an early warning system. $\gamma$-ray flares, often associated with the energisation of new components near the jet base, could effectively trace the physical events that lead to apparent position shifts of the radio sources in the reference frames.

Traditionally, geodesy has treated these structural changes as noise sources to be mitigated. However, the wealth of accumulated astrometric data offers a powerful tool for astrophysics. By linking the stability of the ICRF to the physics of the sources, we can probe the mechanisms of jet production. Specifically, if astrometric ``wobbles'' are driven by the ejection of bright, optically thick jet components, the evolution of component flux densities will shift the centroid of emission. High-energy emission may serve as an early warning system for such events. $\gamma$-ray flares, often associated with the energisation of new components near the jet base, could effectively trace the physical events that lead to apparent position shifts of the radio sources in the reference frames.

It is well known that the position of the brightest component in a VLBI image which is usually referred to as the ``core'', may (or may not) shift over time with respect to the launching point of the jet \citep{niinuma15,koyama19}. This position also often changes as a function of frequency: the well-known ``core-shift'' effect. This is a consequence of the synchrotron self-absorbed (SSA) emission occurring close to a blazar jet base having its surface of last scattering ($\tau=1$) surface change as a function of frequency \citep[e.g.][]{lobanov1998}. Interpreting the ``core-shift'' effect assumes a freely expanding conical jet that, amongst other things, depends on the magnetic field strength and the particle number density \citep{bk79}.

Many studies have been conducted, looking for and finding the ``core-shift'' effect, but have typically been single-epoch studies \citep{kovalev08,OG09,pushkarev12,fromm15,mohan15,algaba17,pushkarev19}.  While more recently, variability of the ``core-shift'' effect as a function of time has been investigated and found by \citet{lisakov17,plavin19,sharma22,chamani23}. The favoured interpretation of this variability is either changes in the magnetic field strength or the particle number density, with perhaps changes in particle number densities preferred \citep{plavin19}. 

It is important to note that the SSA interpretation of the brightest component in VLBI maps is likely true only at cm wavelengths. At mm-wave, it has been suggested that the brightest component is a standing shock \citep[e.g.][]{jor17}. If this is the case, the ``core-shift'' effect would not be expected and may not be observed \citep{hodgson17,dodson17}.

While this variability is undesirable for geodesy \citep[see: ][for a detailed discussion]{chamani23}, this variability also allows us to explore the emission mechanisms of higher energy light by exploring correlations with this temporal variability. In this paper, we investigate near-in-time $\gamma$-ray fluxes with time-dependent changes of the brightest component relative to the ICRF.

\section{Methods and Observations\label{sec:methods}}

We began with a sample of 92 sources that were being monitored as part of the K-band Astro2Geo VLBI observing project \citep{hodgson2020,hodgson23,alet23}. The sources were selected for being easily detectable and with large amounts of existing data. In addition, we added selected sources with large astrometric offset variability (with weighted standard deviation of position time series of several tenths of mas) as reported e.g. by \citet{Cigan2024, Krasna2025} for which K-band data were available ($0119+115$, $0229+131$, $0642+449$). This leads to a natural bias and thus any results should not be interpreted as typical of the general AGN population.

The position offsets with respect to the third realisation of the ICRF at 8.4~GHz \citep[ICRF3-SX;][]{Charlot20} were calculated for both S/X band (2.3/8.4~GHz) and K-band (24~GHz) VLBI data in a geodetic/astrometric analysis. We analysed all 24-h VLBI S/X sessions provided through the International VLBI Service for Geodesy and Astrometry \citep[IVS;][]{Nothnagel2017} Data Centers \citep{Noll2010} starting from 1980.0 until 2024.0. This is a heterogeneous dataset of global VLBI sessions consisting of several programmes with different observation goals, such as providing rapid Earth orientation parameters, maintenance of terrestrial reference frame, or maintenance and expansion of celestial reference frame. Therefore, the uncertainty of estimated parameters, such as source coordinates, varies from session to session based mainly on the terrestrial network involved. On the other hand, analysed 24-h K-band sessions were provided through the Astro2Geo\footnote{\href{https://sites.google.com/sarao.ac.za/k-bandastrogeovlbi}{The K-band Astro2Geo VLBI Project}} project, including the earlier pilot sessions \citep{Lanyi2010, Petrov2011} run with Very Long Baseline Array (VLBA) between 2002 and 2008. Thirteen years later, building from this foundation, a new collaboration accomplished a renewal of regular observations at 24~GHz with VLBA comprising ten 25-m dishes on the U.S. territory providing homogenous VLBI sessions with the primary aim of enhancing the source density, sky coverage and improving the astrometric accuracy of the K-band position catalogue. Further stations (HartRAO in South Africa, Hobart in Tasmania, Yebes in Spain, Mopra in Australia) have joined the K-band Astro2Geo VLBI Project, observing with a core network built from Korean VLBI telescopes since 2022. We processed the VLBI group delays, fundamental observables of geodetic and global astrometric VLBI, which are provided after correlation, fringe fitting and basic pre-processing. The analysis was done with the Vienna VLBI and Satellite Software \citep[VieVS;][]{Boehm18} following the current International Earth Rotation and Reference Systems Service (IERS) Conventions 2020 \citep{iers10} with their updates. A priori models and parametrisation of the single session solution followed the setup described in \citet{Krasna2023sx} for S/X sessions and in \citet{Krasna2023k} for K-band sessions. The a priori positions of the antennas were modelled in International Terrestrial Reference Frame 2020 (ITRF2020) \citep{Altamimi23} with the position offset at epoch 2015.0, the linear velocity accounting for continental drift, and with post-seismic deformation model at stations subjected to major earthquakes. The a priori locations of radio sources were taken from ICRF3-SX accounting for the solar system barycentre acceleration with the recommended amplitude of 5.8~$\mu$as/y in direction to Galactic centre. In order to avoid any distortion of the source position estimates of interest, we did not follow the standard procedure for the definition of the celestial reference frame datum, which suggests applying the no-net-rotation condition \citep{Jacobs2010} on radio source positions with respect to the defining sources of the a priori catalogue. Instead, we estimated the session-wise corrections to source positions in right ascension and declination ($\Delta \rm RA$, $\Delta \rm Dec$) of the selected 92 sources without any constraints. The positions of the remaining sources in the session were fixed to their a priori values.  The magnitude of the position offset (i.e., the angular separation) was calculated by $((\Delta \rm RA\cdot \cos{\rm Dec})^{2} + \Delta \rm Dec^{2})^{1/2}$. The formal errors of the position estimates were taken into account and propagated through the analysis.

It is critical to note the systematic factors influencing these formal errors from single-session adjustment. The astrometric uncertainty is dependent mainly on: i) Flux Density: Brighter sources yield higher signal-to-noise ratios (SNR), resulting in smaller formal position uncertainties; ii) Declination: The distribution of VLBI stations heavily favours the Northern Hemisphere. Consequently, sources with low or negative declinations suffer from poorer UV-coverage, resulting in elongated synthesised beams and larger position uncertainties in the declination component; iii) Number of Observations: 
Higher number of observations reduces statistical error, but it varies between individual sessions;
%Frequent monitoring reduces statistical error, but our sample varies in observational cadence.
and iv) Station network geometry: Short baseline leads to a low angle of intersection, which causes a high uncertainty in the source position estimates.

Despite measuring K-band flux densities, we lack contemporaneous S/X band flux density information for all epochs. This limitation prevents a full spectral index analysis at this stage. It may be analysed in the future if full flux density data become available.

We obtained Fermi-\emph{LAT} light curves from the Fermi Light-Curve Repository \citep{fermi_lc_repository}. The \emph{LAT} continually scans the entire sky, covering an energy range from $20$~MeV to $>300$~GeV, providing regular monitoring of our target sources. After performing the recommended validation work on these data \citep{fermi_lc_repository}, we extracted 30-day cadence light curves to roughly align with the VLBI observation cadence, applying a 2-$\sigma$ detection threshold and leaving the spectral index free in the fit. We then matched near-in-time $\gamma$-ray fluxes that were within $\pm$30 days of the VLBI observations. If multiple $\gamma$-ray fluxes fell within this window, the nearest-in-time observation was selected. While shorter windows (e.g., 14 days) were considered for high-cadence sources, the 30-day window provided the most robust overlap for the full sample. 

To account for the left-censoring of the $\gamma$-ray data and avoid truncation bias, upper limits were retained. We implemented an Expectation-Maximisation (EM) algorithm adapted for Orthogonal Distance Regression (ODR). Instead of dropping non-detections, this approach treats them as left-censored data points, iteratively calculating the expected value of the true flux using a truncated normal distribution based on the radio offsets and the upper limits \citep[see][for a detailed explanation]{isobe86}. Data with relative errors $>100\%$ were excluded unless they were designated as upper limits. Sources with at least one bin detected within the window (with the nearest-in-time used for the analysis) are given in Table \ref{bigtable} along with their redshift and correlation slopes and $p$-values. The fit was performed taking into account errors in both variables. Significant ($p<0.05$) correlations are bolded. In total, 57 sources were found.

In Figure \ref{fig:correlations1} we show an example plot of the analysis. Further plots are given in Appendix \ref{Extended_plots}. In the top left is the S/X band correlation plot in log scale, now displaying both detections and upper limits. The original linear fit is shown alongside the new EM-corrected fit. Below is the same for K-band. 

We then investigated any possible time-delay between variations in the positional offsets with respect to the ICRF3-SX and $\gamma$-ray fluxes by computing the Discrete Correlation Function (DCF) using the open zDCF package \citep{zdcf}. Because the DCF relies mathematically on known variances and discrete pairings, this temporal analysis was restricted strictly to $\gamma$-ray detections to avoid introducing model-dependent variance from imputed limits. Significances were calculated by simulating the $\gamma$-ray light curves 10,000 times using the DELCgen code, recomputing the zDCF for each simulated light curve, and taking the 68, 95 and 99.7\% percentiles for each lag bin. Example results of this analysis are shown in Figure 1b and 1c with 1, 2 and 3 $\sigma$ significance levels indicated. In the bottom row, we present the time series data using twin axes: the monthly-binned $\gamma$-ray fluxes (including upper limits) are plotted alongside the S/X band (left) and K-band (right) positional offsets.

\onecolumn
\setlength{\LTcapwidth}{14.5cm}
\begin{longtable}{@{}llcrcrc@{}}

\hline \hline
IVS name & 4FGL name & $z$ & S/X slope & S/X p-val & K slope & K p-val \\
\hline
\endfirsthead

\multicolumn{7}{c}%
{{\bfseries Table \thetable\ continued from previous page}} \\
\hline \hline
IVS name & 4FGL name & $z$ & S/X slope & S/X p-val & K slope & K p-val \\
\hline
\endhead

\hline \multicolumn{7}{r}{{Continued on next page}} \\ \hline
\endfoot

\hline \hline
\endlastfoot

0016$+$731  &  4FGL J0019.6$+$7327  &  1.781$^{1}$  &  $-$0.36 $\pm$ 0.02  &  $\mathbf{<0.001}$  &  0.02 $\pm$ 0.01  &  0.06 \\
0035$-$252  &  4FGL J0038.2$-$2459  &  0.498$^{2}$  &  0.00 $\pm$ 0.00  &  0.17  &  $-$0.00 $\pm$ 0.03  &  0.97 \\
0109$+$224  &  4FGL J0112.1$+$2245  &  0.36$^{4}$  &  0.54 $\pm$ 0.16  &  $\mathbf{<0.001}$  &  0.56 $\pm$ 0.20  &  $\mathbf{0.01}$ \\
0133$+$476  &  4FGL J0137.0$+$4751  &  0.859$^{1}$  &  $-$0.00 $\pm$ 0.00  &  $\mathbf{<0.001}$  &  $-$0.16 $\pm$ 0.02  &  $\mathbf{<0.001}$ \\
0138$-$097  &  4FGL J0141.4$-$0928  &  0.733$^{3}$  &  $-$0.04 $\pm$ 0.12  &  0.76  &  $-$0.93 $\pm$ 0.10  &  $\mathbf{<0.001}$ \\
0202$+$319  &  4FGL J0205.2$+$3212  &  1.466$^{5}$  &  $-$0.38 $\pm$ 0.03  &  $\mathbf{<0.001}$  &  $-$0.01 $\pm$ 0.01  &  $\mathbf{0.04}$ \\
0234$+$285  &  4FGL J0237.8$+$2848  &  1.206$^{1}$  &  $-$0.69 $\pm$ 0.10  &  $\mathbf{<0.001}$  &  $-$0.41 $\pm$ 0.05  &  $\mathbf{<0.001}$ \\
0235$+$164  &  4FGL J0238.6$+$1637  &  0.94$^{1}$  &  $-$0.53 $\pm$ 0.06  &  $\mathbf{<0.001}$  &  -0.49 $\pm$ 0.11  &  $\mathbf{<0.001}$ \\
3C84      &  4FGL J0319.8$+$4130  &  0.018$^{1}$  &  $-$0.73 $\pm$ 0.32  &  $\mathbf{0.03}$  &  $-$0.34 $\pm$ 0.05  &  $\mathbf{<0.001}$ \\
NRAO140   &  4FGL J0336.4$+$3224  &  1.26$^{6}$  &  3.91 $\pm$ 1.16  &  $\mathbf{<0.001}$  &  $-$0.49 $\pm$ 0.06  &  $\mathbf{<0.001}$ \\
0402$-$362  &  4FGL J0403.9$-$3605  &  1.423$^{3}$  &  $-$0.04 $\pm$ 0.02  &  0.12  &  0.03 $\pm$ 0.11  &  0.76 \\
0405$-$385  &  4FGL J0407.0$-$3826  &  1.285$^{3}$  &  $-$0.01 $\pm$ 0.01  &  0.42  &  $-$0.11 $\pm$ 0.10  &  0.28 \\
0420$-$014  &  4FGL J0423.3$-$0120  &  0.9161$^{3}$  &  0.47 $\pm$ 0.05  &  $\mathbf{<0.001}$  &  0.90 $\pm$ 0.16  &  $\mathbf{<0.001}$ \\
3C120     &  4FGL J0433.0$+$0522  &  0.03301$^{3}$  &  0.23 $\pm$ 0.06  &  $\mathbf{<0.001}$  &  0.00 $\pm$ 0.00  &  0.72 \\
0454$-$234  &  4FGL J0457.0$-$2324  &  1.003$^{3}$  &  0.16 $\pm$ 0.05  &  $\mathbf{<0.001}$  &  $-$0.19 $\pm$ 0.12  &  0.12 \\
0458$-$020  &  4FGL J0501.2$-$0158  &  2.286$^{3}$  &  $-$0.25 $\pm$ 0.03  &  $\mathbf{<0.001}$  &  0.14 $\pm$ 0.18  &  0.46 \\
0528$+$134  &  4FGL J0530.9$+$1332  &  2.07$^{1}$  &  $-$0.03 $\pm$ 0.01  &  $\mathbf{<0.001}$  &  $-$0.23 $\pm$ 0.02  &  $\mathbf{<0.001}$ \\
0552$+$398  &  4FGL J0555.6$+$3947  &  2.363$^{10}$  &  $-$0.48 $\pm$ 0.04  &  $\mathbf{<0.001}$  &  0.01 $\pm$ 0.01  &  0.18 \\
0648$-$165  &  4FGL J0650.2$-$1636  &  0.5$^{11}$  &  $-$0.79 $\pm$ 0.06  &  $\mathbf{<0.001}$  &  1.40 $\pm$ 0.22  &  $\mathbf{<0.001}$ \\
0736$+$017  &  4FGL J0739.2$+$0137  &  0.189410$^{1}$  &  0.16 $\pm$ 0.09  &  0.07  &  $-$0.41 $\pm$ 0.09  &  $\mathbf{<0.001}$ \\
0748$+$126  &  4FGL J0750.8$+$1229  &  0.889$^{1}$  &  $-$0.17 $\pm$ 0.03  &  $\mathbf{<0.001}$  &  $-$0.18 $\pm$ 0.04  &  $\mathbf{<0.001}$ \\
OJ287     &  4FGL J0854.8$+$2006  &  0.3056$^{1}$  &  $-$0.46 $\pm$ 0.03  &  $\mathbf{<0.001}$  &  $-$0.79 $\pm$ 0.12  &  $\mathbf{<0.001}$ \\
0954$+$658  &  4FGL J0958.7$+$6534  &  0.3694$^{1}$  &  $-$0.06 $\pm$ 0.05  &  0.25  &  0.07 $\pm$ 0.17  &  0.70 \\
1044$+$719  &  4FGL J1048.4$+$7143  &  1.14$^{1}$  &  $-$0.07 $\pm$ 0.03  &  $\mathbf{0.01}$  &  0.13 $\pm$ 0.26  &  0.62 \\
1055$+$018  &  4FGL J1058.4$+$0133  &  0.893572$^{1}$  &  0.78 $\pm$ 0.14  &  $\mathbf{<0.001}$  &  1.08 $\pm$ 0.14  &  $\mathbf{<0.001}$ \\
1215$+$303  &  4FGL J1217.9$+$3007  &  0.1305$^{7}$  &  $-$0.17 $\pm$ 0.13  &  0.22  &  $-$0.05 $\pm$ 0.26  &  0.85 \\
1334$-$127  &  4FGL J1337.6$-$1257  &  0.539$^{3}$  &  1.00 $\pm$ 0.07  &  $\mathbf{<0.001}$  &  $-$0.54 $\pm$ 0.09  &  $\mathbf{<0.001}$ \\
1424$-$418  &  4FGL J1427.9$-$4206  &  1.522$^{2}$  &  $-$0.23 $\pm$ 0.03  &  $\mathbf{<0.001}$  &  $-$0.03 $\pm$ 0.07  &  0.71 \\
1510$-$089  &  4FGL J1512.8$-$0906  &  0.36$^{3}$  &  $-$0.14 $\pm$ 0.04  &  $\mathbf{<0.001}$  &  0.25 $\pm$ 0.16  &  0.13 \\
1546$+$027  &  4FGL J1549.5$+$0236  &  0.41421$^{1}$  &  0.82 $\pm$ 0.06  &  $\mathbf{<0.001}$  &  0.46 $\pm$ 0.06  &  $\mathbf{<0.001}$ \\
1548$+$056  &  4FGL J1550.7$+$0528  &  1.4204$^{1}$  &  $-$0.00 $\pm$ 0.00  &  0.45  &  0.11 $\pm$ 0.02  &  $\mathbf{<0.001}$ \\
1611$+$343  &  4FGL J1613.6$+$3411  &  1.398$^{1}$  &  2.94 $\pm$ 0.29  &  $\mathbf{<0.001}$  &  $-$0.49 $\pm$ 0.06  &  $\mathbf{<0.001}$ \\
1705$+$018  &  4FGL J1707.9$+$0016  &  2.568$^{3}$  &  $-$0.47 $\pm$ 0.03  &  $\mathbf{<0.001}$  &  $-$0.00 $\pm$ 0.00  &  $\mathbf{<0.001}$ \\
NRAO530   &  4FGL J1733.0$-$1305  &  0.902$^{8}$  &  $-$0.40 $\pm$ 0.06  &  $\mathbf{<0.001}$  &  0.00 $\pm$ 0.02  &  0.97 \\
1749$+$096  &  4FGL J1751.5$+$0938  &  0.322$^{1}$  &  $-$0.79 $\pm$ 0.06  &  $\mathbf{<0.001}$  &  0.31 $\pm$ 0.08  &  $\mathbf{<0.001}$ \\
1751$+$288  &  4FGL J1753.7$+$2847  &  1.118$^{9}$  &  0.70 $\pm$ 0.04  &  $\mathbf{<0.001}$  &  $-$0.17 $\pm$ 0.03  &  $\mathbf{<0.001}$ \\
1921$-$293  &  4FGL J1924.8$-$2914  &  0.353$^{9}$  &  1.23 $\pm$ 0.07  &  $\mathbf{<0.001}$  &  $-$0.51 $\pm$ 0.10  &  $\mathbf{<0.001}$ \\
1933$-$400  &  4FGL J1937.2$-$3958  &  0.965$^{3}$  &  $-$2.43 $\pm$ 0.51  &  $\mathbf{<0.001}$  &  0.48 $\pm$ 0.11  &  $\mathbf{<0.001}$ \\
1949$-$052  &  4FGL J1951.8$-$0511  &  1.083$^{9}$  &  $-$0.21 $\pm$ 0.03  &  $\mathbf{<0.001}$  &  0.68 $\pm$ 0.17  &  $\mathbf{<0.001}$ \\
1953$-$325  &  4FGL J1957.1$-$3231  &  1.242$^{3}$  &  $-$3.31 $\pm$ 0.00  &  $\mathbf{<0.001}$  &  10.71 $\pm$ 17.35  &  0.55 \\
1954$-$388  &  4FGL J1958.0$-$3845  &  0.626$^{3}$  &  $-$0.25 $\pm$ 0.02  &  $\mathbf{<0.001}$  &  $-$0.34 $\pm$ 0.11  &  $\mathbf{<0.001}$ \\
1958$-$179  &  4FGL J2000.9$-$1748  &  0.652$^{9}$  &  $-$0.64 $\pm$ 0.06  &  $\mathbf{<0.001}$  &  $-$0.41 $\pm$ 0.12  &  $\mathbf{<0.001}$ \\
2029$+$121  &  4FGL J2032.0$+$1219  &  1.11614$^{1}$  &  $-$10.54 $\pm$ 4.64  &  $\mathbf{0.03}$  &  1.01 $\pm$ 0.24  &  $\mathbf{<0.001}$ \\
3C418      &  4FGL J2038.7$+$5117  &  1.686$^{12}$  &  $-$0.01 $\pm$ 0.00  &  $\mathbf{<0.001}$  &  0.39 $\pm$ 0.07  &  $\mathbf{<0.001}$ \\
2121$+$053  &  4FGL J2123.6$+$0535  &  1.94$^{1}$  &  $-$1.44 $\pm$ 0.18  &  $\mathbf{<0.001}$  &  $-$1.38 $\pm$ 0.16  &  $\mathbf{<0.001}$ \\
2131$-$021  &  4FGL J2134.2$-$0154  &  1.285$^{3}$  &  1.65 $\pm$ 0.24  &  $\mathbf{<0.001}$  &  0.71 $\pm$ 0.10  &  $\mathbf{<0.001}$ \\
2134$+$00   &  4FGL J2136.2$+$0032  &  1.941251$^{1}$  &  $-$1.44 $\pm$ 0.23  &  $\mathbf{<0.001}$  &  $-$0.00 $\pm$ 0.00  &  0.28 \\
2145$+$067  &  4FGL J2148.6$+$0652  &  1.003$^{1}$  &  $-$1.92 $\pm$ 0.23  &  $\mathbf{<0.001}$  &  $-$0.12 $\pm$ 0.01  &  $\mathbf{<0.001}$ \\
2155$-$304  &  4FGL J2158.8$-$3013  &  0.116$^{13}$  &  $-$0.44 $\pm$ 0.30  &  0.16  &  1.09 $\pm$ 0.52  &  0.06 \\
BLLAC  &  4FGL J2202.7$+$4216  &  0.0686$^{14}$  &  $-$0.30 $\pm$ 0.24  &  0.22  &  $-$0.23 $\pm$ 0.09  &  $\mathbf{0.01}$ \\
3C446      &  4FGL J2225.7$-$0457  &  1.404$^{3}$  &  1.29 $\pm$ 0.08  &  $\mathbf{<0.001}$  &  $-$1.95 $\pm$ 0.34  &  $\mathbf{<0.001}$ \\
2227$-$088  &  4FGL J2229.7$-$0832  &  1.56$^{3}$  &  $-$0.60 $\pm$ 0.05  &  $\mathbf{<0.001}$  &  0.40 $\pm$ 0.08  &  $\mathbf{<0.001}$ \\
CTA102    &  4FGL J2232.6$+$1143  &  1.037$^{15}$  &  0.09 $\pm$ 0.09  &  0.30  &  0.37 $\pm$ 0.36  &  0.33 \\
2245$-$328  &  4FGL J2248.7$-$3235  &  2.268$^{3}$  &  $-$0.91 $\pm$ 0.17  &  $\mathbf{<0.001}$  &  $-$1.00 $\pm$ 0.26  &  $\mathbf{<0.001}$ \\
3C454.3   &  4FGL J2253.9$+$1609  &  0.859$^{1}$  &  $-$0.00 $\pm$ 0.01  &  0.84  &  0.06 $\pm$ 0.06  &  0.31 \\
2255$-$282  &  4FGL J2258.1$-$2759  &  0.92584$^{3}$  &  $-$0.00 $\pm$ 0.00  &  0.78  &  0.09 $\pm$ 0.04  &  $\mathbf{0.04}$ \\
2320$-$035  &  4FGL J2323.5$-$0317  &  1.411$^{9}$  &  0.18 $\pm$ 0.11  &  0.25  &  0.46 $\pm$ 0.19  &  $\mathbf{0.02}$ \\
\hline
\caption{Results of the correlation analysis for the cross-matched IVS and Fermi-LAT (4FGL) source sample. Superscript numbers in the redshift column correspond to the literature references. Redshift references: 
[1] \citet{purismo13}; [2]; \citet{jones09}; [3] \citet{drinkwater97}; [4] \citet{magic_0109_z2018}; [5] \citet{burbidge1970}; [6] \citet{de_Veny_1971}; [7] \citet{furniss19}; [8] \citet{Junkkarinen1984}; [9] \citet{Healey2008}; [10] \citet{mccintosh99}; [11] photometric estimate from OCARS: \citet{Flesch2023}; [12] \citet{SmithSpinrad1980}; [13] \citet{Falomo1993}; [14] \citet{vermeulen1995}; [15] \citet{Schmidt1965}. } \label{bigtable} \\

\end{longtable}

\twocolumn

\begin{figure*}[ht!]
    \centering
    \includegraphics[width=0.95\textwidth]{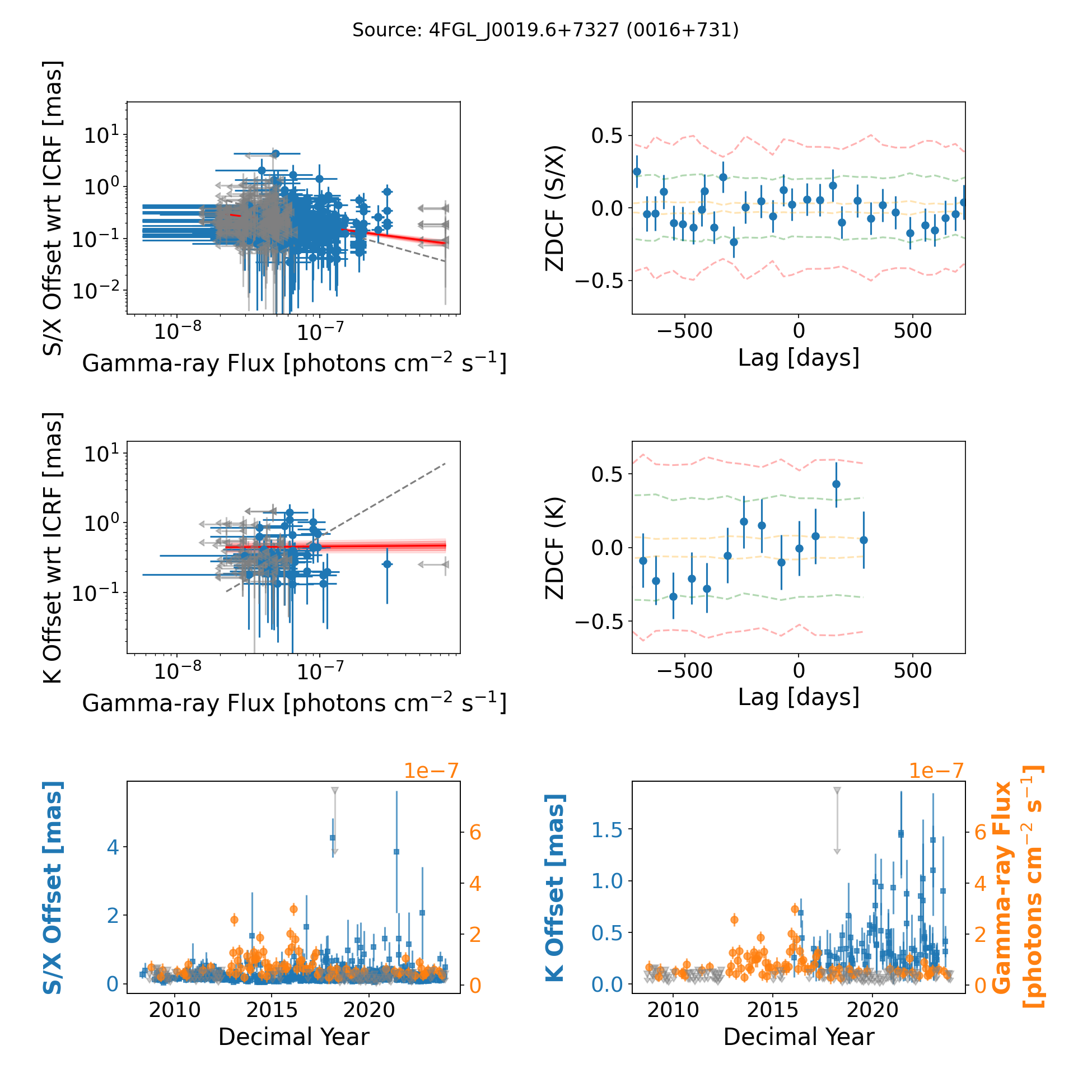}
    \caption{Example plot for 0016+731. Top left: S/X band correlation plot. The Survivor-bias (SC) corrected best fit is in red with 1, 2 and 3 $\sigma$  errors in grey. The grey dashed line is the non corrected fit. Blue marks are detection and grey marks are upper limits. Top right: zDCF at S/X band with 1 (yellow), 2 (green) and 3 (red) $\sigma$ significance levels indicated. Note that zDCF panels may be empty (particularly at K-band) for sources with little data. Additionally, the 1, 2, and 3$\sigma$ significance contours (derived simulated light curves) may span a wider range of lag times than the real data; hence, these bounds can extend across the plot even in cases where the real data only yields one or two valid zDCF points. Middle row: Same as top row but for K-band. Bottom:  monthly-binned $\gamma$-ray flux (blue) overplotted with the S/X band offsets (left) and with the K-band offsets (right) with respect to the ICRF3-SX.}
    \label{fig:correlations1}
\end{figure*}

\begin{figure*}[ht!]
    \centering
    \includegraphics[width=0.55\linewidth]{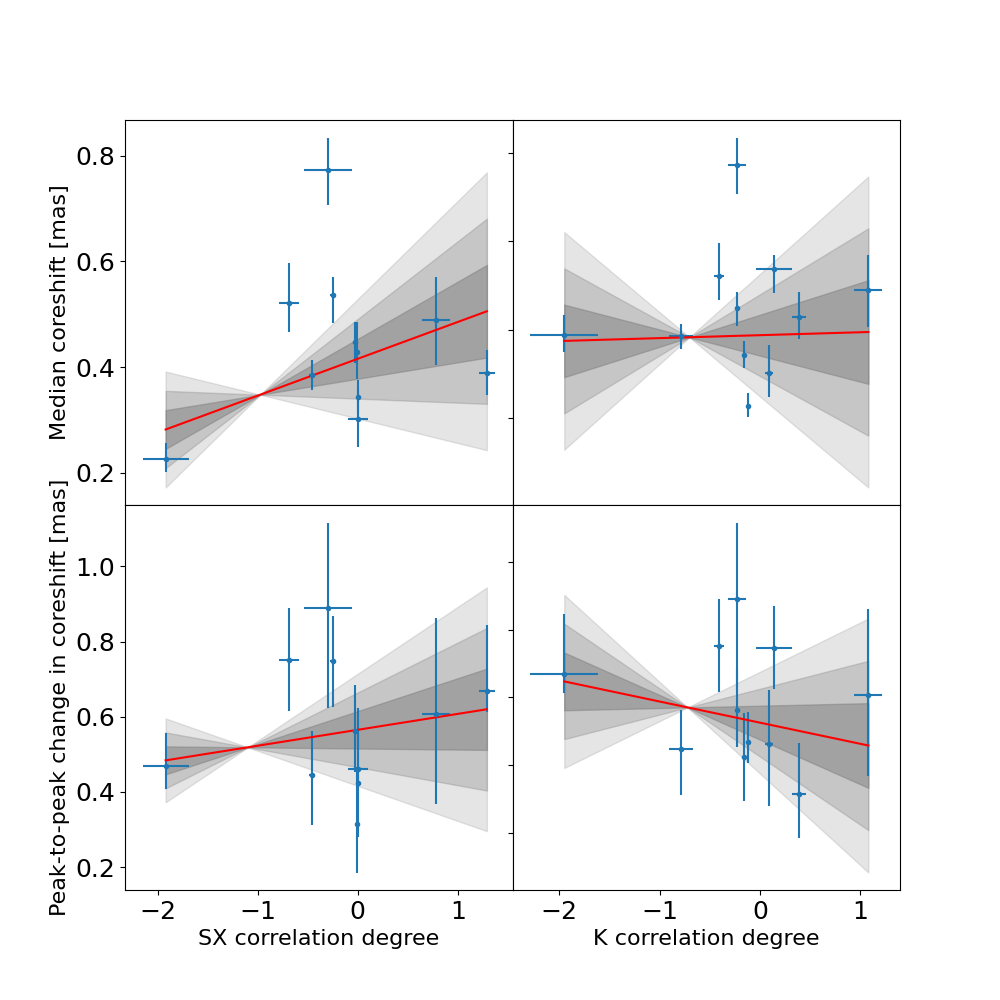}
    \includegraphics[width=0.3\linewidth]{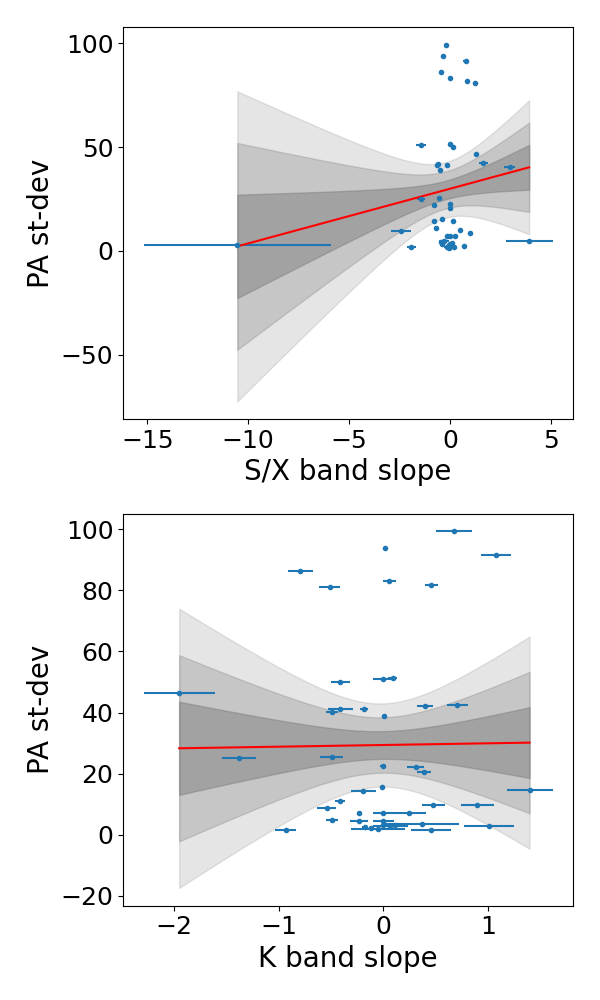}
    \caption{Left: Correlation plots between the median core-shift (as reported by \citet{plavin19}) with the strength of the ICRF3-SX offset correlation at S/X-band (top-left) and K-band (top right). The same but with the peak-to-peak core-shift variability for S/X band (bottom-left) and K-band (bottom right). Right, upper: ICRF3-SX offset vs the standard deviation of the PA from \citet{alet23} at S/X band. Right, lower: same as top but for K band. In all panels, the red solid line represents the line of best fit, while the three progressively lighter shades of gray indicate the 1$\sigma$, 2$\sigma$, and 3$\sigma$ uncertainty ranges (confidence intervals) of the fit, respectively.}
    \label{fig:plavin_corrs}
\end{figure*}

\section{Results and Discussion}\label{results_discussion}

We find a high proportion of statistically significant correlation between radio core astrometric offsets and $\gamma$-ray fluxes within $\sim$30 days of VLBI observations. Approximately 90\% of sources show a significant power-law relationship ($p<0.05$) in at least one of S/X or K-bands. We emphasise that the sample is biased towards sources with extensive K-band astrometric observations, with a few additional sources included because they exhibit large position offsets, and therefore the results should not be interpreted as representative of the entire blazar or broader AGN population.

\subsection{Comparison with the OCARS catalogue}

To investigate this bias, we compared our 92 selected sources against the full Optical Characteristics of Astrometric Radio Sources (OCARS) catalogue \citep{OCARS2018}. We performed two-sample Kolmogorov-Smirnov (K-S) tests on the observational parameters supplied in the OCARS catalogue in order to determine if the subset distributions differ significantly from the parent population. The results are summarised in Table \ref{tab:ocars_comparison}.

The most significant deviations are found in astrometric precision and optical brightness. The subset exhibits a median semi-major axis error of 0.032\,mas compared to the parent median of 1.377\,mas ($p < 10^{-78}$), this tells us that our sample represents the astrometrically cleaner sample. Similarly, the subset is significantly brighter in the optical regime, with a median magnitude of 17.1 compared to the parent median of 19.4 ($p < 10^{-14}$). Spatially, we find a statistically significant difference in the distribution of Galactic Latitudes ($p \approx 0.001$).

However, we note that the redshift distributions show no statistically significant difference between our subset and the parent catalogue. This implies that while our sample is biased towards brighter and more compact radio sources, it remains representative of the general AGN population in terms of cosmological distance and epoch.

\begin{table}[ht!]
    \centering
    \caption{K-S test results comparing the study subset (N=92) with the parent OCARS catalogue.}
    \label{tab:ocars_comparison}
    \begin{tabular}{lcccc}
    \hline \hline
    Parameter & Subset & OCARS median & K-S Stat & $p$-value \\
    \hline
    $\sigma_{\text{pos, maj}}$ (mas) & 0.032 & 1.377 & 0.96 & $< 10^{-78}$ \\
    Optical Mag. & 17.10 & 19.40 & 0.52 & $7.6 \times 10^{-15}$ \\
    Gal. Lat. ($^\circ$) & $-19.60$ & 5.72 & 0.25 & $1.2 \times 10^{-3}$ \\
    Gal. Lon. ($^\circ$) & 120.64 & 158.36 & 0.21 & $0.012$ \\
    $z$ & 1.00 & 0.88 & 0.16 & 0.11 \\
    % $z_{\text{simbad}}$ & 1.00 & 0.88 & 0.15 & 0.15 \\
    % $z_{\text{sdss}}$ & 1.40 & 1.09 & 0.22 & 0.81 \\
    \hline
    \end{tabular}
\end{table}

%Assuming independence, selecting 43 and 38 sources from a population of 57 yields an expected overlap of roughly 29 sources. We observe 32 sources that are statistically significant in both bands, indicating that astrometric-y-ray correlation is an intrinsic property of these active sources across frequencies. Most revealingly, of these 32 co-significant sources, 14 (nearly 45\%) exhibit opposite correlation signs between the S/X and K bands (e.g., positive at S/X but negative at K-band). Such complex, frequency-dependent behaviour suggests that no single, universal mechanism explains the correlations across the population. We now explore several potential mechanisms.

% , but the consistent ratio of positive to negative correlations across both bands ($\sim$30\% in S/X and $\sim$45\% in K-band) while this suggests a population-wide connection,

\subsection{Physical interpretation}

The physical nature of this correlation is complex. Out of the 57 analysed sources, 43 exhibit a statistically significant correlation ($p < 0.05$) in the S/X-band (13 positive, 30 negative), while 38 are significant in the K-band (16 positive, 22 negative). The slightly lower incidence of significant correlations at K-band is observationally expected; the S/X-band data goes back to 1980, whereas regular K-band observations only began in 2016. The resulting lower number of epochs in K-band reduces statistical power, making it harder to cross the $p < 0.05$ threshold.

Assuming independence, selecting 43 and 38 sources from a population of 57 yields an expected overlap of roughly 29 sources. We observe 32 sources that are statistically significant in both bands. Because this excess is marginal ($\sim 2 \sigma$) and while this might be partially driven by the limited K-band time baseline, the current data does not support a strongly predictive, universal link between the two bands. Furthermore, among the 32 overlapping sources, we observe opposite correlation signs in approximately 40\% (13/32) of the cases. 

That we see opposite correlations within the same source could be due to the synchrotron self-absorbed (SSA) turnover frequency being sometimes above 8~GHz and below 22~GHz, therefore leading to the brightest position in the source being different at different frequencies, but the $\gamma$-ray emission site is the same. When the $\gamma$-ray emission site is located further downstream from the 'normal' ICRF position, this will be reflected as a positive correlation. If the 'normal' location is already downstream and then the $\gamma$-rays are then coming 'upstream' of this position, we would observe a negative correlation. When the SSA turnover frequency is above 8~GHz and below 22~GHz, the correlation will go in opposite directions. Indeed, such a situation was recently reported by \citet{alet23b}.

\subsubsection{Multiple $\gamma$-ray emission sites}

%Blazar jets having $\gamma$-rays occuring in multple locations (e.g. in the "core" and downstream quasi-stationary features) is well documented in sources such as CTA\,102, OJ\,287 and 3C\,84 \citep{schinzel12,hodgson17,hodgson18}. Such shifts in $\gamma$-ray production sites would naturally explain a correlation between $\gamma$-ray flux and the position of the radio core.

Blazar jets having $\gamma$-rays occurring in multiple locations (e.g. in the "core" and downstream quasi-stationary features) is well documented in sources such as CTA\,102, OJ\,287, 3C\,84 and PKS\,1510-089 \citep{schinzel12,hodgson17,hodgson18,orienti11,marscher10}. Such shifts in $\gamma$-ray production sites would naturally explain a correlation between $\gamma$-ray flux and the position of the radio core.

The four sources above are included in our analysis. In OJ\,287, we find a significant correlation, but with no time-delay. In contrast, 3C\,84 shows no significant near-in-time correlation, but hints of a time-delayed correlation. CTA\,102 shows a strong correlation at K-band, but inconclusive evidence of a time-delayed correlation. 

From this analysis alone, we cannot provide unambiguous evidence of this mechanism and detailed source-by-source analysis of VLBI maps would be required, which is beyond the scope of this work.  

\subsubsection{Changes in radio core opacity}
Changes in the physical conditions at the base of the jet, such as the particle density or magnetic field can alter the opacity of the radio core. This would manifest itself as a changing position of the $\tau=1$ surface - commonly observed as core-shift \citep{plavin19}. If these changes are linked to $\gamma$-ray flaring, a correlation would be expected. \citet{chamani23} found evidence that core-shift variability is connected to core-flux variability consistent with the predictions of \citet{bk79} and similarly would naturally explain the correlations.

To test this, we compared our correlation strengths with the core-shift measurements from \citet{plavin19} for the 11 sources common to both samples, with the results shown in Figure \ref{fig:plavin_corrs} (left). We find no statistically significant correlation ($p\approx 0.11$) between the median core-shift magnitude and the strength of the $\gamma$-ray-offset correlation. There is similarly no evidence for a link between the $\gamma$-ray-offset and the peak-to-peak variability of the core-shift. This suggests that simple changes in average core-shift properties, at least for the 11 sources in common with \citet{plavin19}, are not the likely to be the primary mechanism for the correlation.

\subsubsection{Changes in jet Position Angle}
%If the jet physically changes its orientation on the sky (i.e. its PA), the measured position of the core could change relative to its average ICRF position. If this jet wobbling is connected to the events causing the $\gamma$-ray flaring, a correlation could arise as has been suggested by \citet{rani14}.

If the jet physically changes its orientation on the sky (i.e. its Position Angle, PA), the measured position of the core could change relative to its average ICRF position. If this jet wobbling is connected to the events causing the $\gamma$-ray flaring, a correlation could arise as has been suggested by \citet{rani14} and  \citet{raiteri2017}.

To test this, we used the jet PA measurements from \citet{alet23} and compared the standard deviation of the PA (as a proxy for variability) against our measured correlation slopes (see Figure \ref{fig:plavin_corrs}, right). We find no significant correlations at S/X band or K-band. Therefore we conclude that PA changes are likely not driving the correlations.

\subsubsection{Time delayed correlations}

A time-lag between astrometric shifts and $\gamma$-ray flux variability could constrain the relative locations of their emission regions. For example, $\gamma$-ray flares are often observed to lead radio flares by several months, suggesting that they occur closer to the jet base than the radio core \citep[e.g.,][]{MaxMoerbeck2014, Fuhrmann2014, Ramakrishnan2015}. The simplistic view suggests that an event causes a $\gamma$-ray flare close to the core, followed by a rising radio flux as a new jet component emerges and moves downstream. To search for a similar relationship with astrometric positions, we performed a cross-correlation analysis using zDCF. We find tentative evidence ($>2 \sigma$) for time offsets in only 5 sources (0133+476, 0235+164, 1424-418, 3C\,120, and 3C\,84). However, these detections are not robust and in some cases imply an astrophysically challenging delay on the order of years. The fact that we only see this in five sources suggests that the simplistic view of delayed structural evolution is likely incomplete or easily masked by multiple flaring locations. The dominant signal across the population (consistent across both S/X and K-bands) appears to be from correlations within our 60 day window. However, an analysis in this way is obviously limited and future analysis with radio flux density information would be valuable.

\section{Conclusions}

We have found a high incidence ($\sim$90\%) of correlation between the changes in time series of radio core astrometric positions and $\gamma$-ray fluxes in our biased sample. We have explored four potential interpretations, including multiple/changing $\gamma$-ray emission sites, changes in nuclear opacity, changes in radio core flux densities, and changes in PA. We find no evidence or inconclusive results for a single mechanism driving the correlations. We therefore interpret the correlations as not arising from any single universal mechanism and point towards a complex interplay of many physical factors that likely change from source to source. For some objects, shifts in the $\gamma$-ray emission sites may dominate while in others subtle opacity or structural changes in the radio core may cause the correlations. Detailed source-by-source analysis of VLBI images may be required to disentangle the mechanisms.

It is important to note the limitations of this study - especially that it is a biased sample and should not be considered a representative result for the entire blazar (or AGN) population. 

In future work we will more directly investigate the effects of radio emission on the correlations and furthermore will strive to do it with a more complete and unbiased sample in order to find how universal such an effect may be. Further, we will investigate effects such as source type (e.g. BL Lac vs FSRQ) and/or other properties such as jet apparent speeds.

\begin{acknowledgements}
J.A.H. acknowledges the support of the National Research Foundation of Korea (NRF) (NRF-2021R1C1C1009973). This work was supported by the National Research Foundation of Korea (NRF) grant funded by the Korea government(MSIT) RS-2025-16302968.
We acknowledge use of the VLBA under the USNO’s time allocation program since 2017. This work supports USNO’s ongoing research into the celestial reference frame and geodesy. The VLBA is operated by the National Radio Astronomy Observatory, which is a facility of the National Science Foundation operated under cooperative agreement by Associated Universities, Inc. 
We acknowledge use of the Hobart, HartRAO, Mopra, Yebes and Korean VLBI Network (KVN) K-band observations. We acknowledge the IVS for providing the S/X data and the Astro2Geo VLBI Project for the K-band data. We thank David Gordon (USNO) for the fringe-fitting of the VLBI K-band observations and creation of the databases.
\end{acknowledgements}

\section*{Data Availability}
The VLBI S/X data used in this work are available from the IVS Data Centers. The K-band data are available upon request from the Astro2Geo VLBI Project. Fermi-LAT data are publicly available from the Fermi Light-Curve Repository.

% Using pas style bibliography
\bibliographystyle{pas}
\bibliography{biblio}{}

\clearpage

\appendix
\setcounter{figure}{0}
\renewcommand{\thefigure}{A\arabic{figure}}

\section{Extended plots}\label{Extended_plots}

\begin{figure}[ht!]
    \centering
    \includegraphics[width=\textwidth]{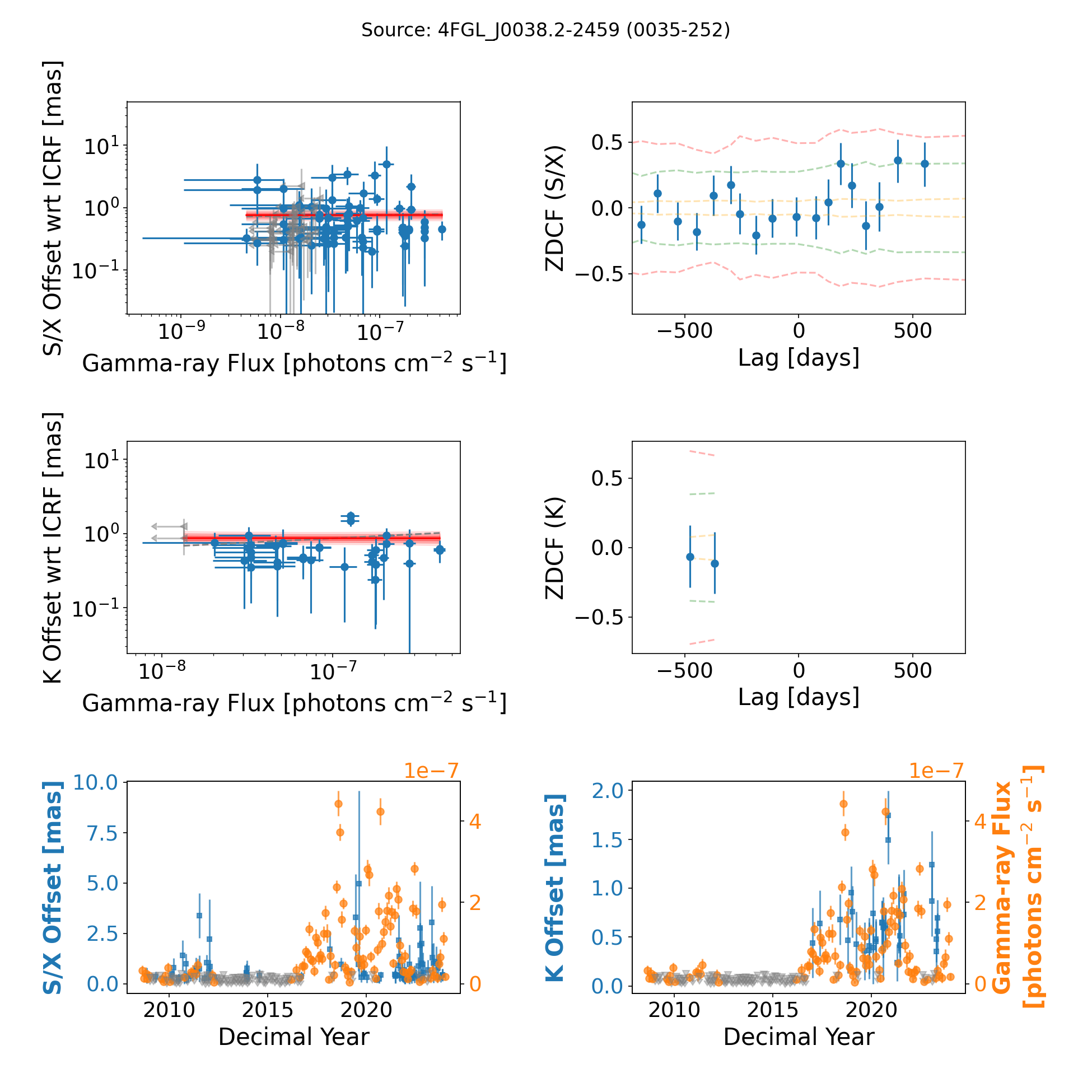}
    \caption{Same as Figure \ref{fig:correlations1}, but for the source 0035$-$252.}
    \label{fig:correlationsJ0038}
\end{figure}

\clearpage

\begin{figure}[ht!]
    \centering
    \includegraphics[width=\textwidth]{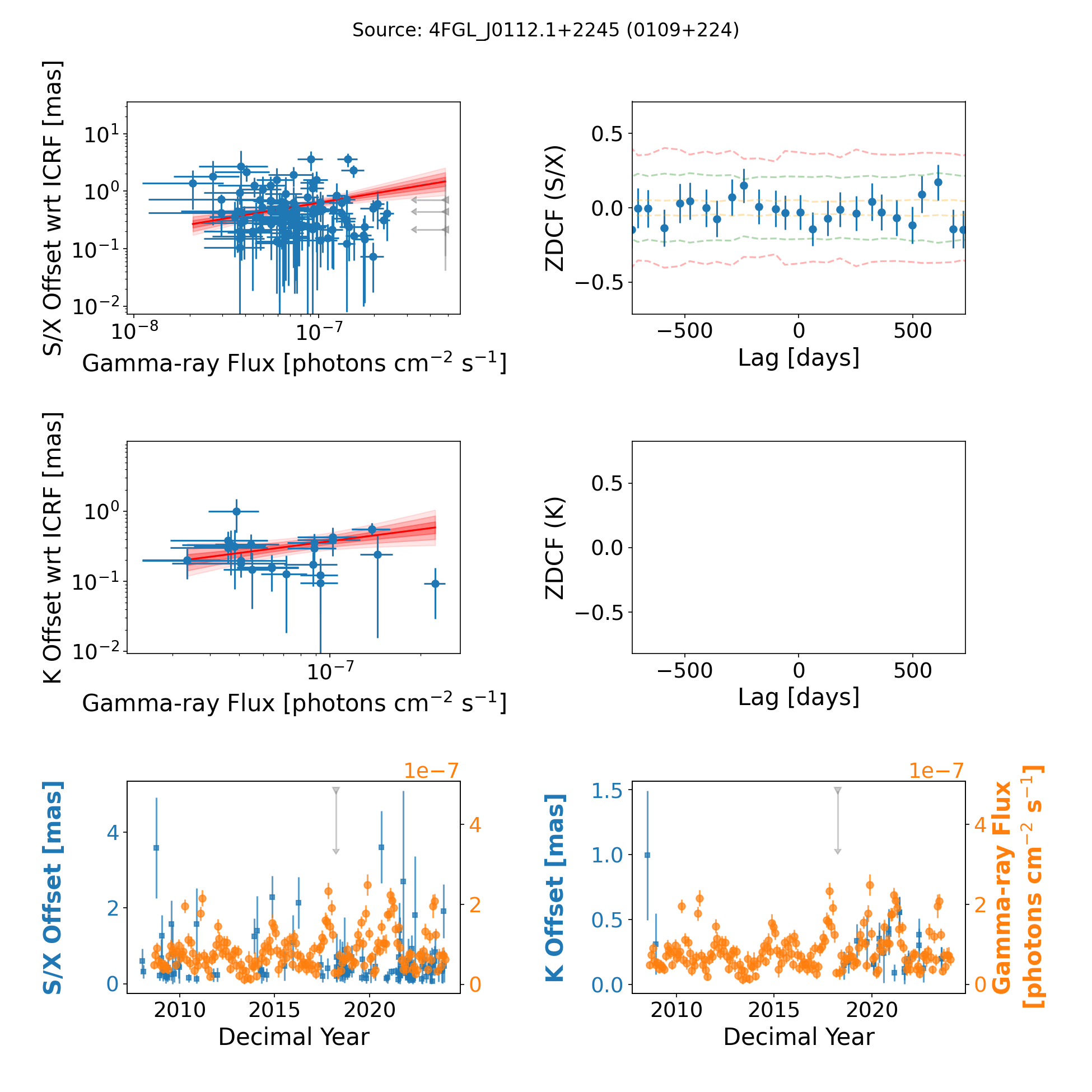}
    \caption{Same as Figure \ref{fig:correlations1}, but for the source 0109$+$224.}
    \label{fig:correlationsJ0112}
\end{figure}

% \end{document}

\clearpage

\begin{figure}[ht!]
    \centering
    \includegraphics[width=\textwidth]{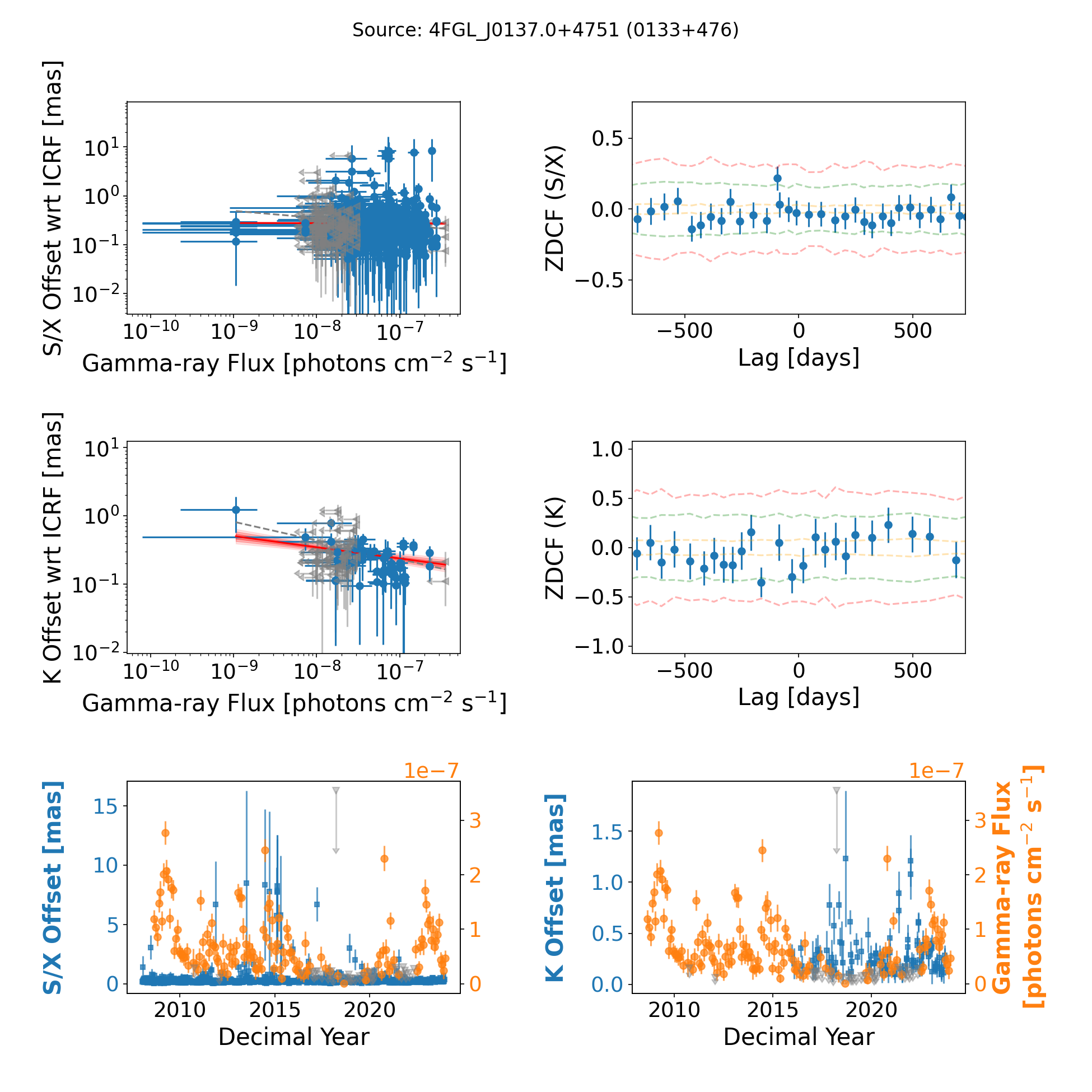}
    \caption{Same as Figure \ref{fig:correlations1}, but for the source 0133$+$476.}
    \label{fig:correlationsJ0137}
\end{figure}

\clearpage

\begin{figure}[ht!]
    \centering
    \includegraphics[width=\textwidth]{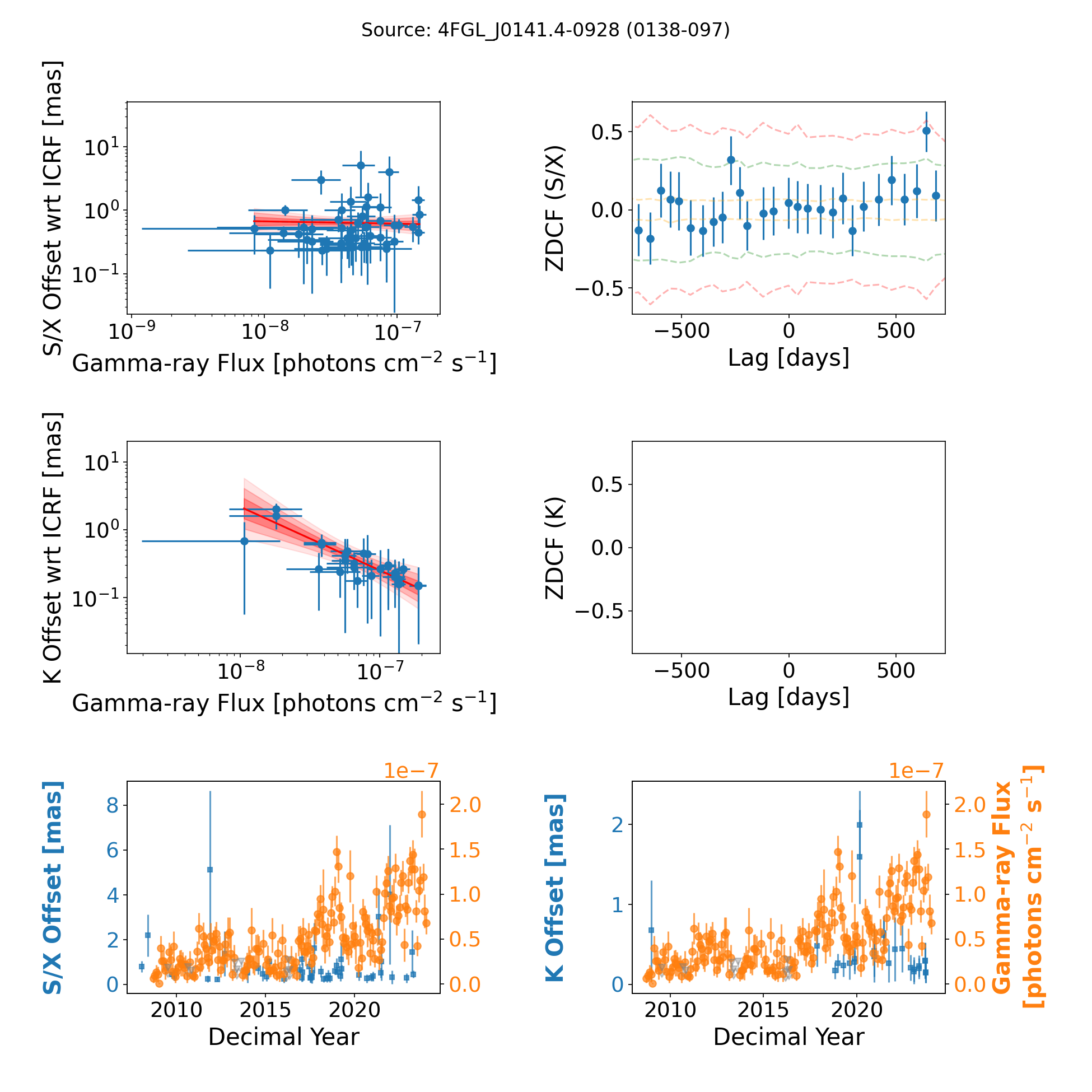}
    \caption{Same as Figure \ref{fig:correlations1}, but for the source 0138$-$097.}
    \label{fig:correlationsJ0141}
\end{figure}

\clearpage

\begin{figure}[ht!]
    \centering
    \includegraphics[width=\textwidth]{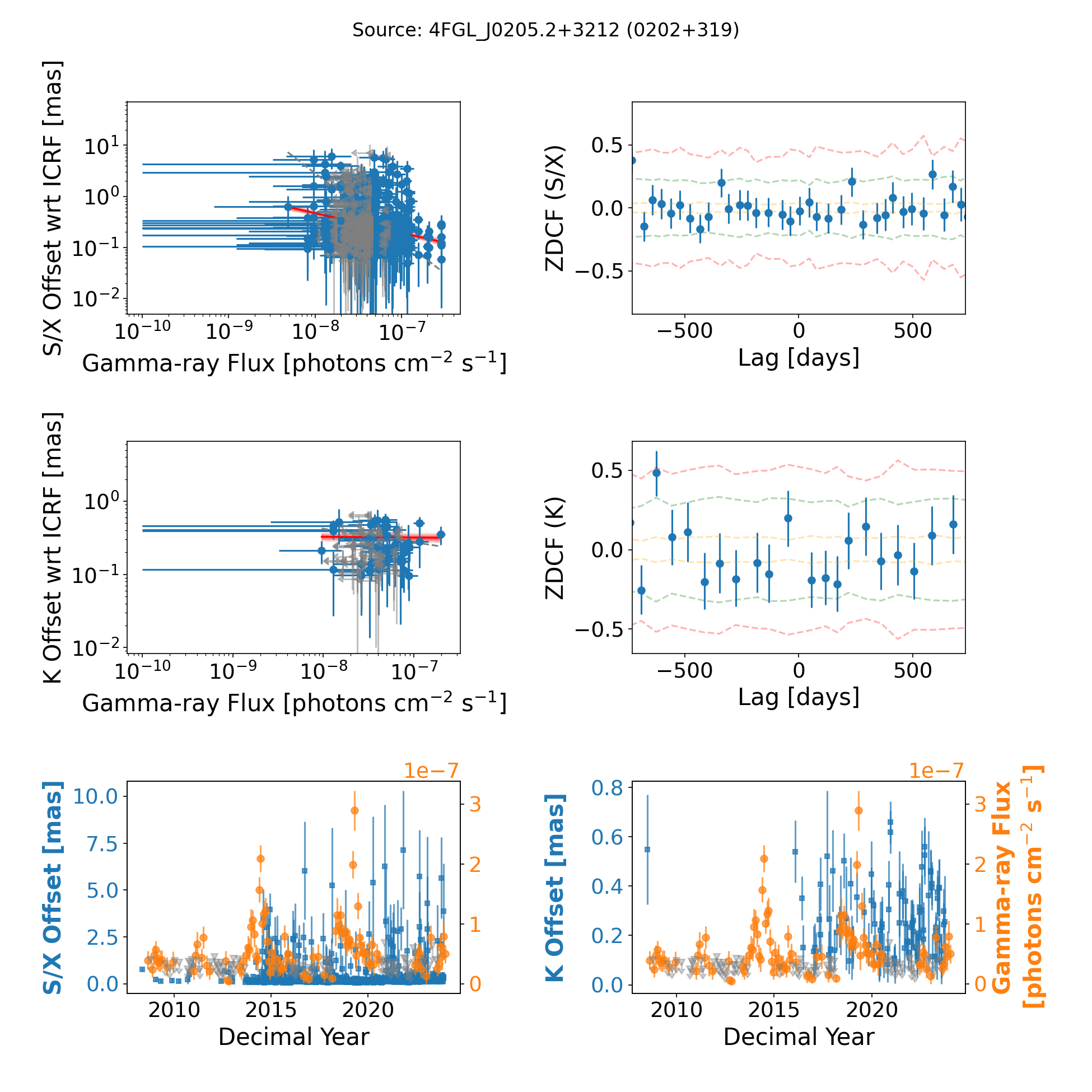}
    \caption{Same as Figure \ref{fig:correlations1}, but for the source 0202$+$319.}
    \label{fig:correlationsJ0205}
\end{figure}

\clearpage

\begin{figure}[ht!]
    \centering
    \includegraphics[width=\textwidth]{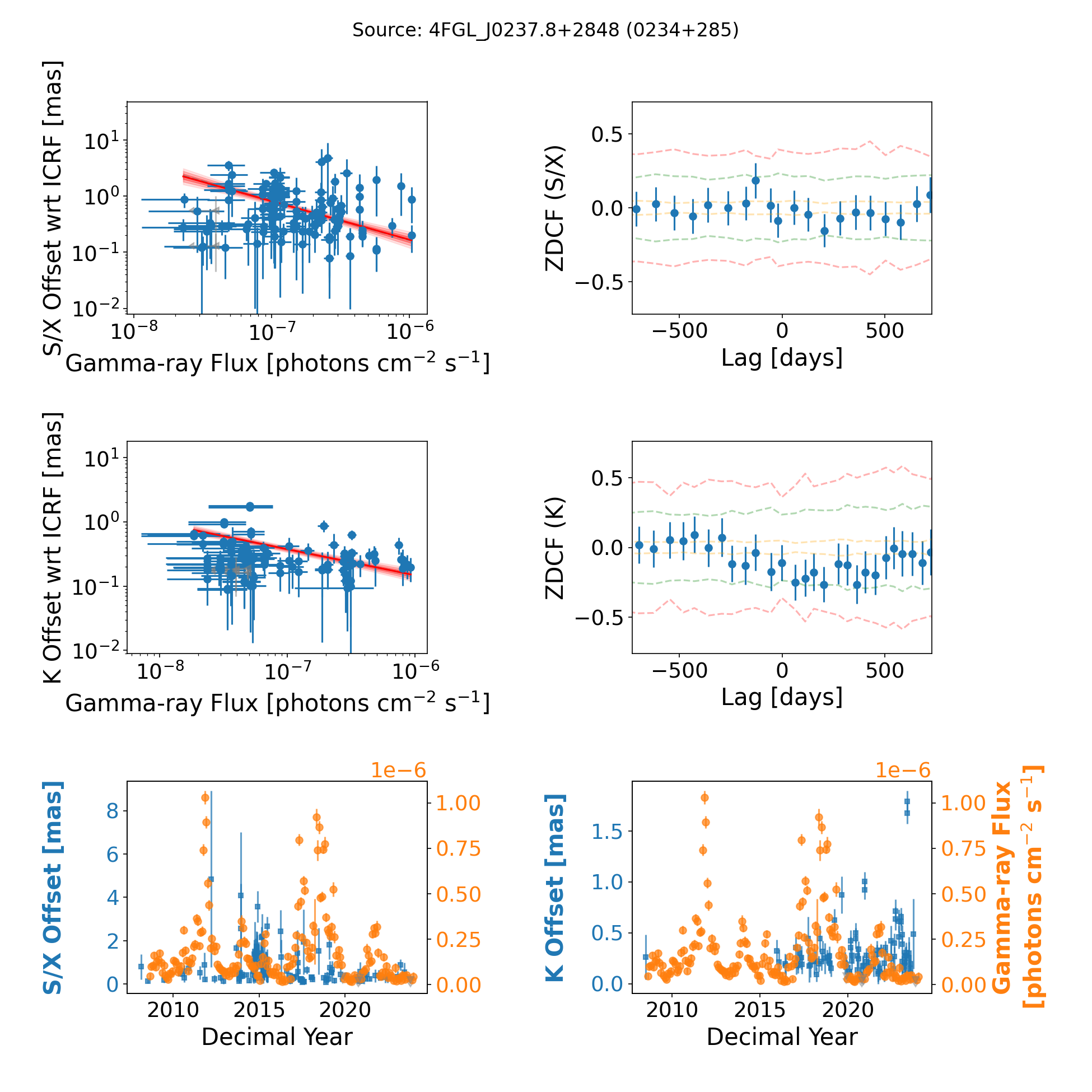}
    \caption{Same as Figure \ref{fig:correlations1}, but for the source 0234$+$285.}
    \label{fig:correlationsJ0237}
\end{figure}

\clearpage

\begin{figure}[ht!]
    \centering
    \includegraphics[width=\textwidth]{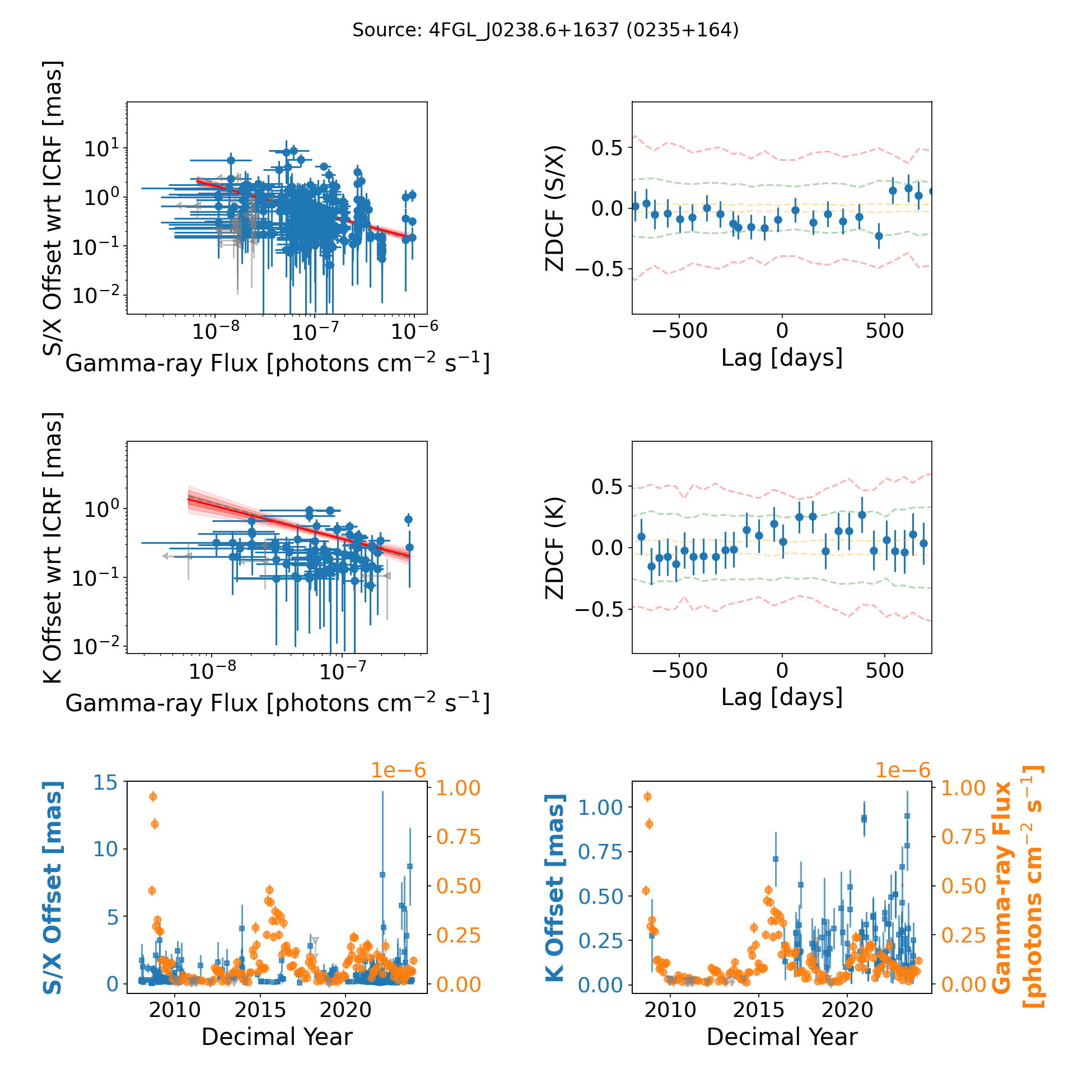}
    \caption{Same as Figure \ref{fig:correlations1}, but for the source 0235$+$164.}
    \label{fig:correlationsJ0238}
\end{figure}

\clearpage

\begin{figure}[ht!]
    \centering
    \includegraphics[width=\textwidth]{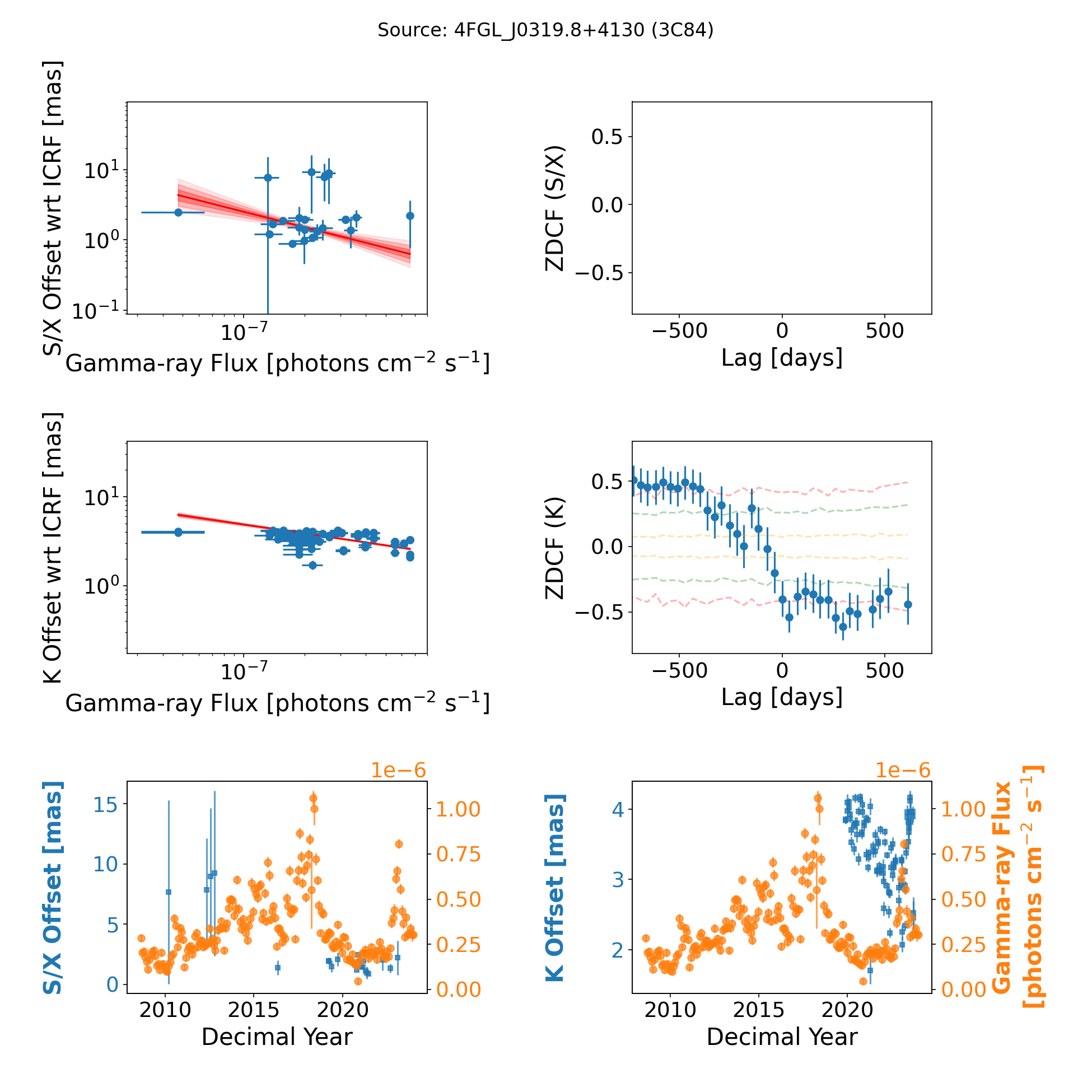}
    \caption{Same as Figure \ref{fig:correlations1}, but for the source 3C84.}
    \label{fig:correlationsJ0319}
\end{figure}

\clearpage

\begin{figure}[ht!]
    \centering
    \includegraphics[width=\textwidth]{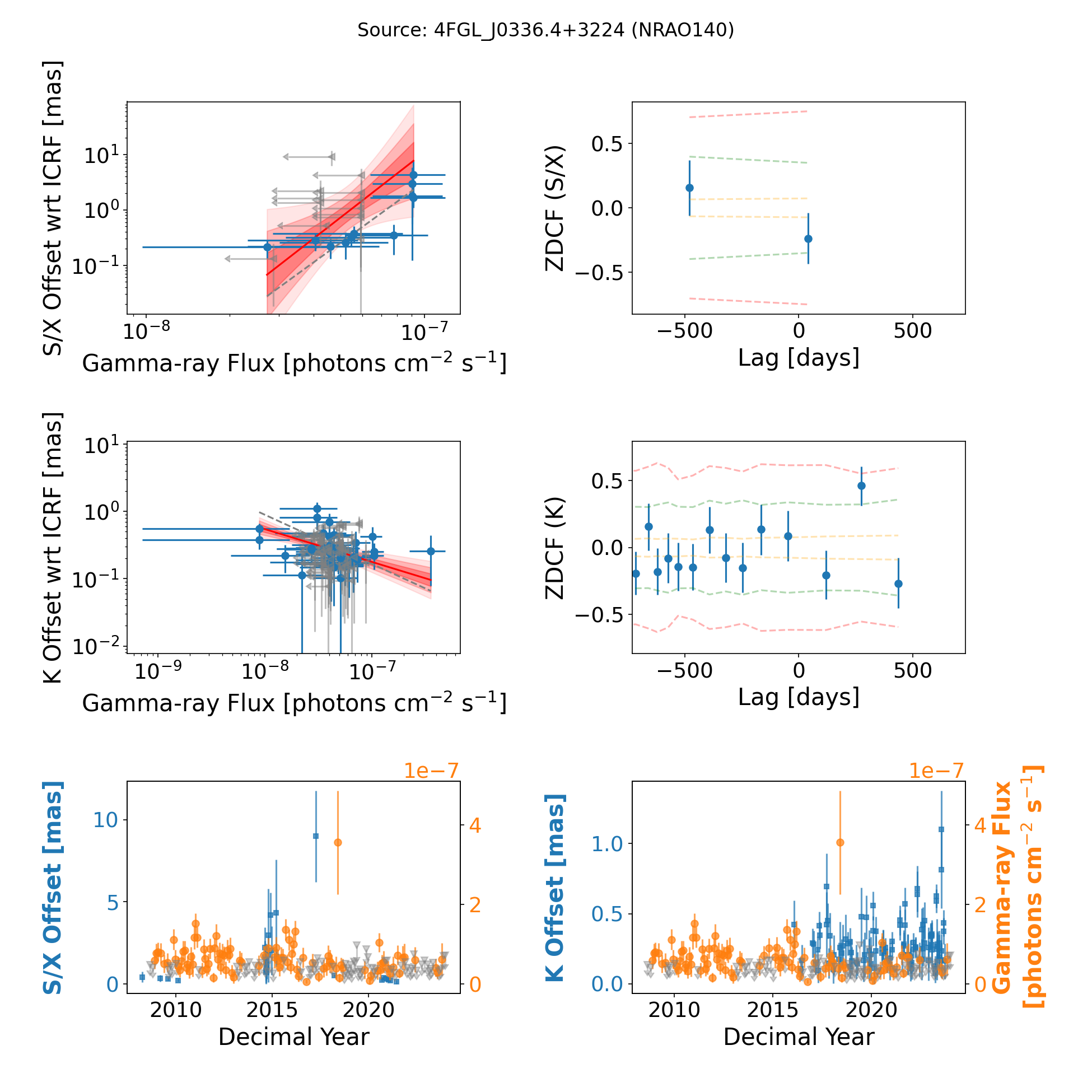}
    \caption{Same as Figure \ref{fig:correlations1}, but for the source NRAO140.}
    \label{fig:correlationsJ0336}
\end{figure}

\clearpage

\begin{figure}[ht!]
    \centering
    \includegraphics[width=\textwidth]{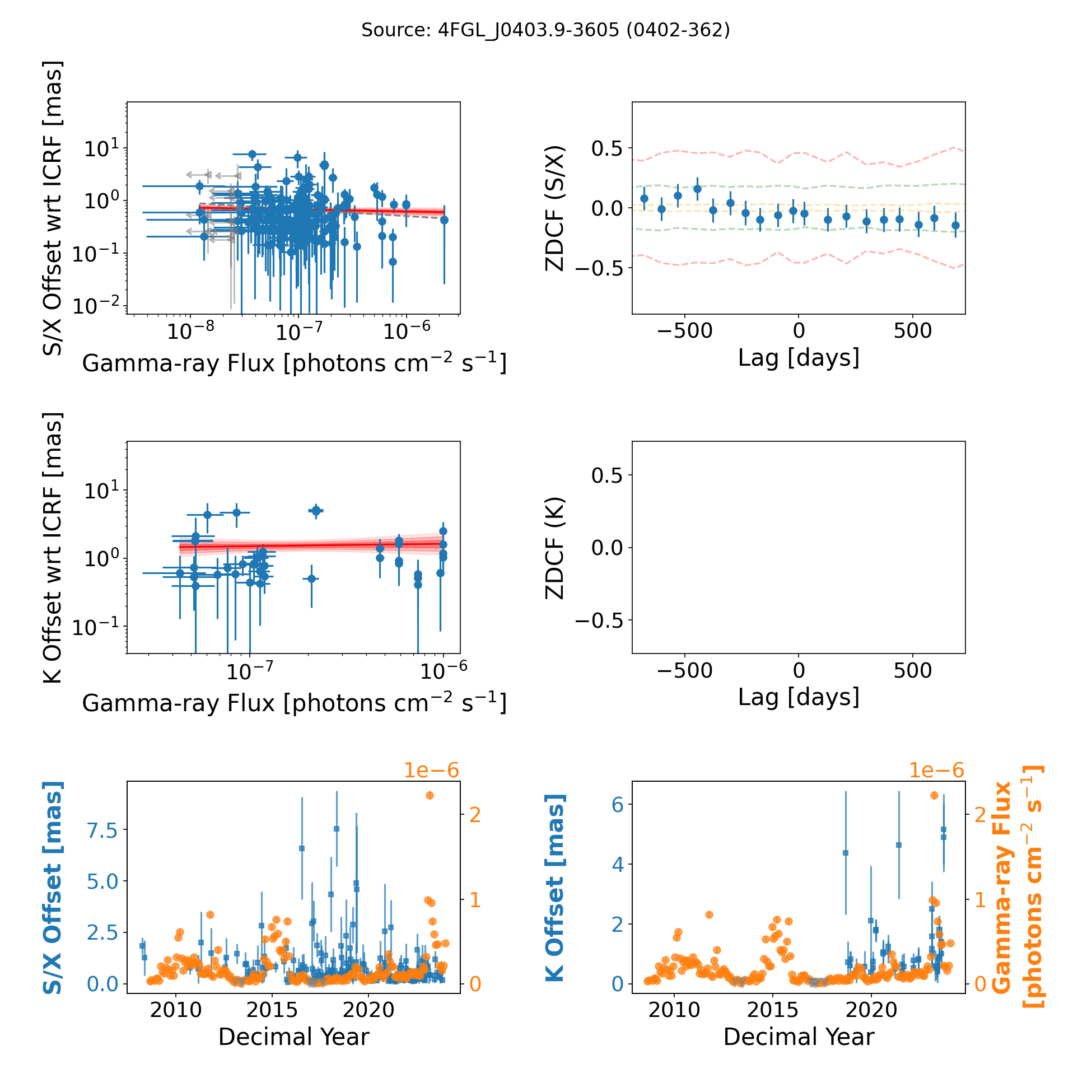}
    \caption{Same as Figure \ref{fig:correlations1}, but for the source 0402$-$362.}
    \label{fig:correlationsJ0403}
\end{figure}

\clearpage

\begin{figure}[ht!]
    \centering
    \includegraphics[width=\textwidth]{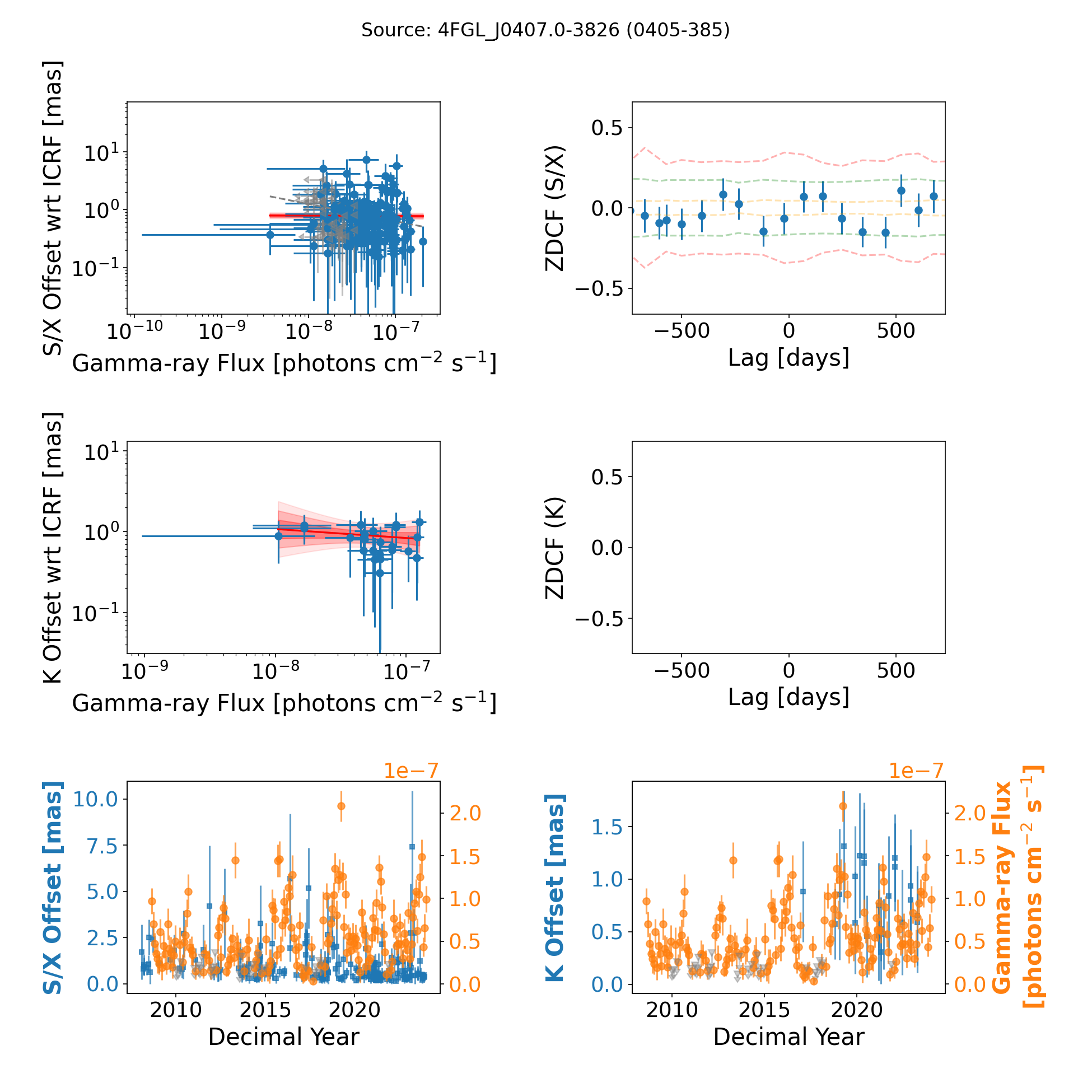}
    \caption{Same as Figure \ref{fig:correlations1}, but for the source 0405$-$385.}
    \label{fig:correlationsJ0407}
\end{figure}

\clearpage

\begin{figure}[ht!]
    \centering
    \includegraphics[width=\textwidth]{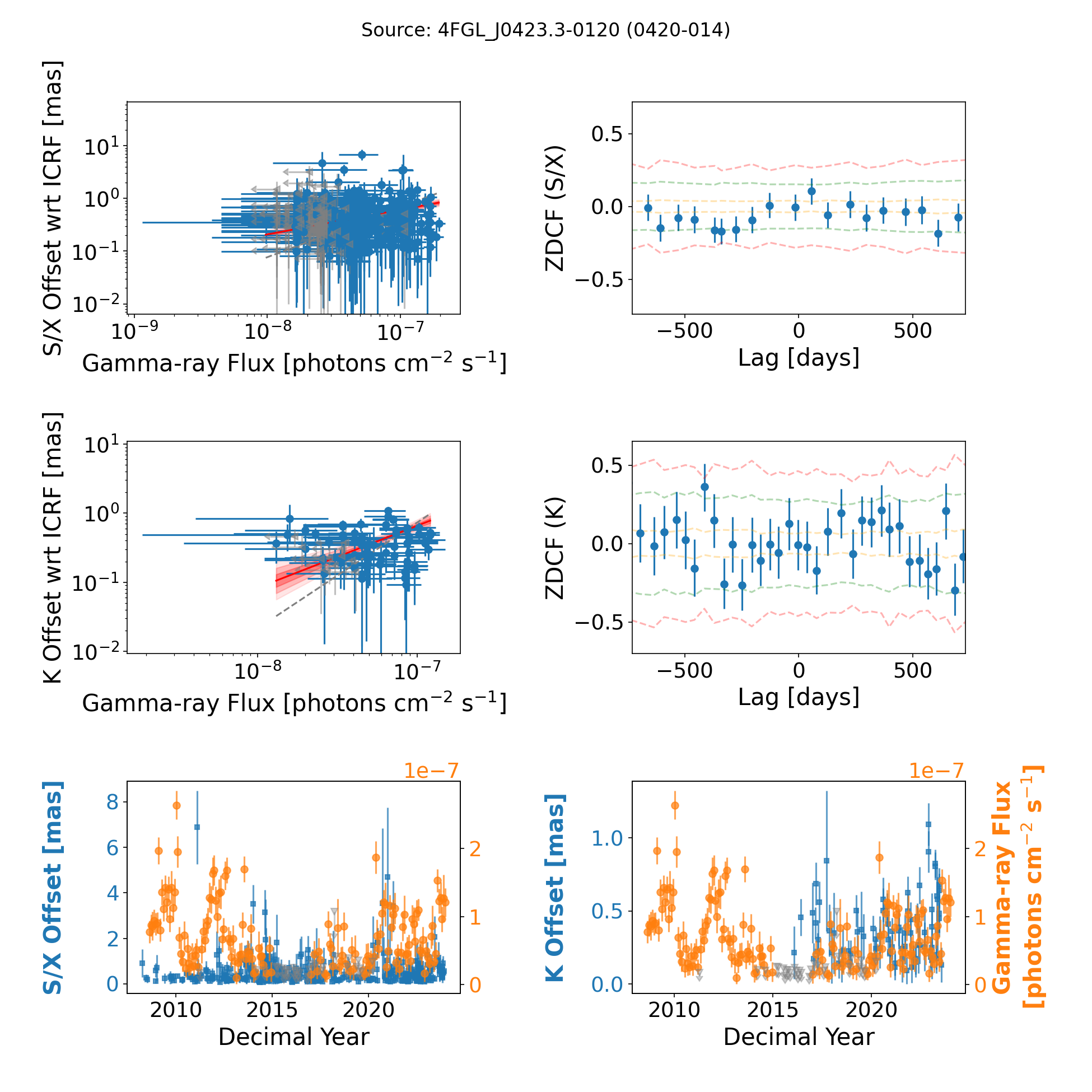}
    \caption{Same as Figure \ref{fig:correlations1}, but for the source 0420$-$014.}
    \label{fig:correlationsJ0423}
\end{figure}

\clearpage

\begin{figure}[ht!]
    \centering
    \includegraphics[width=\textwidth]{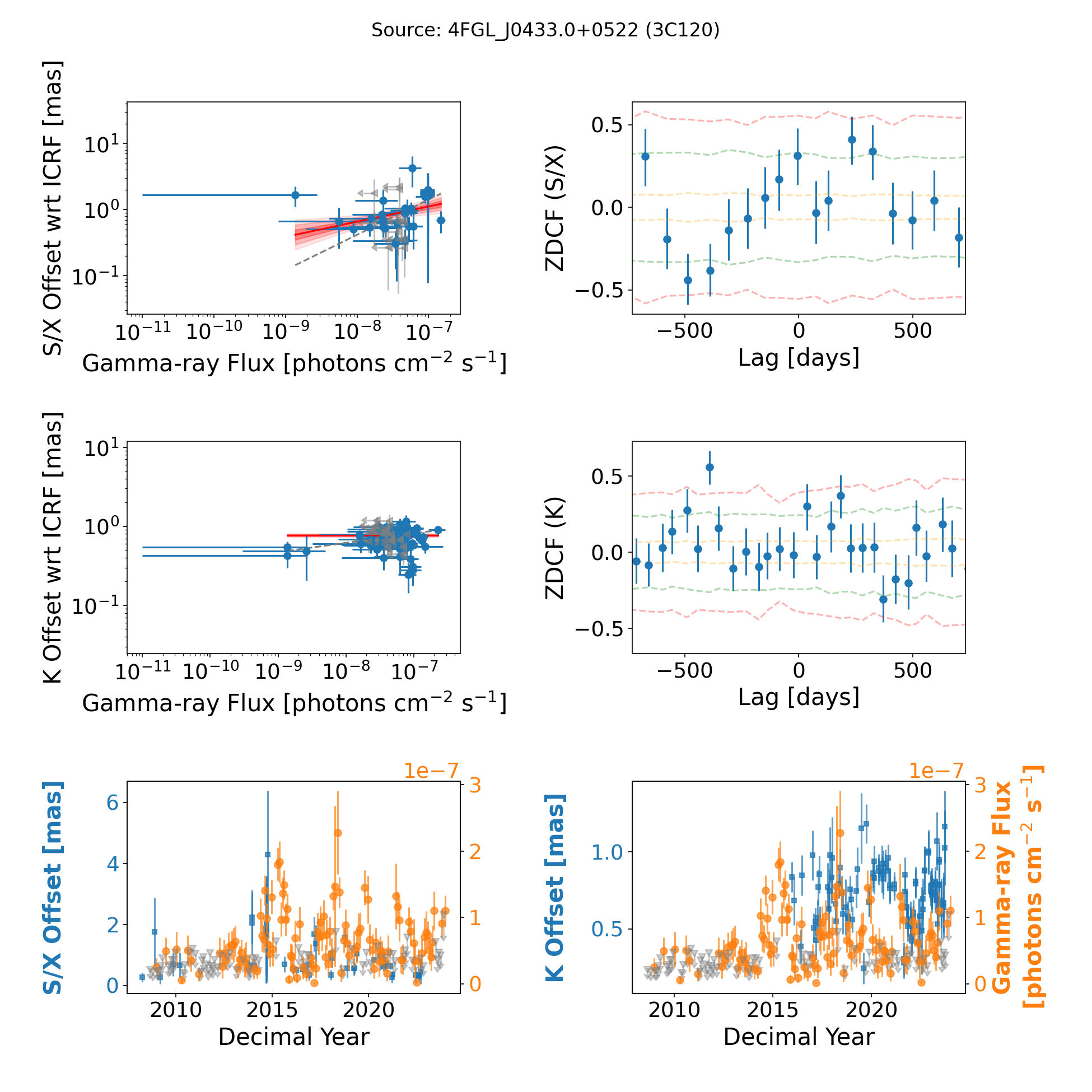}
    \caption{Same as Figure \ref{fig:correlations1}, but for the source 3C120.}
    \label{fig:correlationsJ0433}
\end{figure}

\clearpage

\begin{figure}[ht!]
    \centering
    \includegraphics[width=\textwidth]{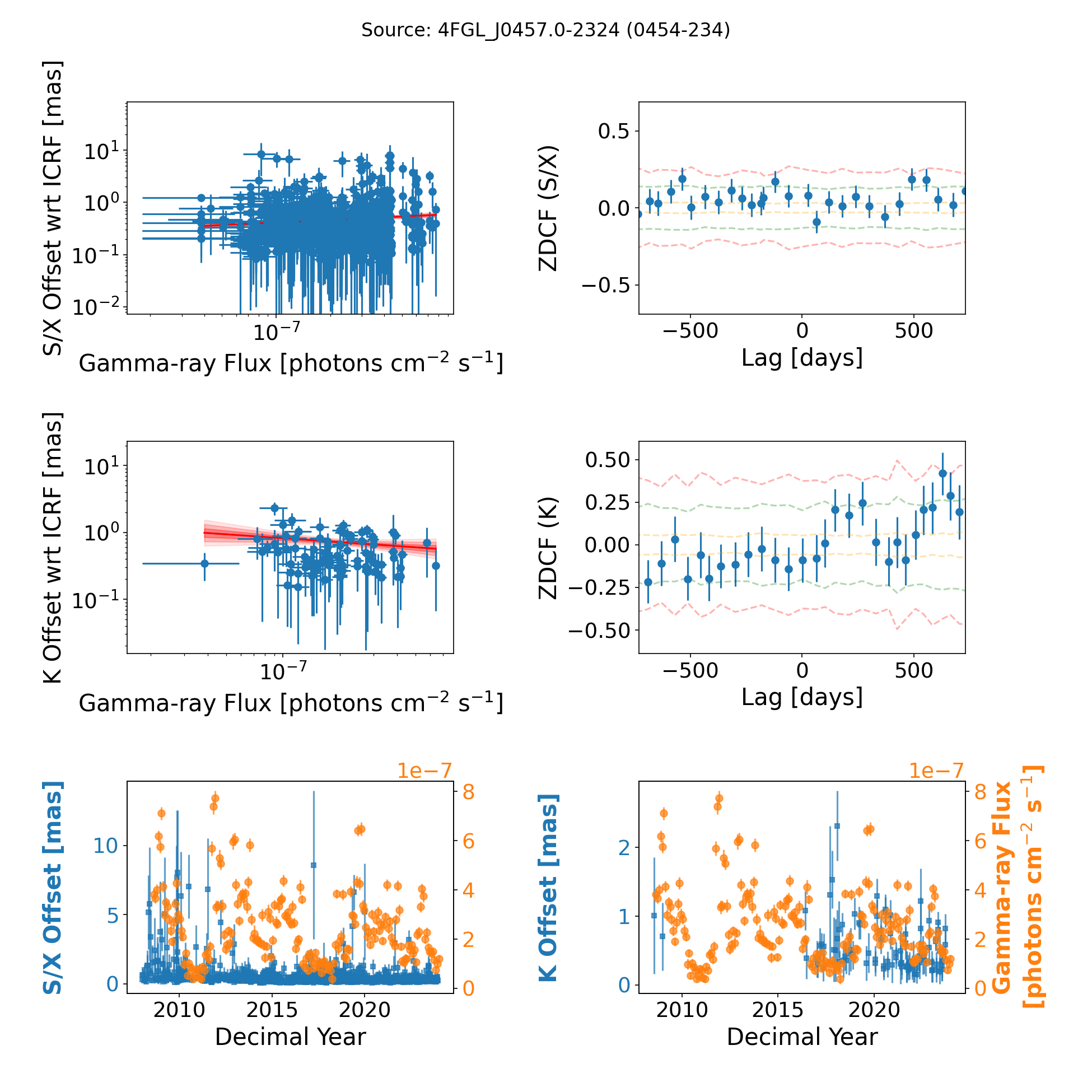}
    \caption{Same as Figure \ref{fig:correlations1}, but for the source 0454$-$234.}
    \label{fig:correlationsJ0457}
\end{figure}

\clearpage

\begin{figure}[ht!]
    \centering
    \includegraphics[width=\textwidth]{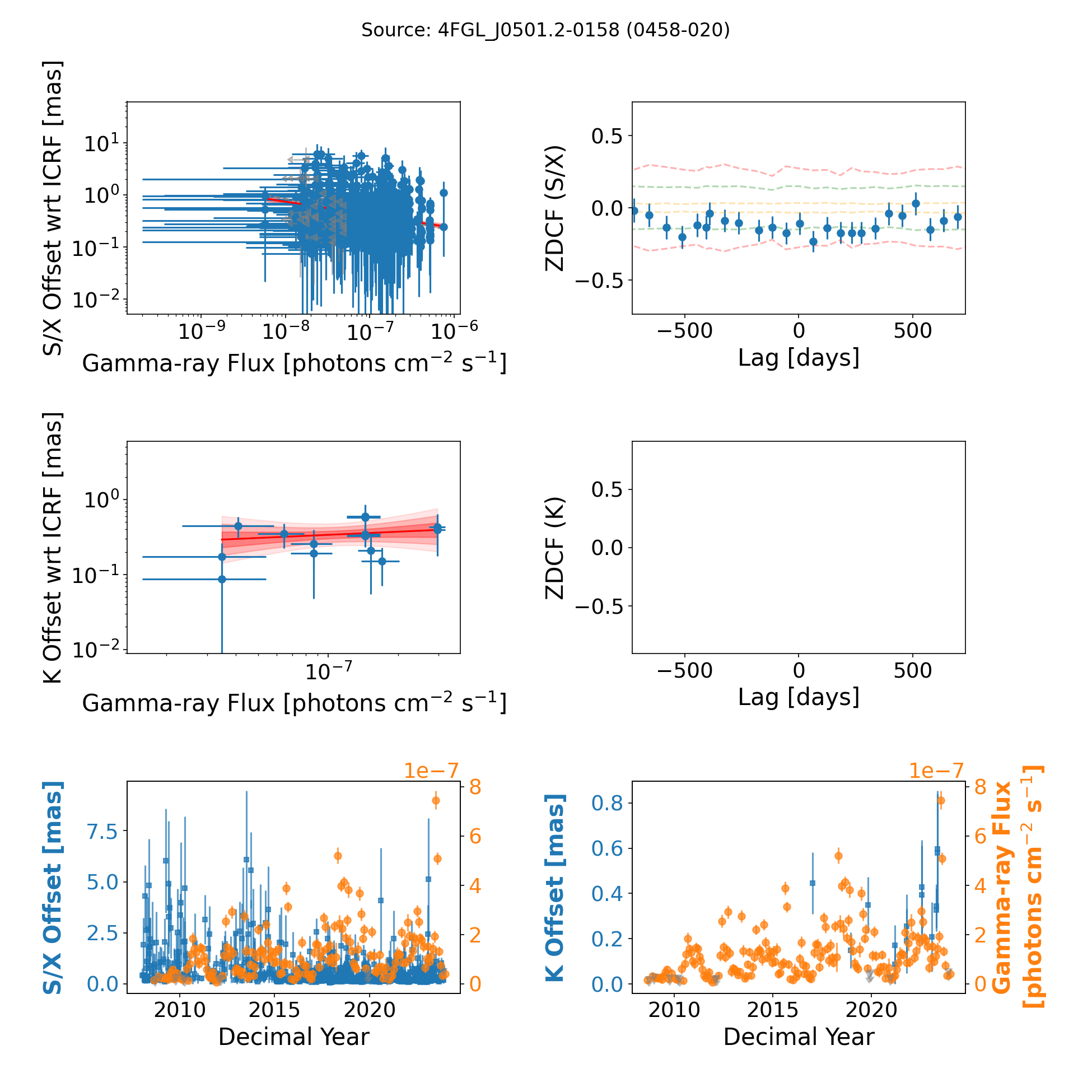}
    \caption{Same as Figure \ref{fig:correlations1}, but for the source 0458$-$020.}
    \label{fig:correlationsJ0501}
\end{figure}

\clearpage

\begin{figure}[ht!]
    \centering
    \includegraphics[width=\textwidth]{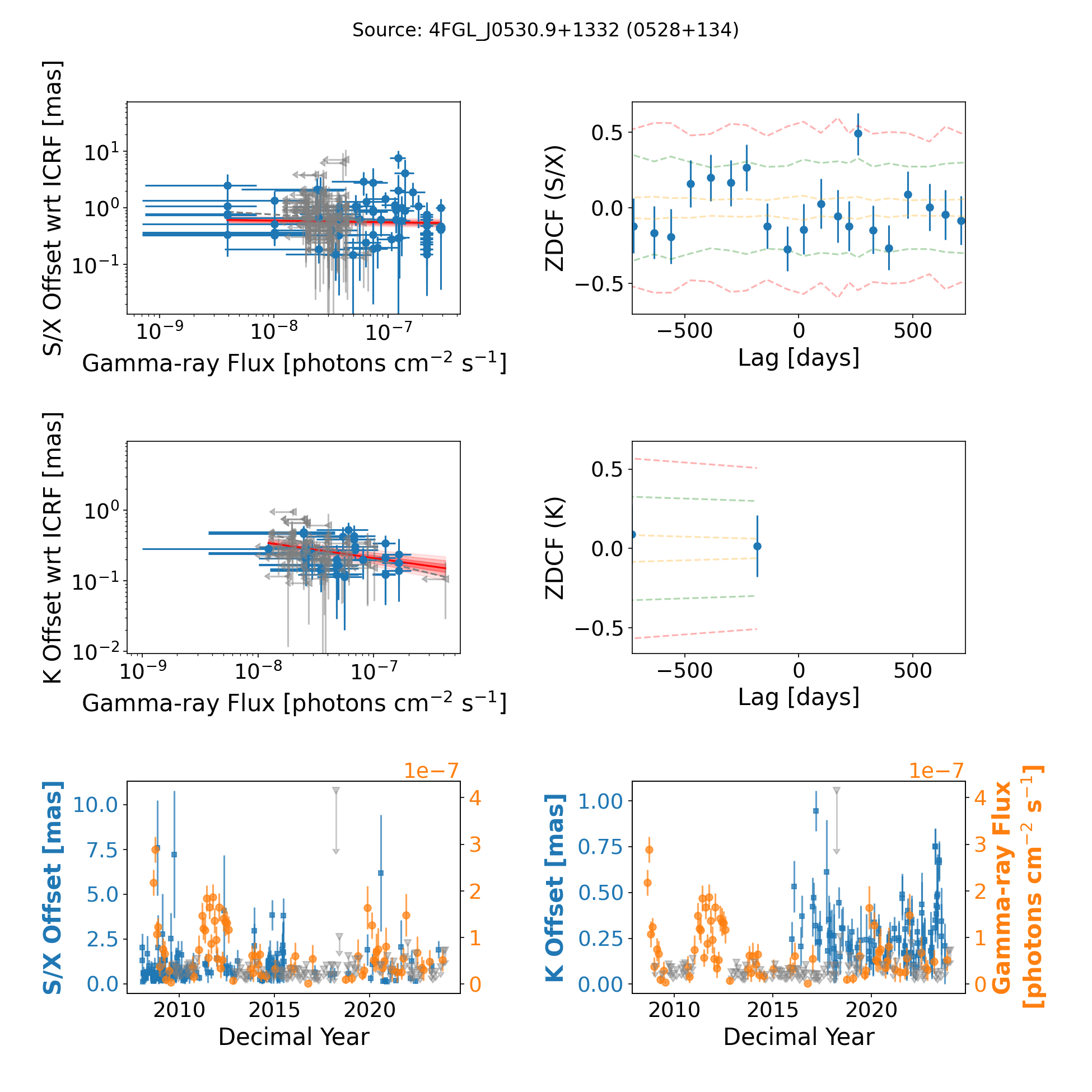}
    \caption{Same as Figure \ref{fig:correlations1}, but for the source 0528$+$134.}
    \label{fig:correlationsJ0530}
\end{figure}

\clearpage

\begin{figure}[ht!]
    \centering
    \includegraphics[width=\textwidth]{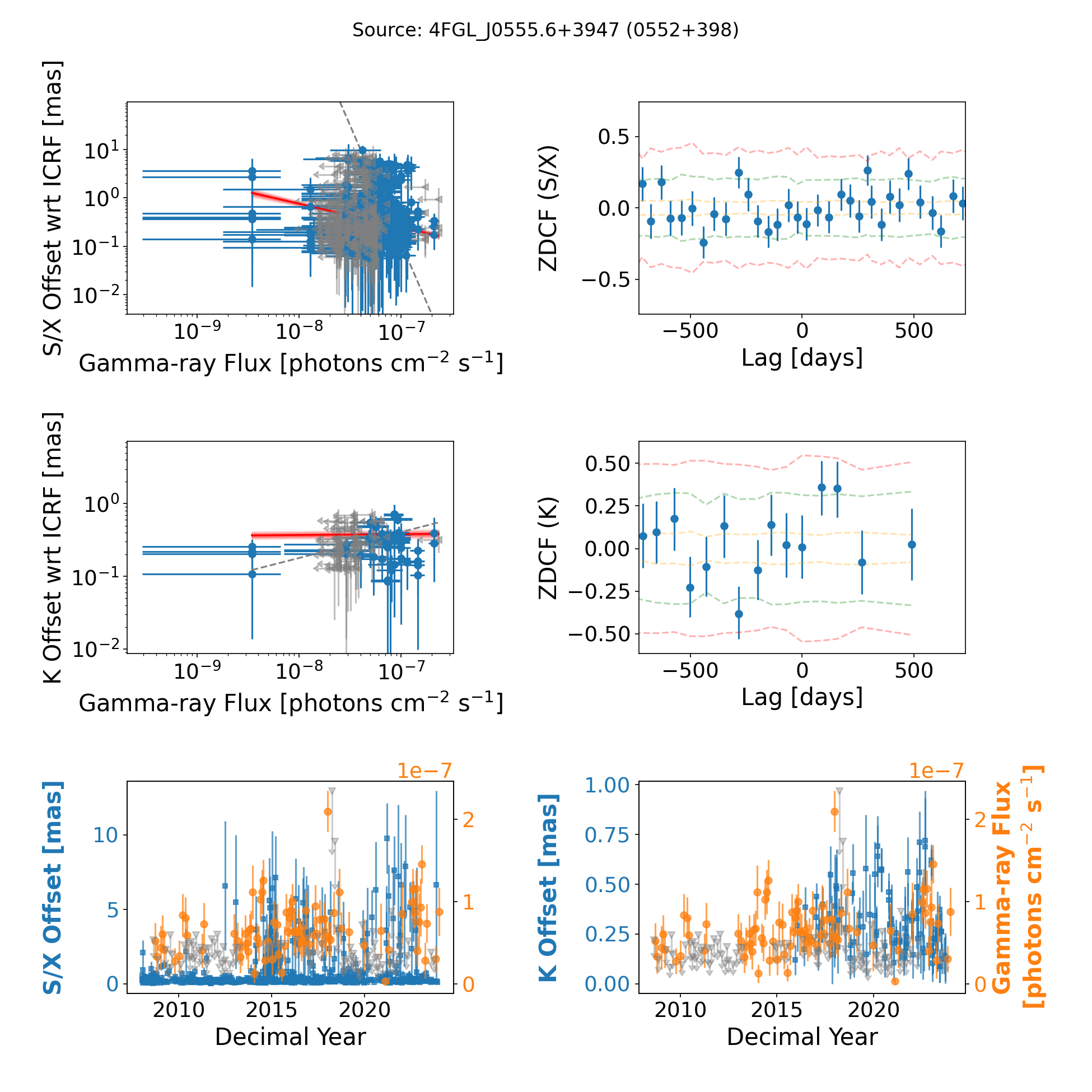}
    \caption{Same as Figure \ref{fig:correlations1}, but for the source 0552$+$398.}
    \label{fig:correlationsJ0555}
\end{figure}

\clearpage

\begin{figure}[ht!]
    \centering
    \includegraphics[width=\textwidth]{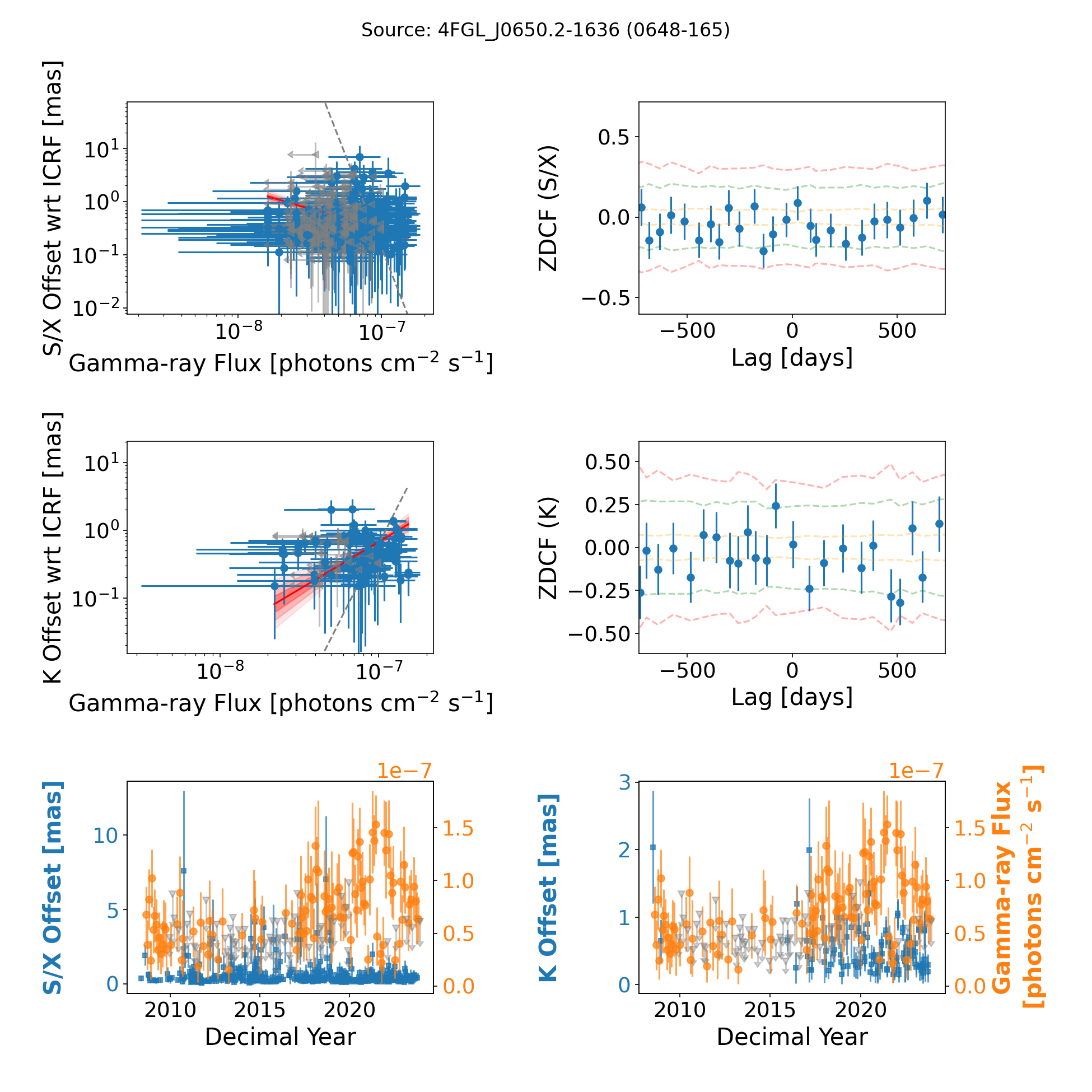}
    \caption{Same as Figure \ref{fig:correlations1}, but for the source 0648$-$165.}
    \label{fig:correlationsJ0650}
\end{figure}

\clearpage

\begin{figure}[ht!]
    \centering
    \includegraphics[width=\textwidth]{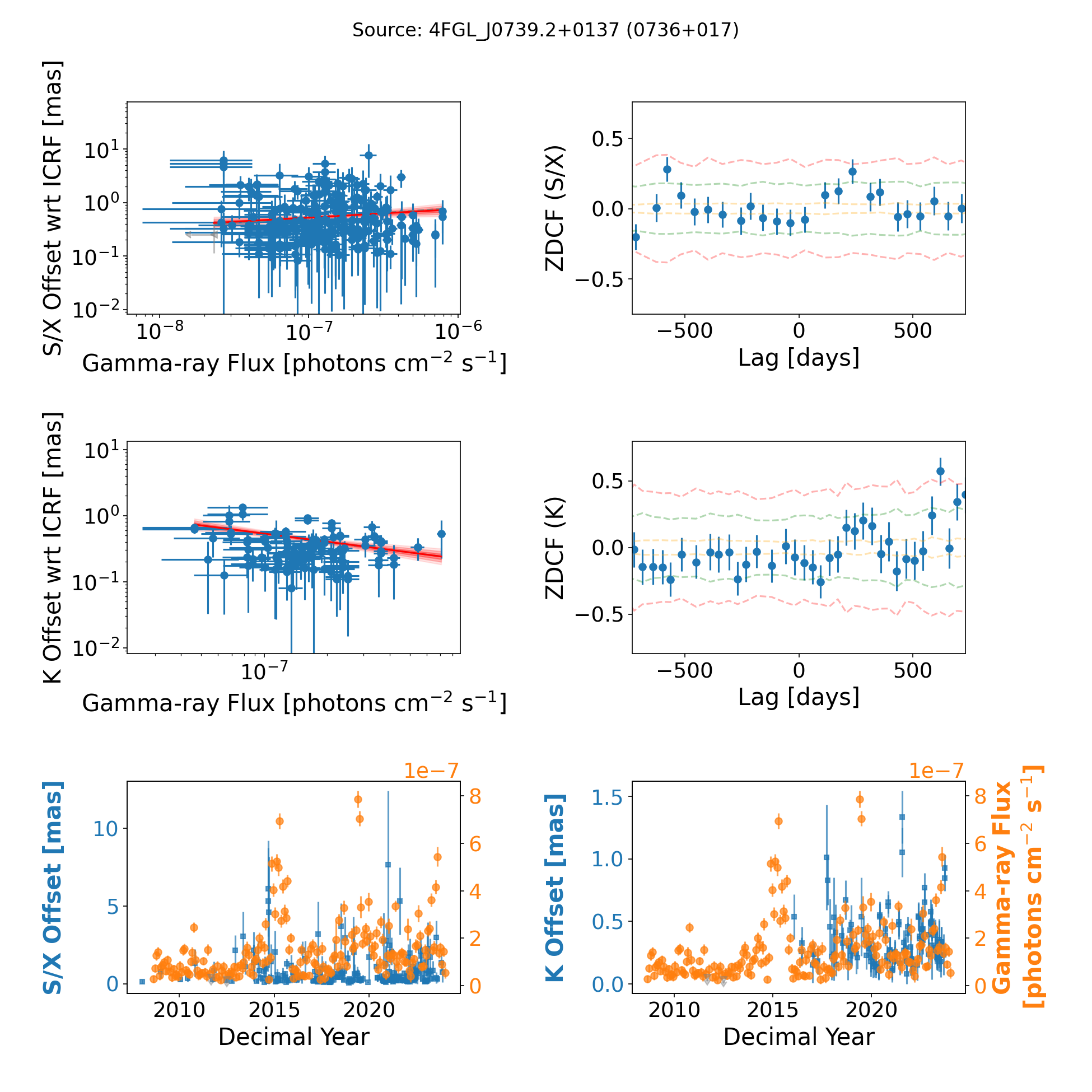}
    \caption{Same as Figure \ref{fig:correlations1}, but for the source 0736$+$017.}
    \label{fig:correlationsJ0739}
\end{figure}

\clearpage

\begin{figure}[ht!]
    \centering
    \includegraphics[width=\textwidth]{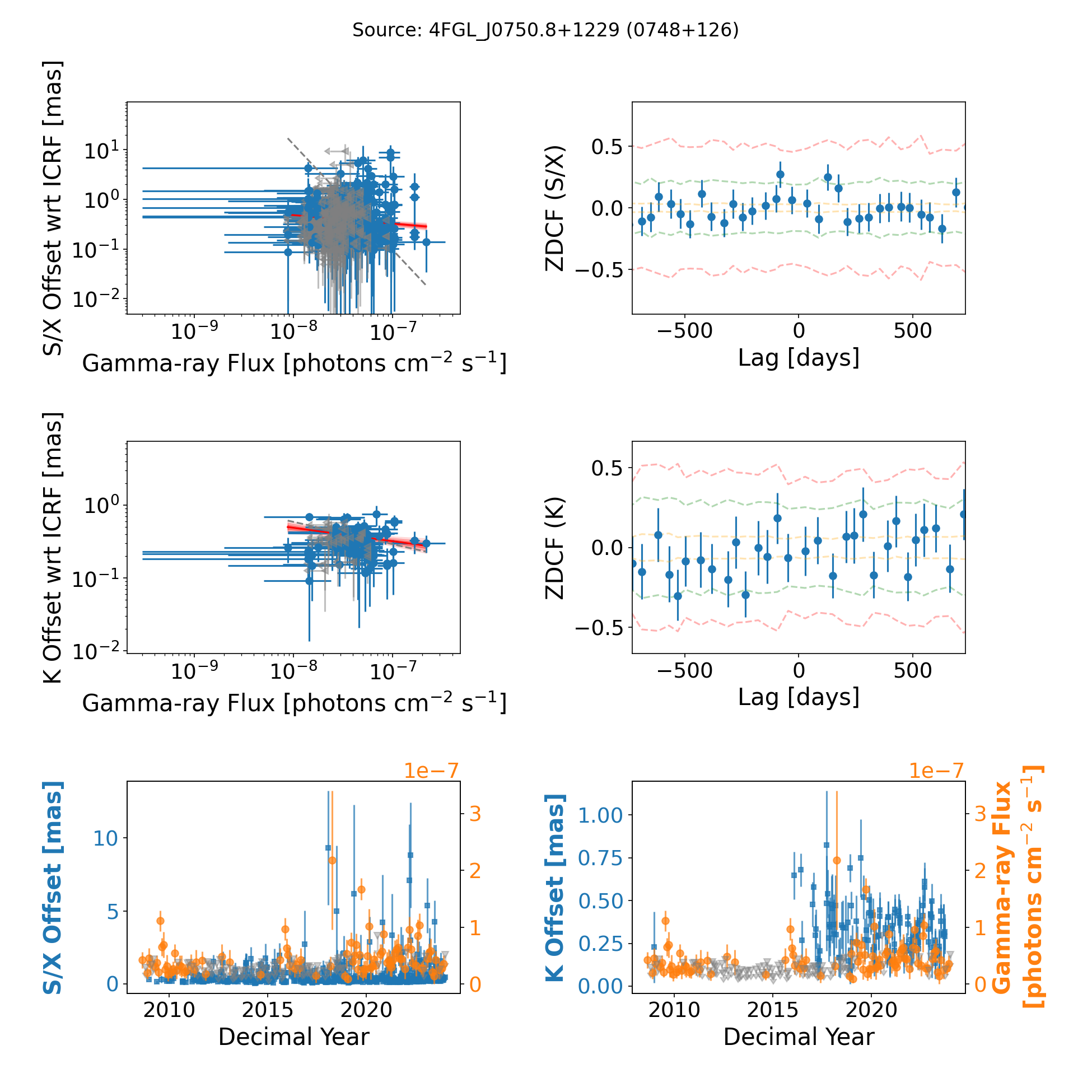}
    \caption{Same as Figure \ref{fig:correlations1}, but for the source 0748$+$126.}
    \label{fig:correlationsJ0750}
\end{figure}

\clearpage

\begin{figure}[ht!]
    \centering
    \includegraphics[width=\textwidth]{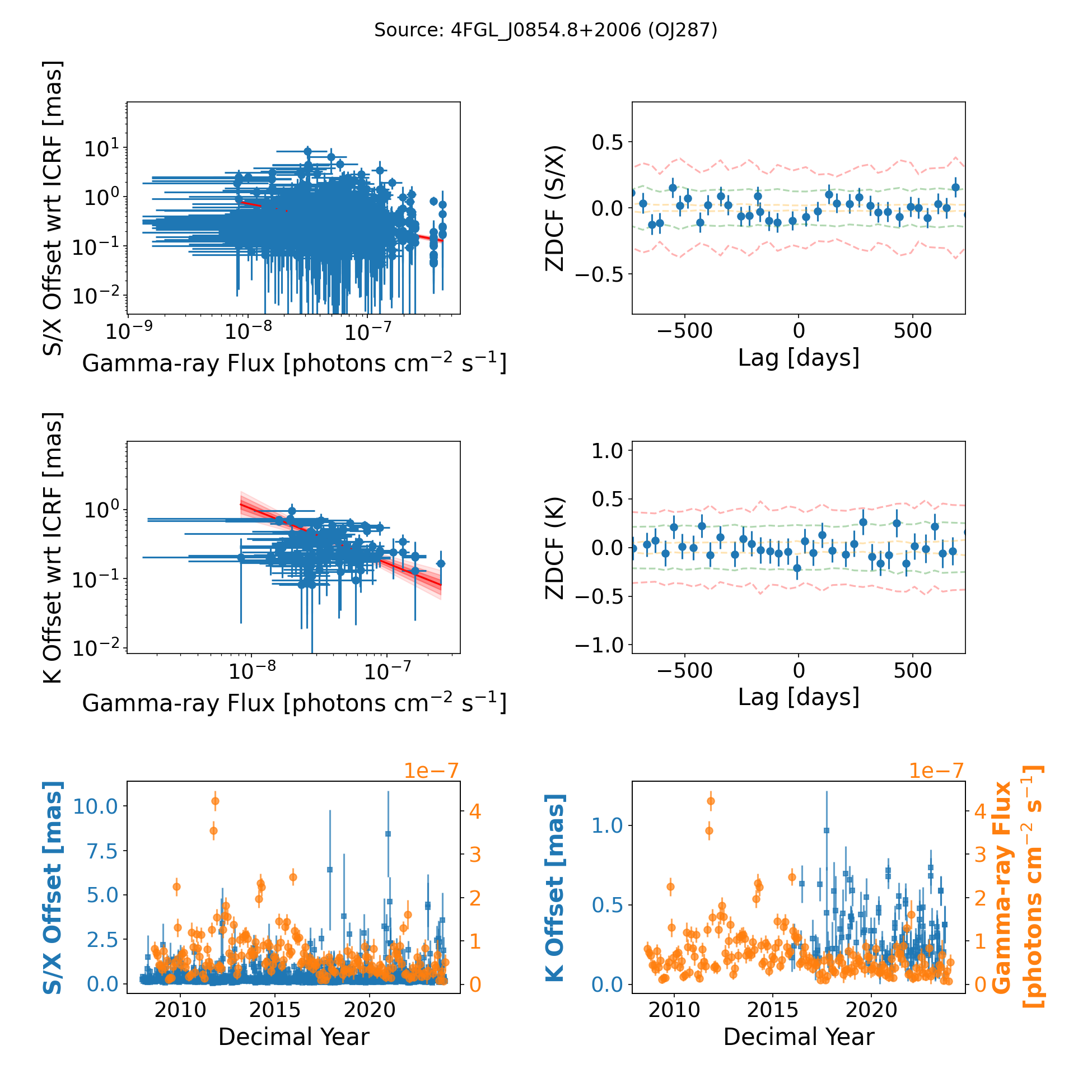}
    \caption{Same as Figure \ref{fig:correlations1}, but for the source OJ287.}
    \label{fig:correlationsJ0854}
\end{figure}

\clearpage

\begin{figure}[ht!]
    \centering
    \includegraphics[width=\textwidth]{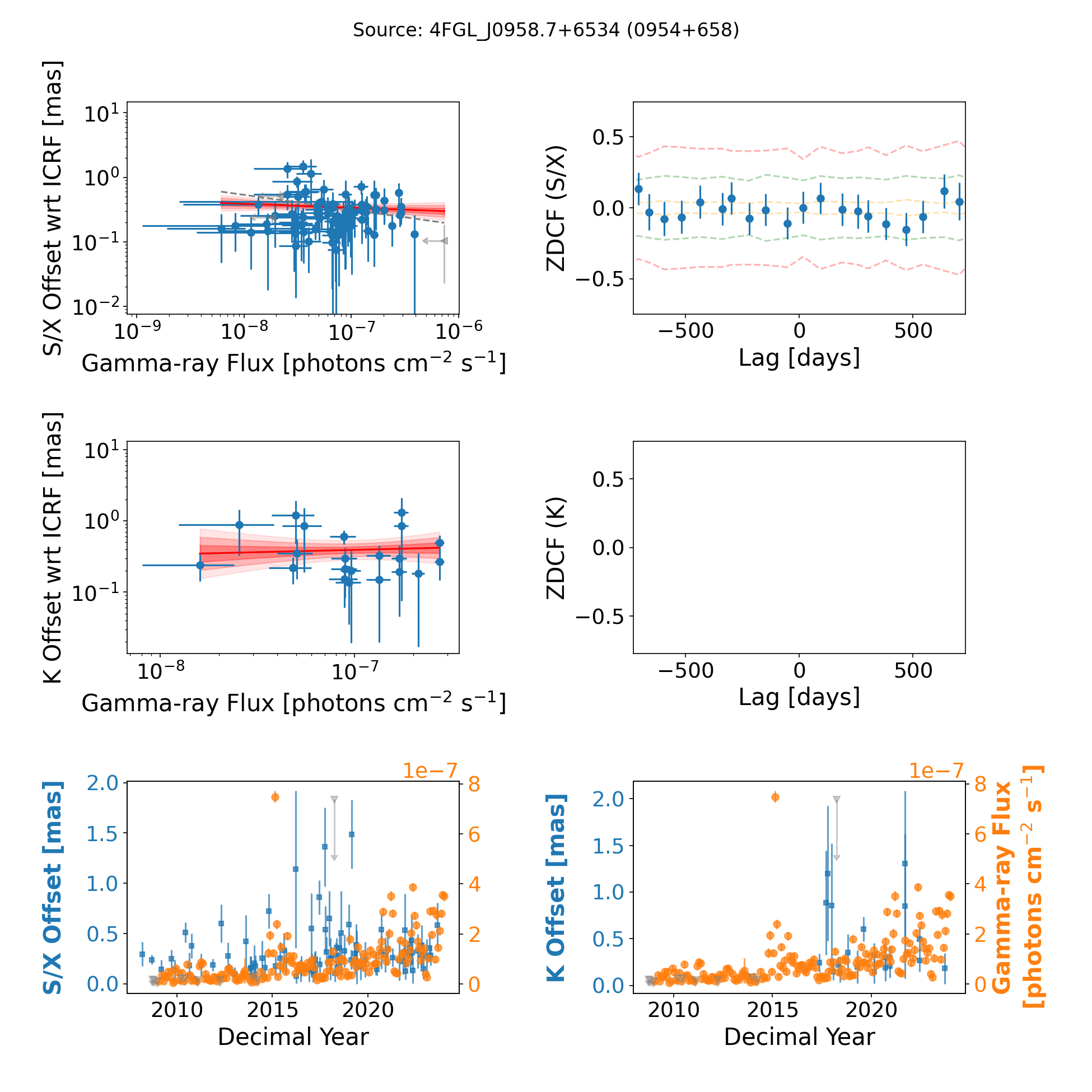}
    \caption{Same as Figure \ref{fig:correlations1}, but for the source 0954$+$658.}
    \label{fig:correlationsJ0958}
\end{figure}

\clearpage

\begin{figure}[ht!]
    \centering
    \includegraphics[width=\textwidth]{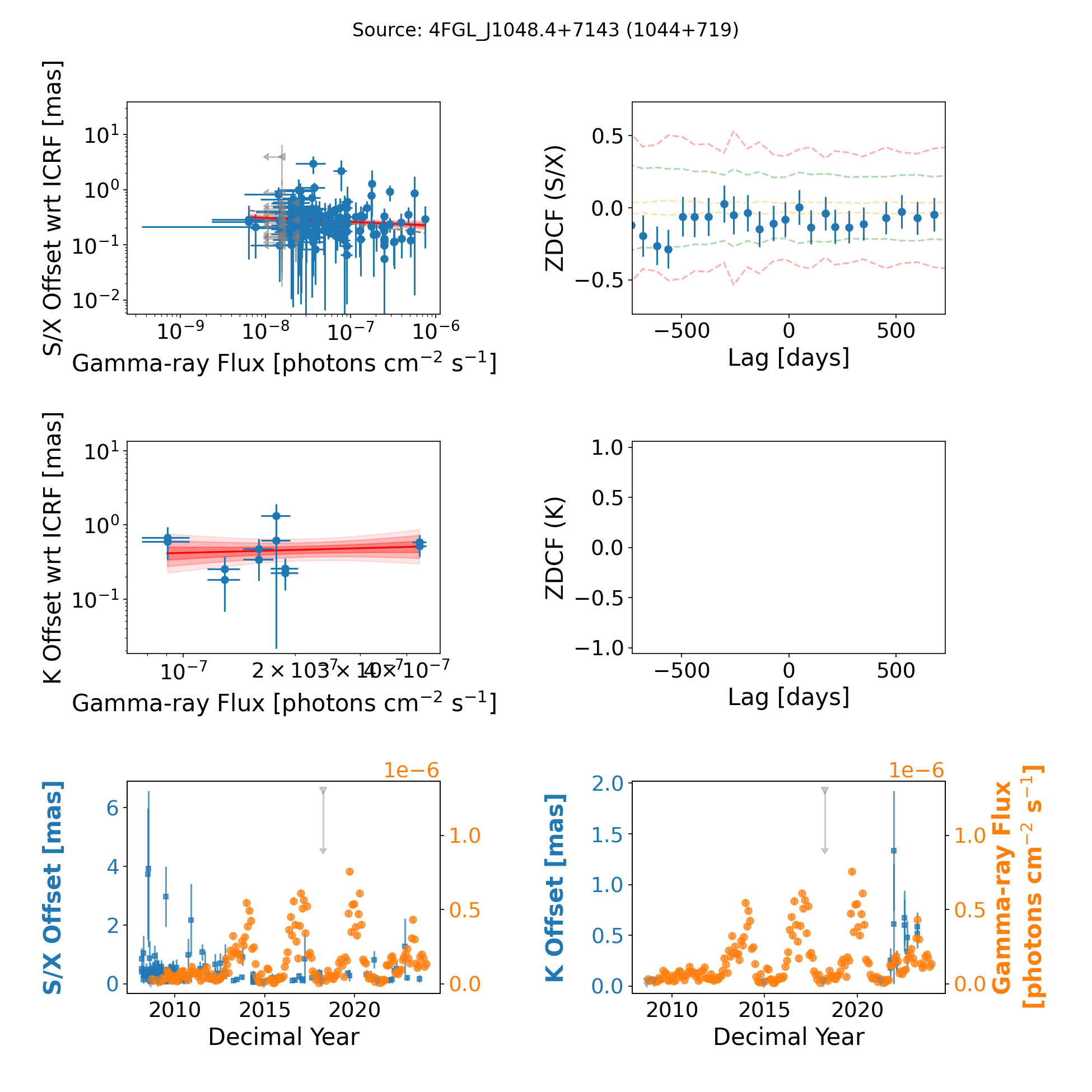}
    \caption{Same as Figure \ref{fig:correlations1}, but for the source 1044$+$719.}
    \label{fig:correlationsJ1048}
\end{figure}

\clearpage

\begin{figure}[ht!]
    \centering
    \includegraphics[width=\textwidth]{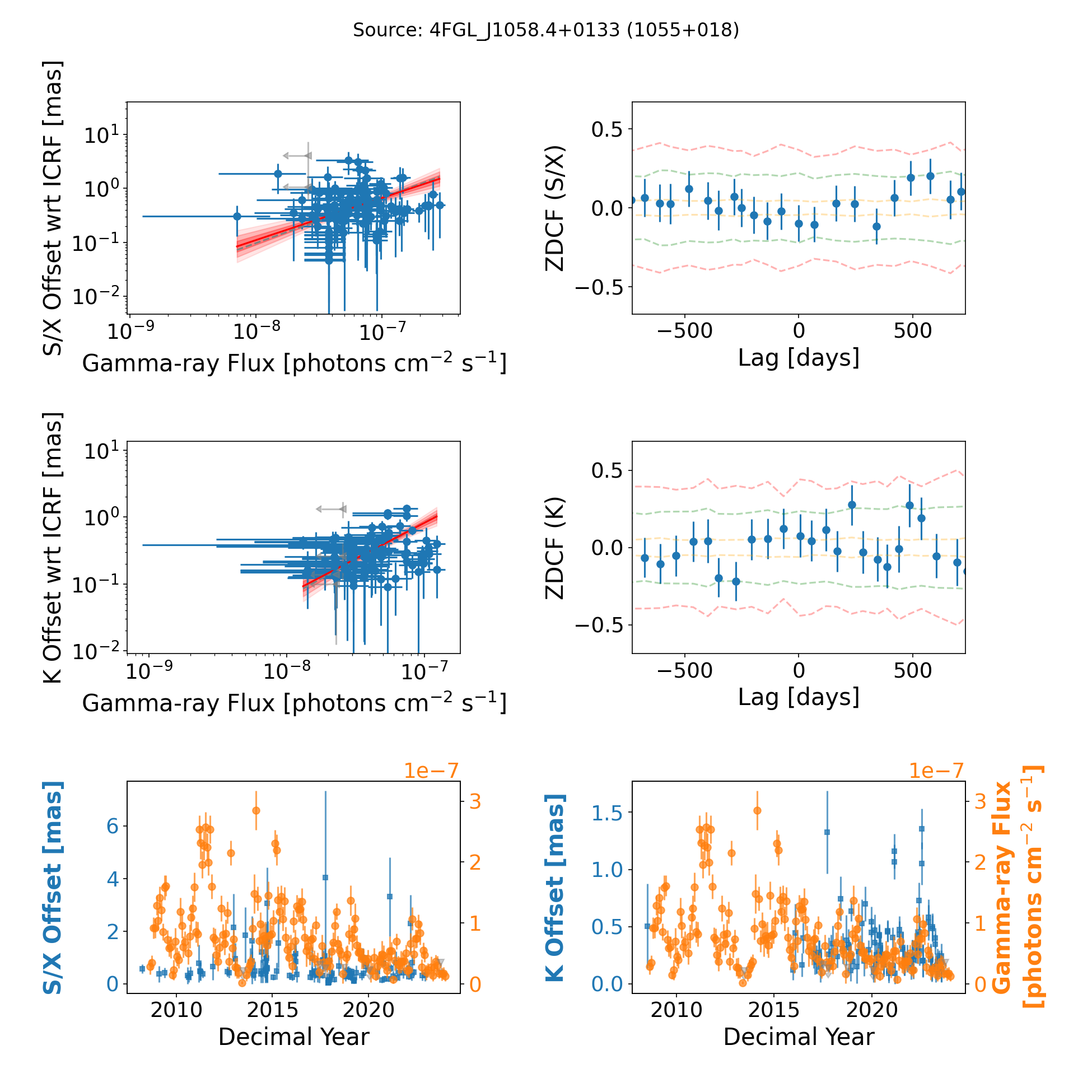}
    \caption{Same as Figure \ref{fig:correlations1}, but for the source 1055$+$018.}
    \label{fig:correlationsJ1058}
\end{figure}

\clearpage

\begin{figure}[ht!]
    \centering
    \includegraphics[width=\textwidth]{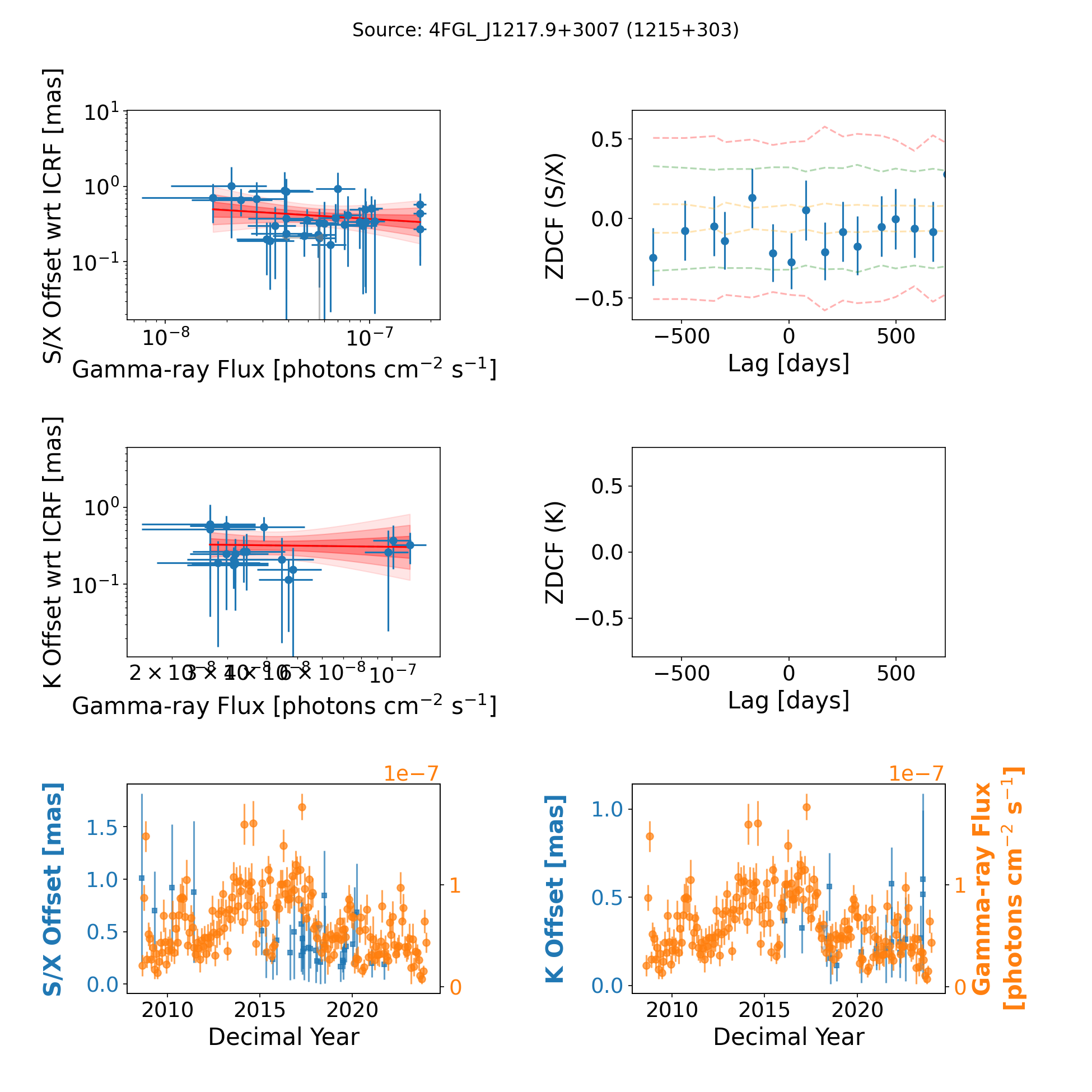}
    \caption{Same as Figure \ref{fig:correlations1}, but for the source 1215$+$303.}
    \label{fig:correlationsJ1217}
\end{figure}

\clearpage

\begin{figure}[ht!]
    \centering
    \includegraphics[width=\textwidth]{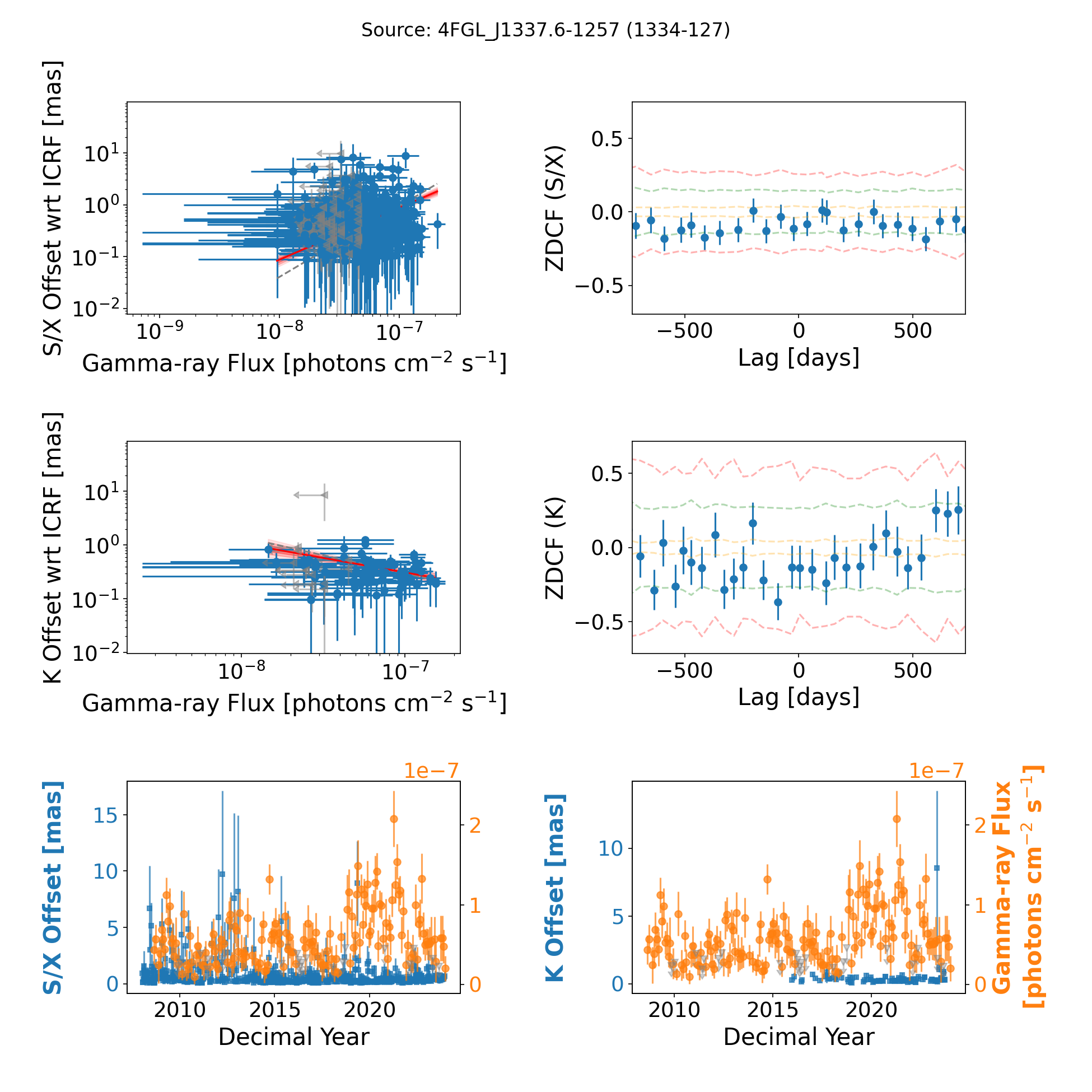}
    \caption{Same as Figure \ref{fig:correlations1}, but for the source 1334$-$127.}
    \label{fig:correlationsJ1337}
\end{figure}

\clearpage

\begin{figure}[ht!]
    \centering
    \includegraphics[width=\textwidth]{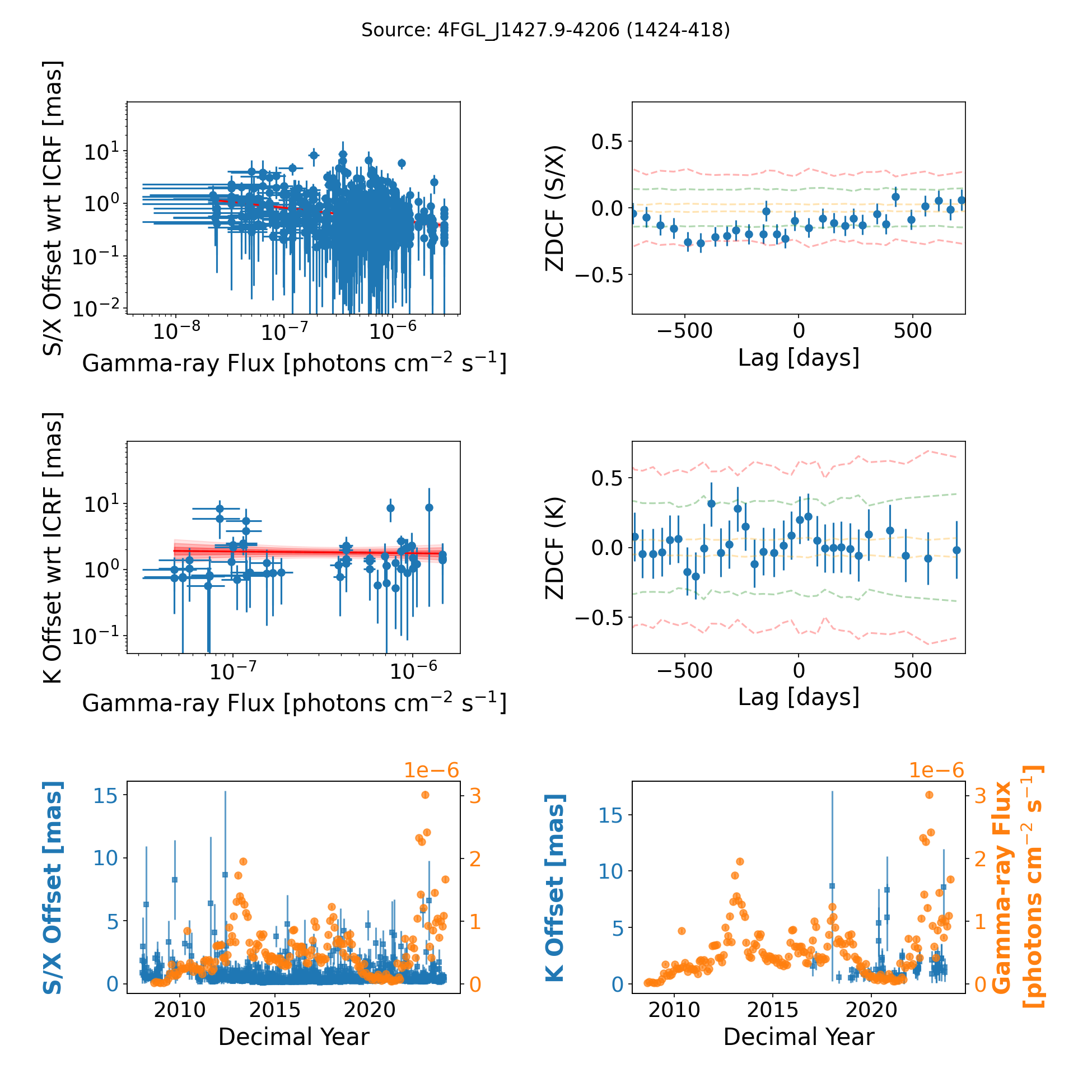}
    \caption{Same as Figure \ref{fig:correlations1}, but for the source 1424$-$418.}
    \label{fig:correlationsJ1427}
\end{figure}

\clearpage

\begin{figure}[ht!]
    \centering
    \includegraphics[width=\textwidth]{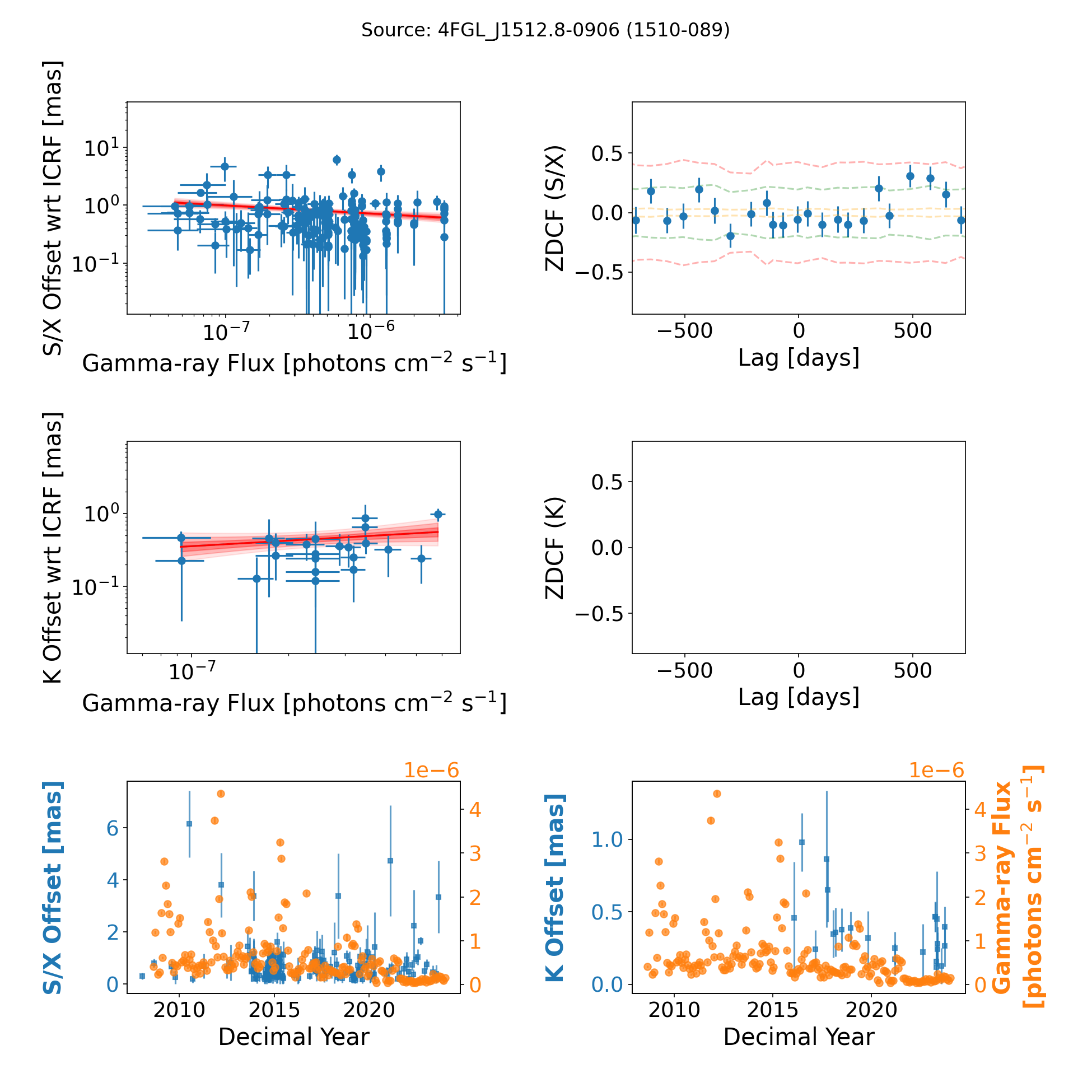}
    \caption{Same as Figure \ref{fig:correlations1}, but for the source 1510$-$089.}
    \label{fig:correlationsJ1512}
\end{figure}

\clearpage

\begin{figure}[ht!]
    \centering
    \includegraphics[width=\textwidth]{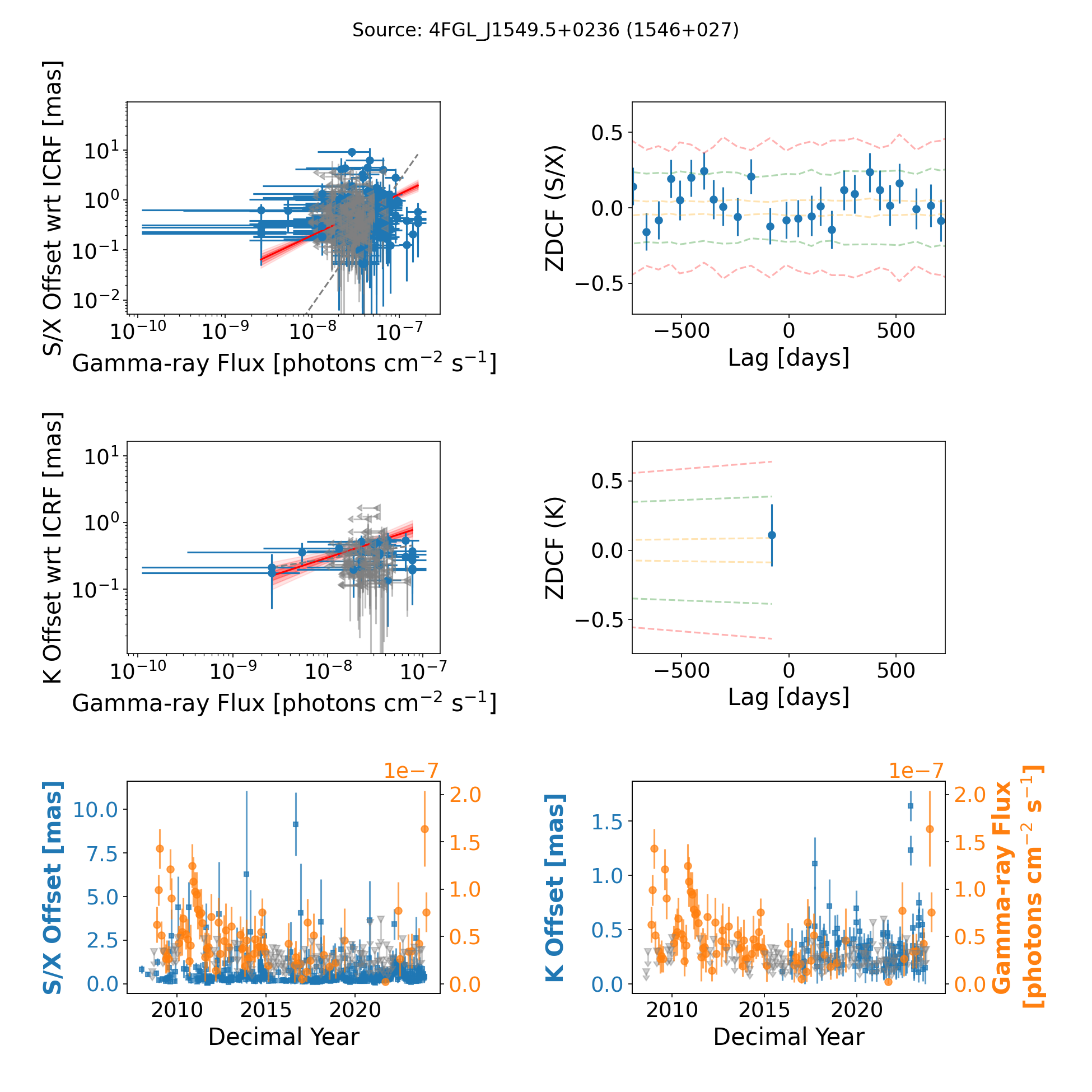}
    \caption{Same as Figure \ref{fig:correlations1}, but for the source 1546$+$027.}
    \label{fig:correlationsJ1549}
\end{figure}

\clearpage

\begin{figure}[ht!]
    \centering
    \includegraphics[width=\textwidth]{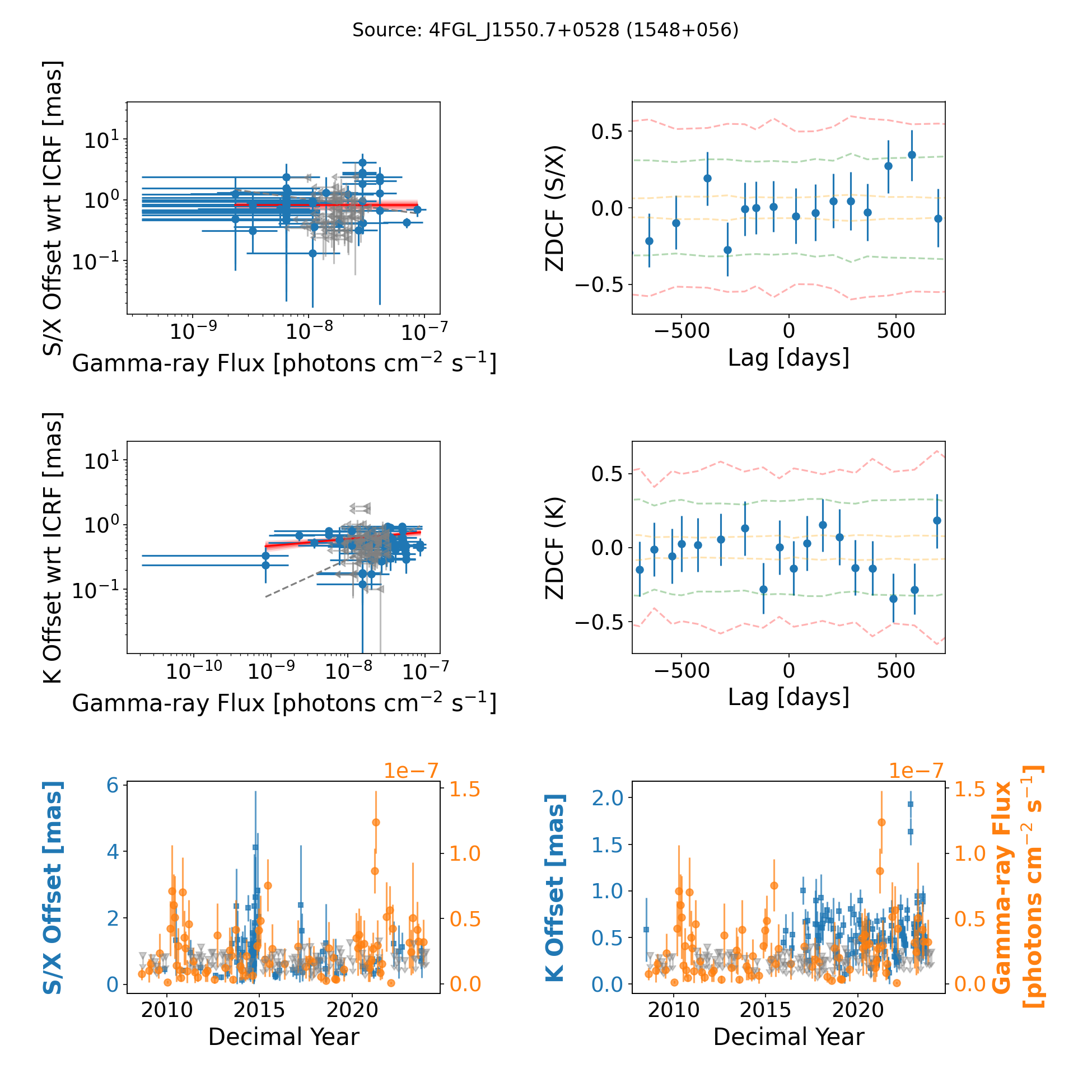}
    \caption{Same as Figure \ref{fig:correlations1}, but for the source 1548$+$056.}
    \label{fig:correlationsJ1550}
\end{figure}

\clearpage

\begin{figure}[ht!]
    \centering
    \includegraphics[width=\textwidth]{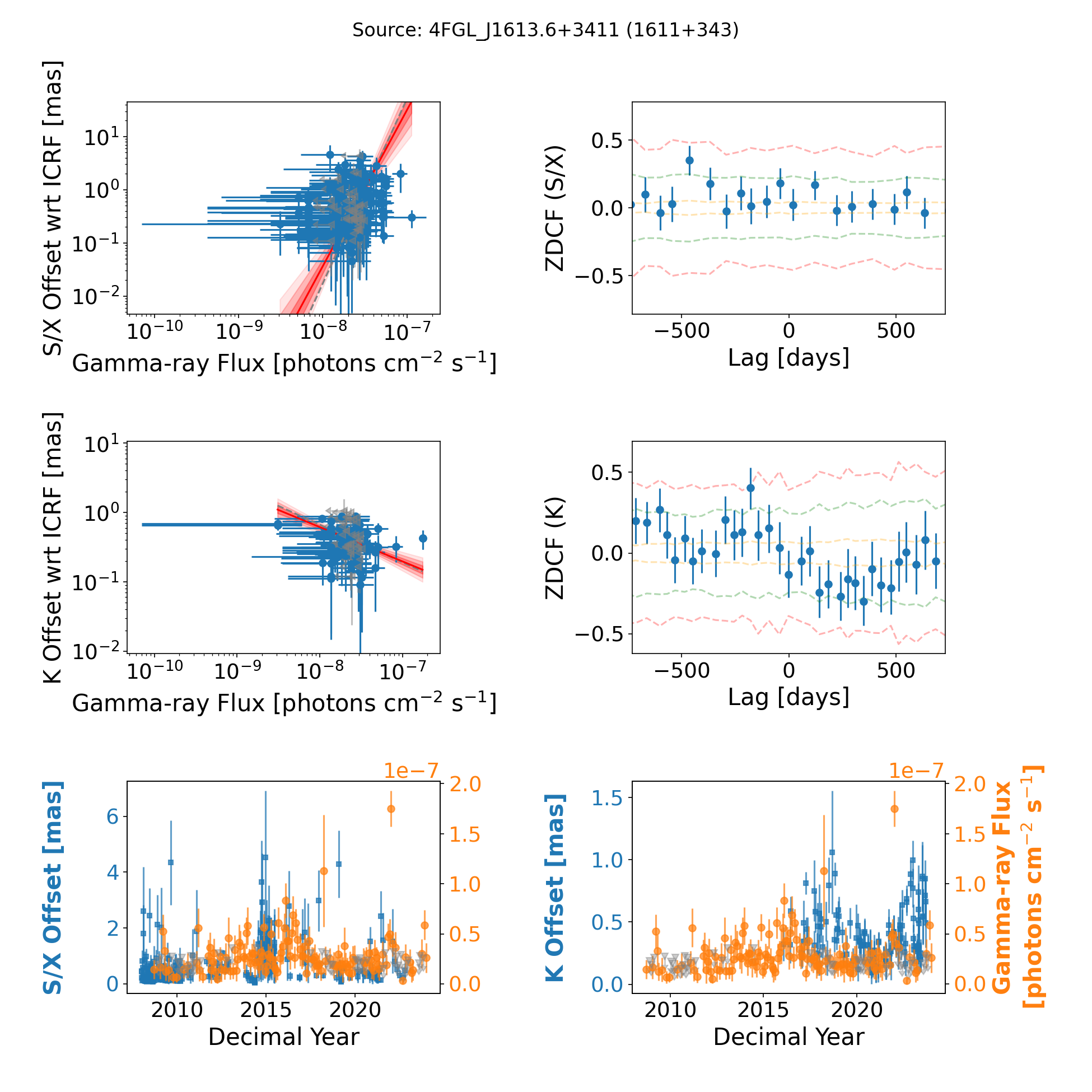}
    \caption{Same as Figure \ref{fig:correlations1}, but for the source 1611$+$343.}
    \label{fig:correlationsJ1639}
\end{figure}

\clearpage

\begin{figure}[ht!]
    \centering
    \includegraphics[width=\textwidth]{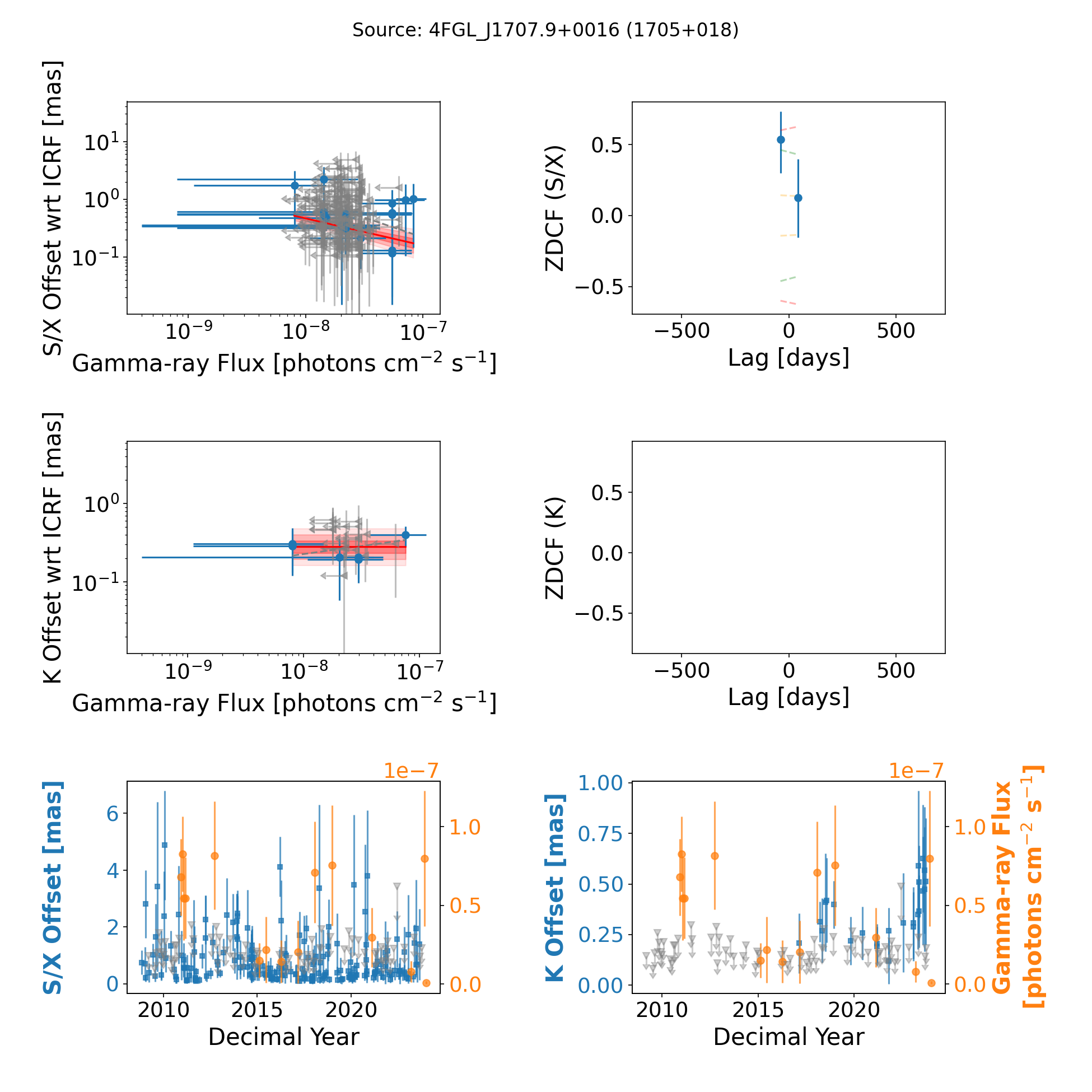}
    \caption{Same as Figure \ref{fig:correlations1}, but for the source 1705$+$018.}
    \label{fig:correlationsJ1707}
\end{figure}

\clearpage

\begin{figure}[ht!]
    \centering
    \includegraphics[width=\textwidth]{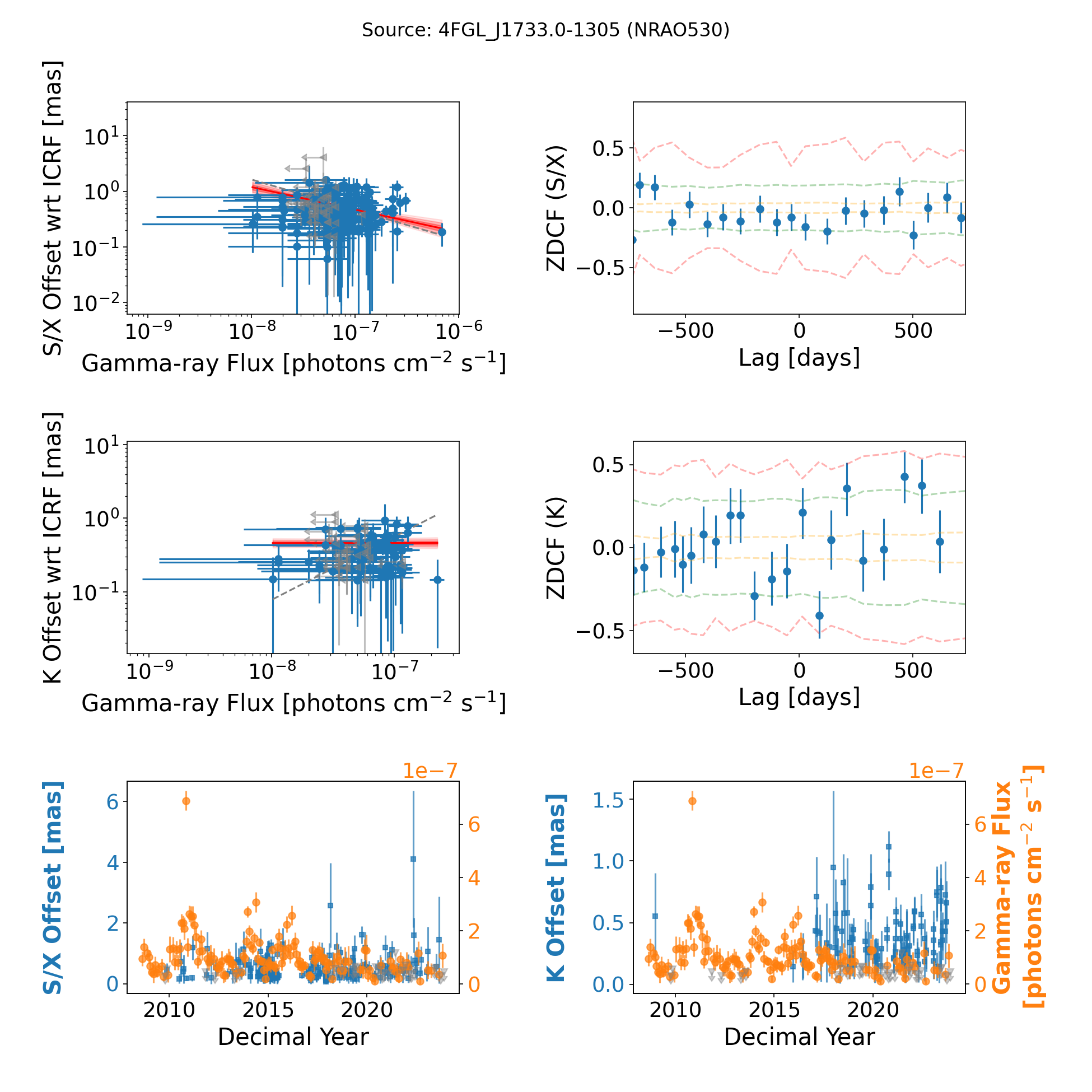}
    \caption{Same as Figure \ref{fig:correlations1}, but for the source NRAO530.}
    \label{fig:correlationsJ1733}
\end{figure}

\clearpage

\begin{figure}[ht!]
    \centering
    \includegraphics[width=\textwidth]{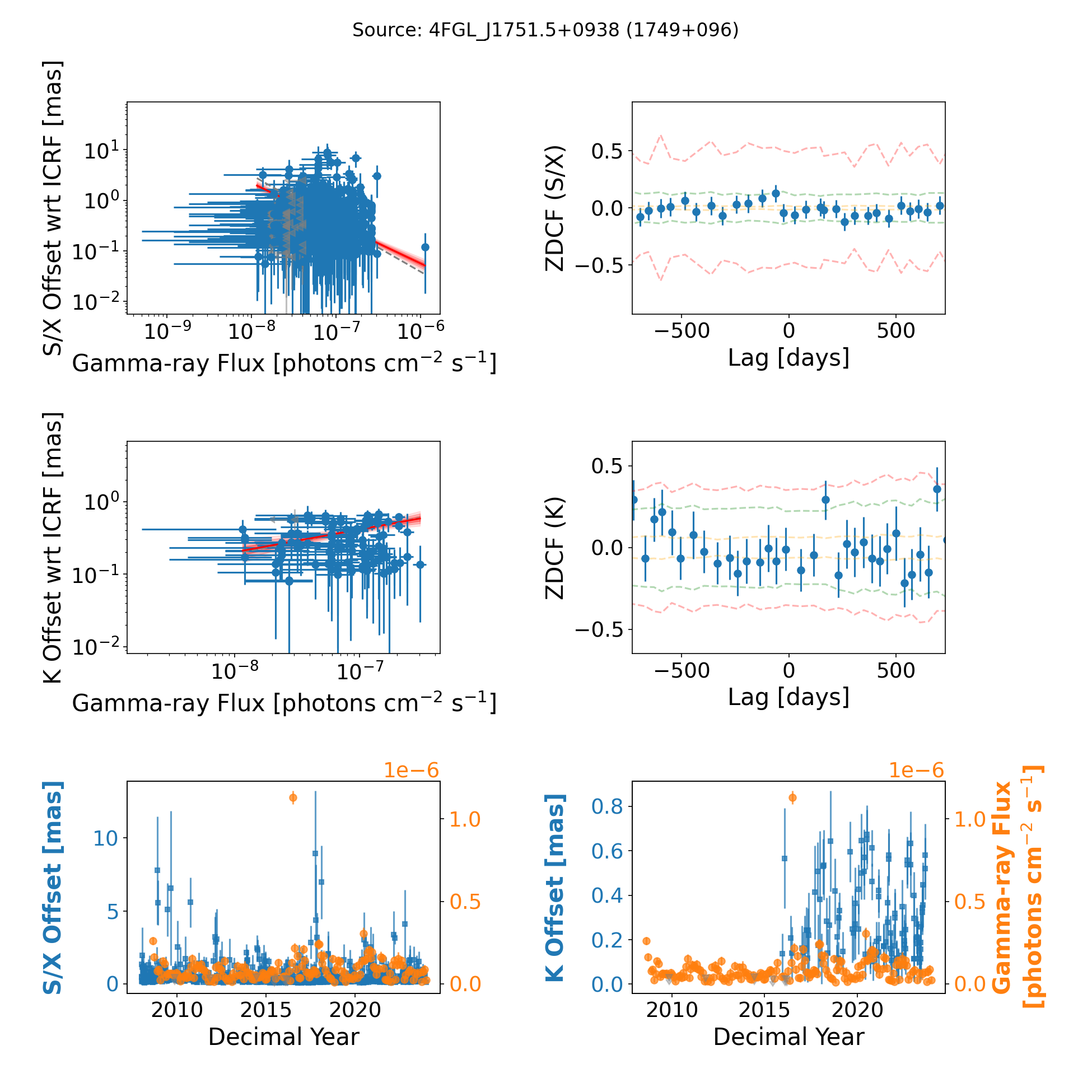}
    \caption{Same as Figure \ref{fig:correlations1}, but for the source 1749$+$096.}
    \label{fig:correlationsJ1751}
\end{figure}

\clearpage

\begin{figure}[ht!]
    \centering
    \includegraphics[width=\textwidth]{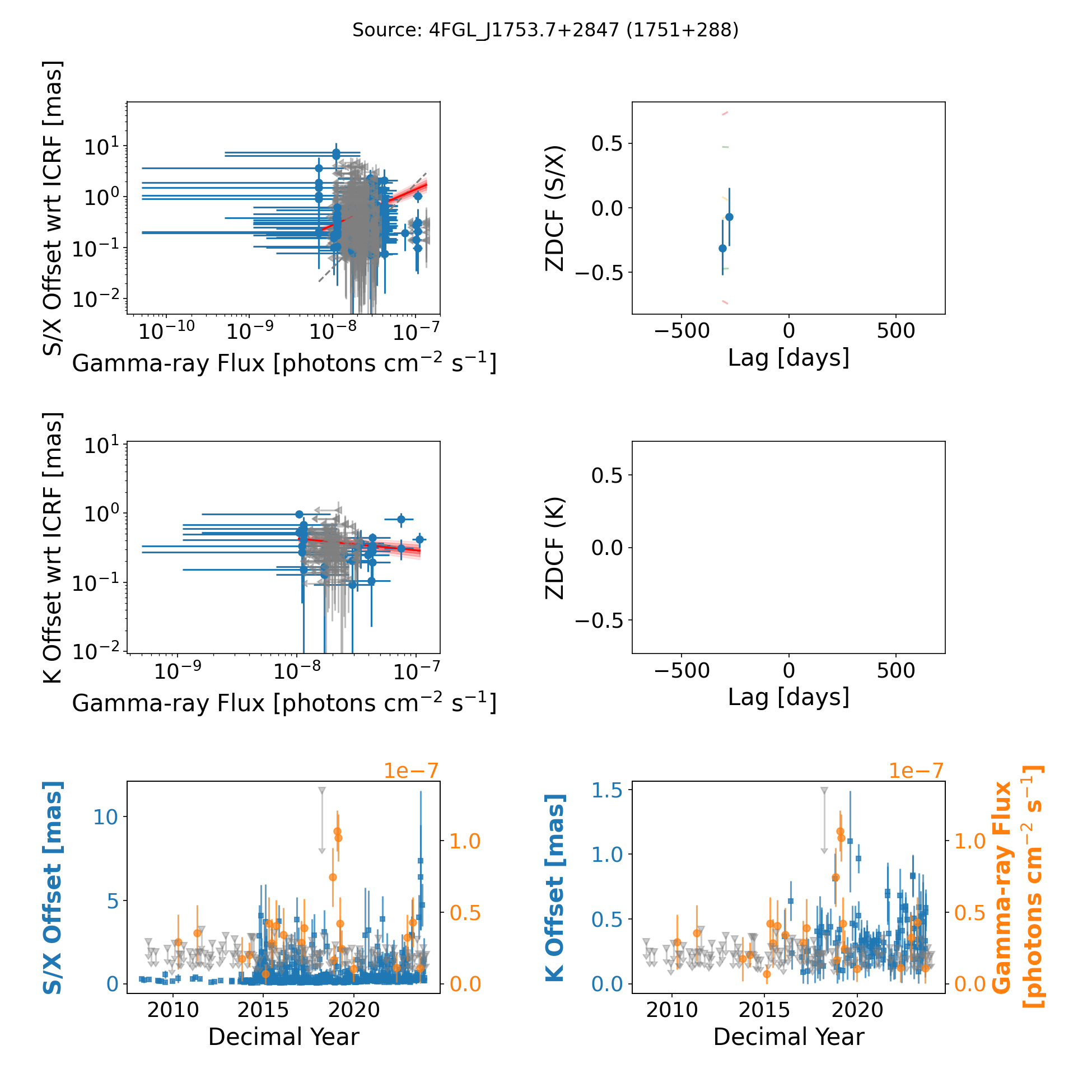}
    \caption{Same as Figure \ref{fig:correlations1}, but for the source 1751$+$288.}
    \label{fig:correlationsJ1753}
\end{figure}

\clearpage

\begin{figure}[ht!]
    \centering
    \includegraphics[width=\textwidth]{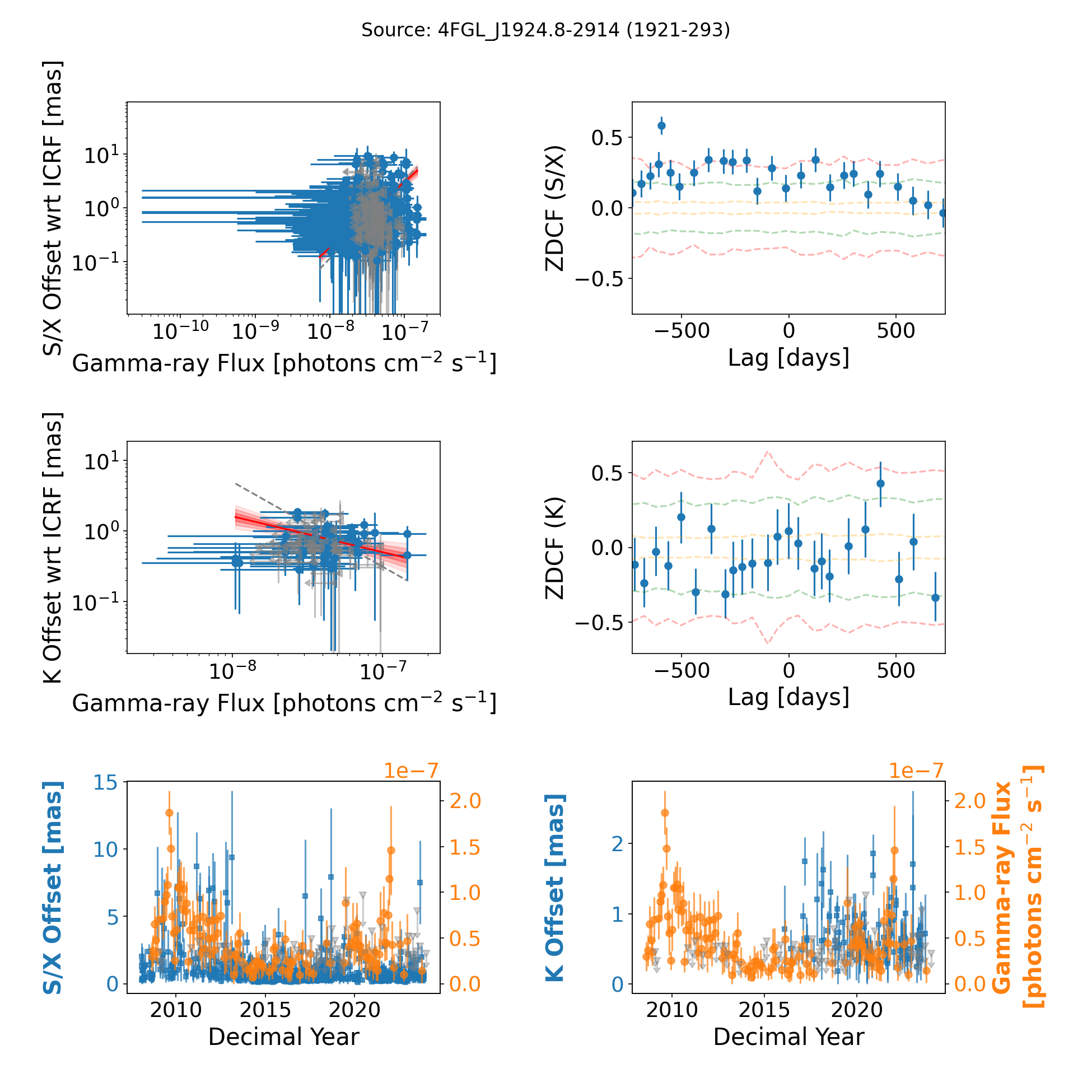}
    \caption{Same as Figure \ref{fig:correlations1}, but for the source 1921$-$293.}
    \label{fig:correlationsJ1924}
\end{figure}

\clearpage

\begin{figure}[ht!]
    \centering
    \includegraphics[width=\textwidth]{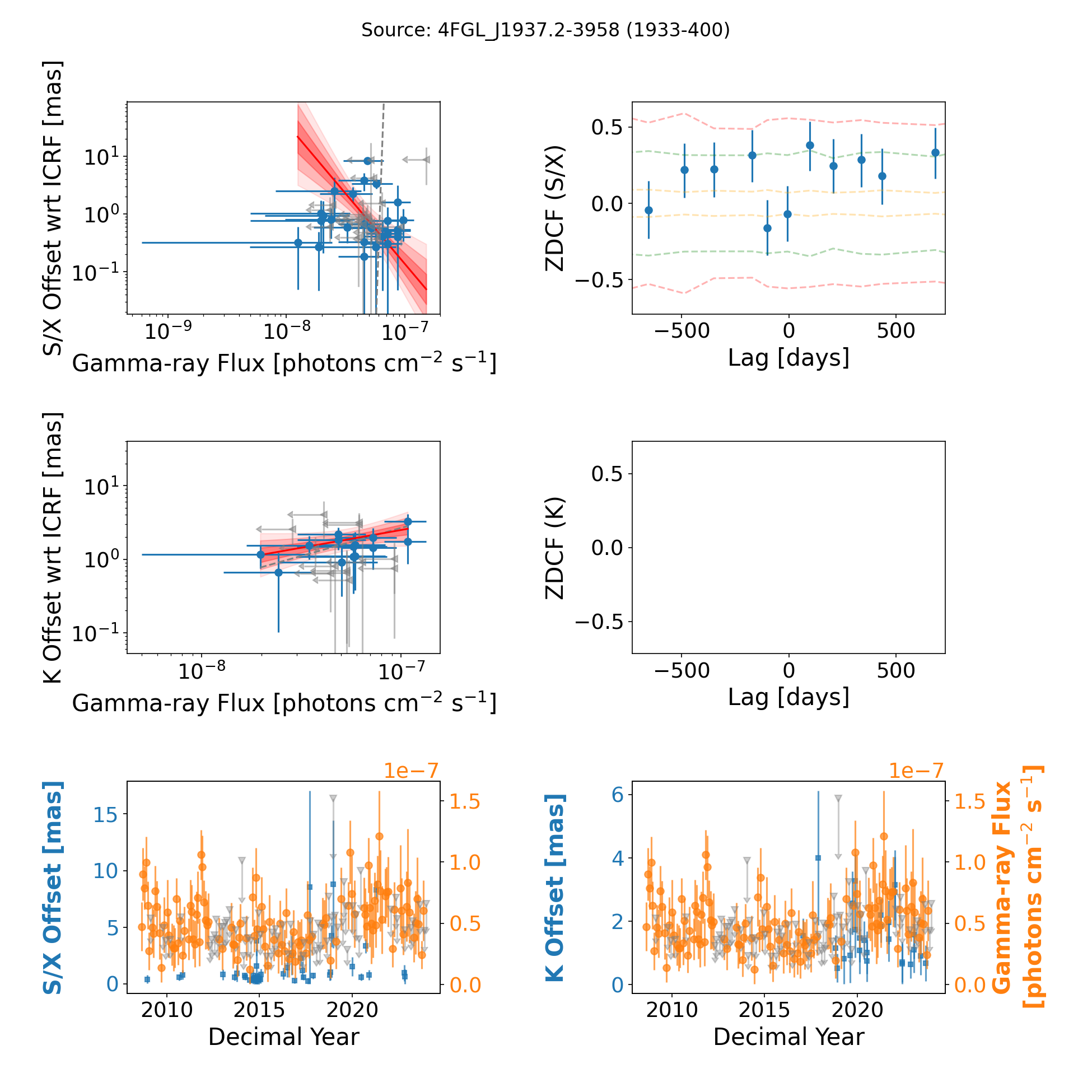}
    \caption{Same as Figure \ref{fig:correlations1}, but for the source 1933$-$400.}
    \label{fig:correlationsJ1737}
\end{figure}

\clearpage

\begin{figure}[ht!]
    \centering
    \includegraphics[width=\textwidth]{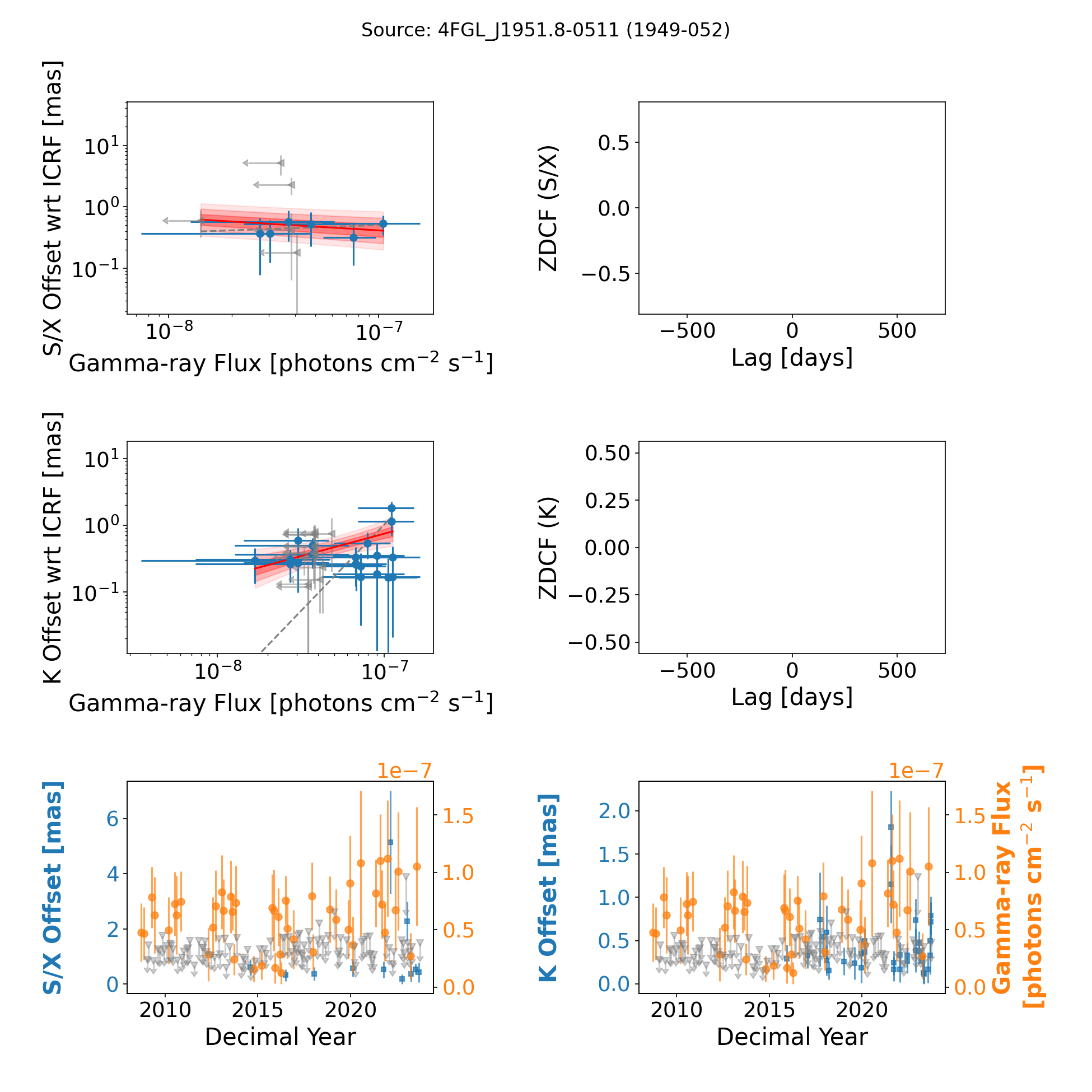}
    \caption{Same as Figure \ref{fig:correlations1}, but for the source 1949$-$052.}
    \label{fig:correlationsJ1951}
\end{figure}

\clearpage

\begin{figure}[ht!]
    \centering
    \includegraphics[width=\textwidth]{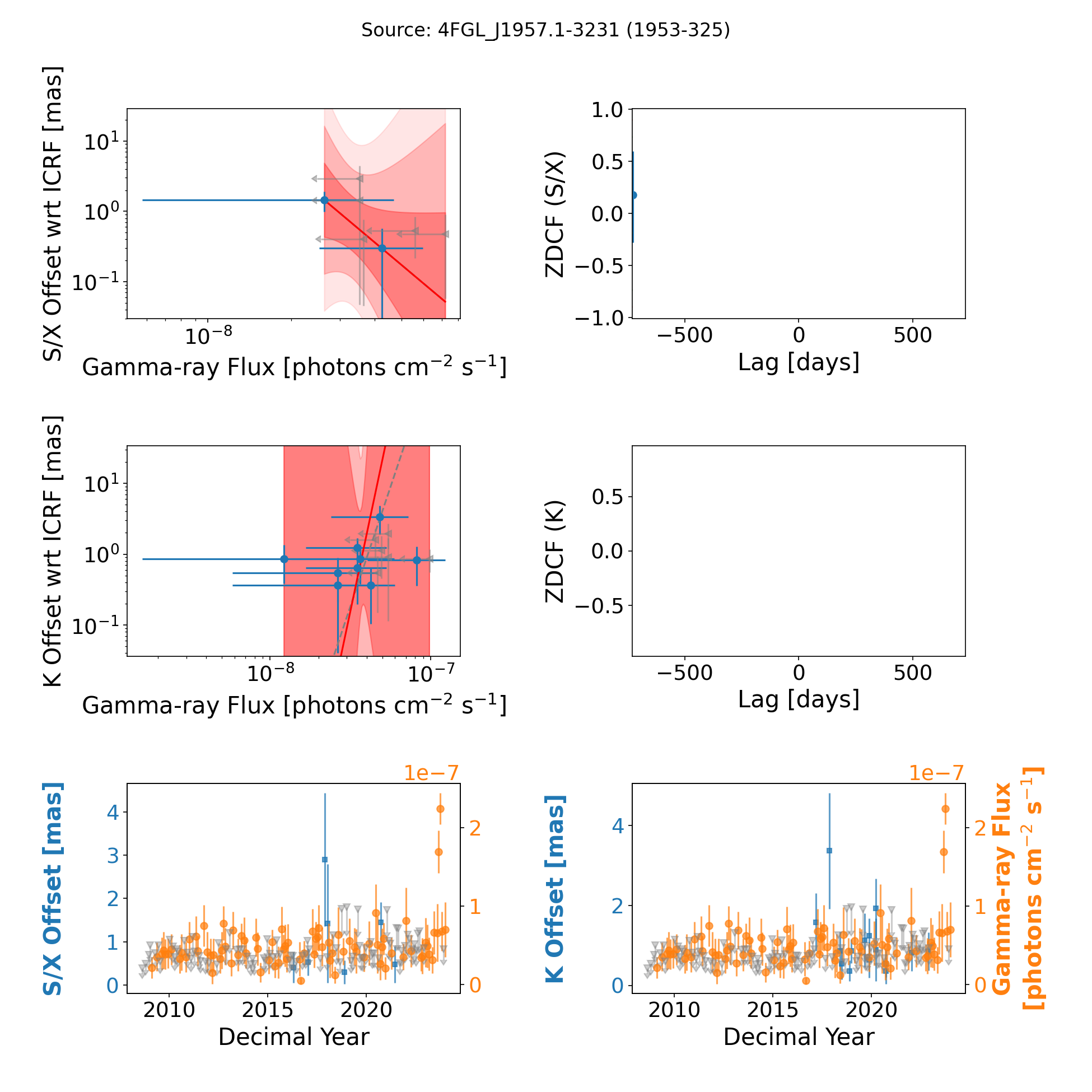}
    \caption{Same as Figure \ref{fig:correlations1}, but for the source 1953$-$325.}
    \label{fig:correlationsJ1957}
\end{figure}

\clearpage

\begin{figure}[ht!]
    \centering
    \includegraphics[width=\textwidth]{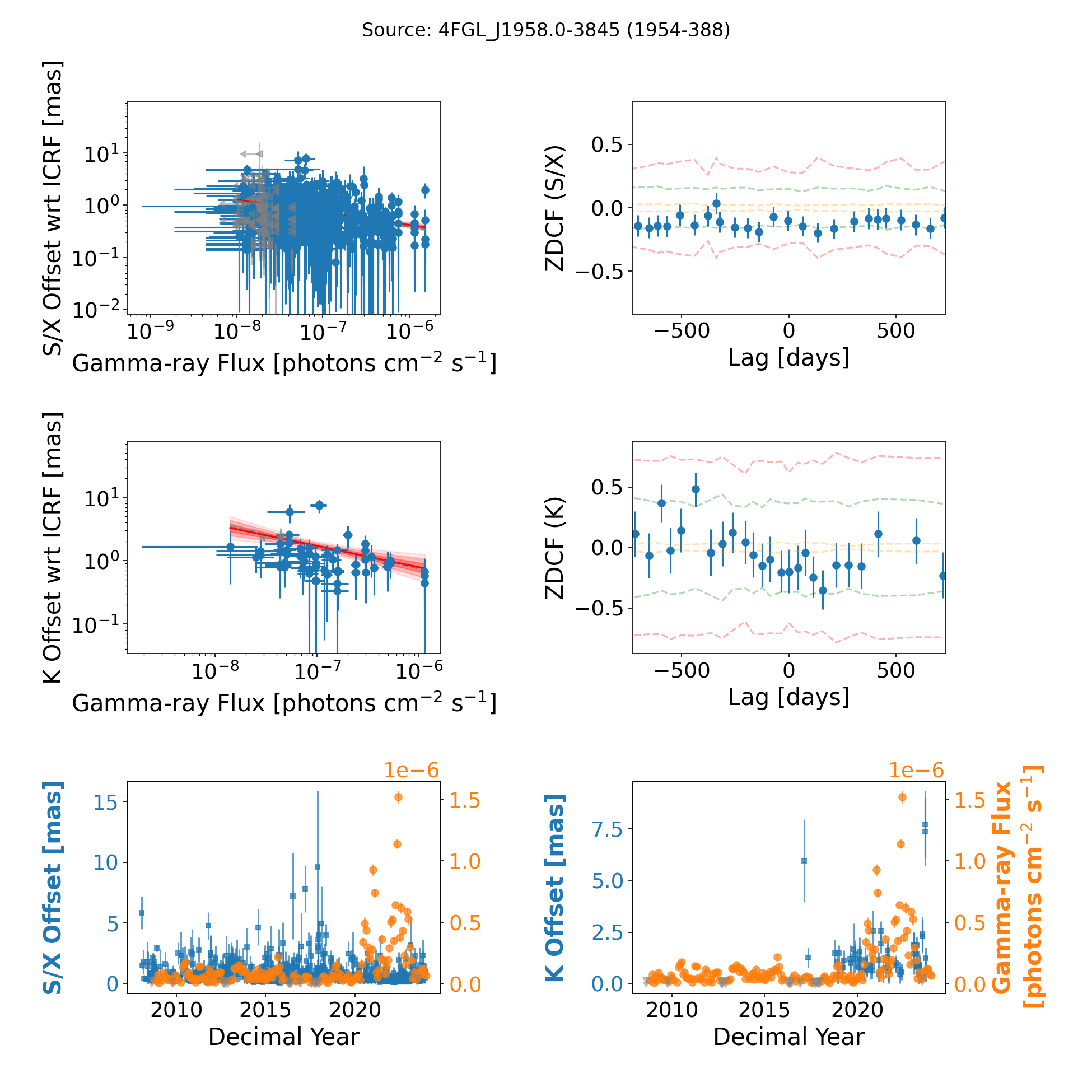}
    \caption{Same as Figure \ref{fig:correlations1}, but for the source 1954$-$388.}
    \label{fig:correlationsJ1958}
\end{figure}

\clearpage

\begin{figure}[ht!]
    \centering
    \includegraphics[width=\textwidth]{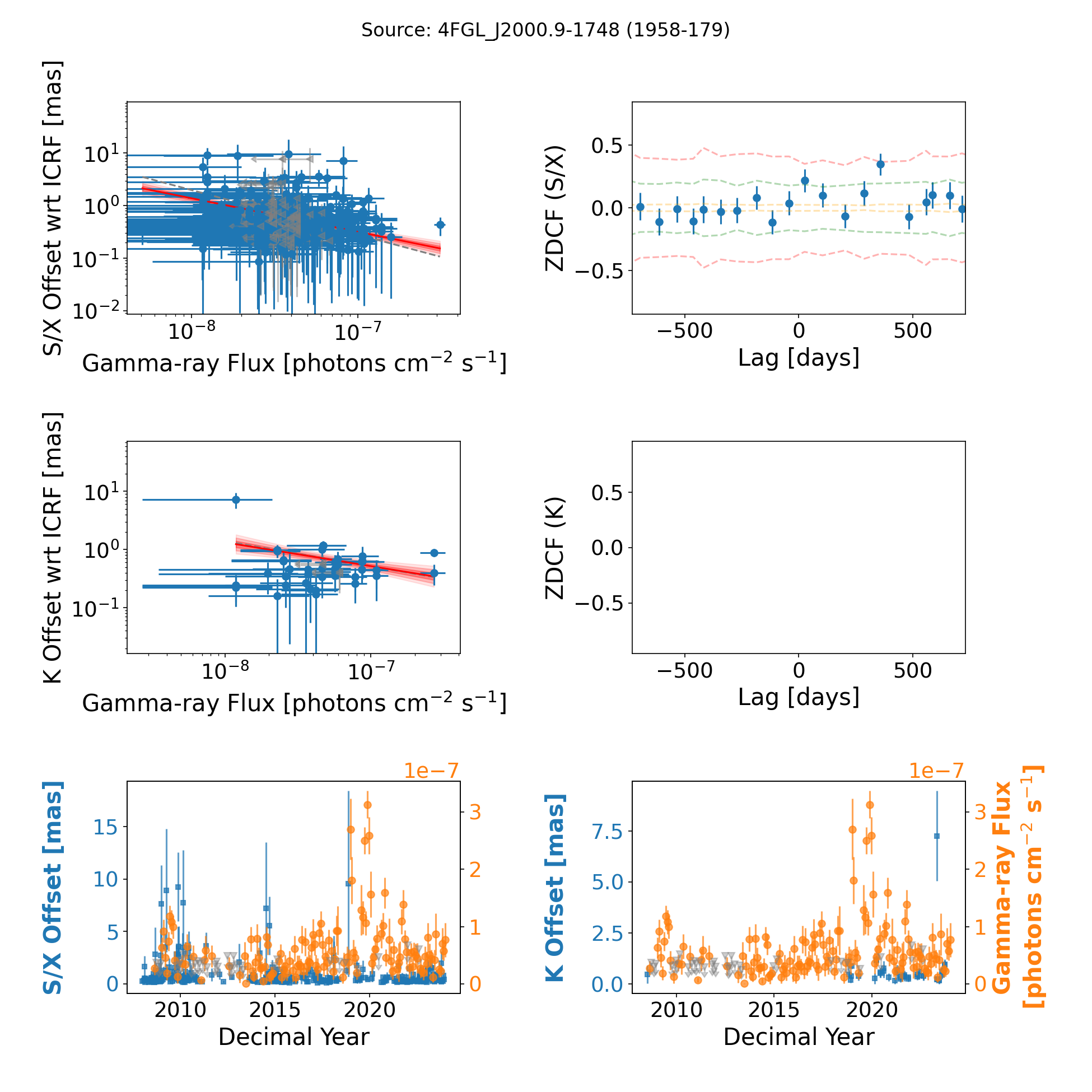}
    \caption{Same as Figure \ref{fig:correlations1}, but for the source 1958$-$179.}
    \label{fig:correlationsJ2000}
\end{figure}

\clearpage

\begin{figure}[ht!]
    \centering
    \includegraphics[width=\textwidth]{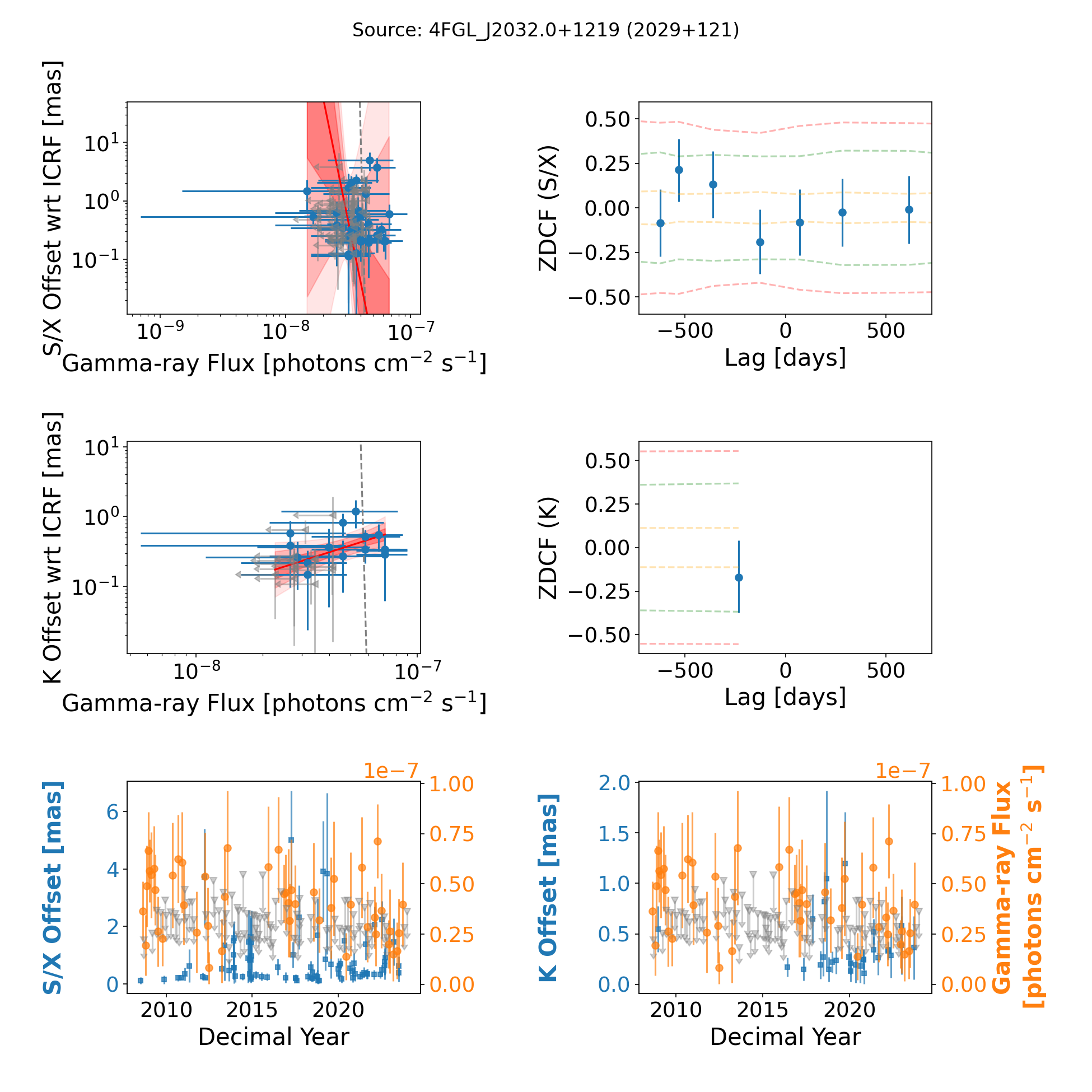}
    \caption{Same as Figure \ref{fig:correlations1}, but for the source 2029$+$121.}
    \label{fig:correlationsJ2032}
\end{figure}

\clearpage

\begin{figure}[ht!]
    \centering
    \includegraphics[width=\textwidth]{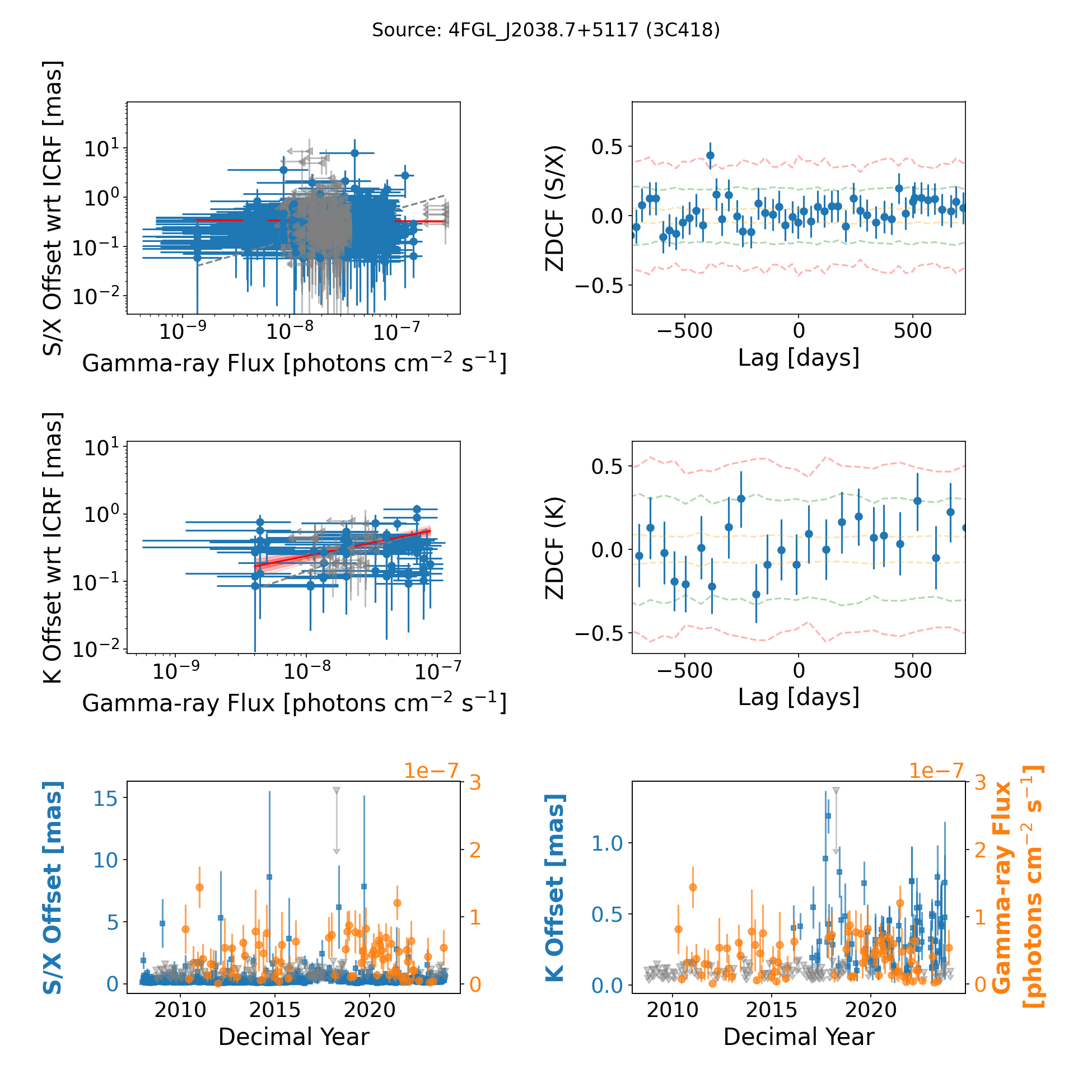}
    \caption{Same as Figure \ref{fig:correlations1}, but for the source 3C418.}
    \label{fig:correlationsJ2038}
\end{figure}

\clearpage

\begin{figure}[ht!]
    \centering
    \includegraphics[width=\textwidth]{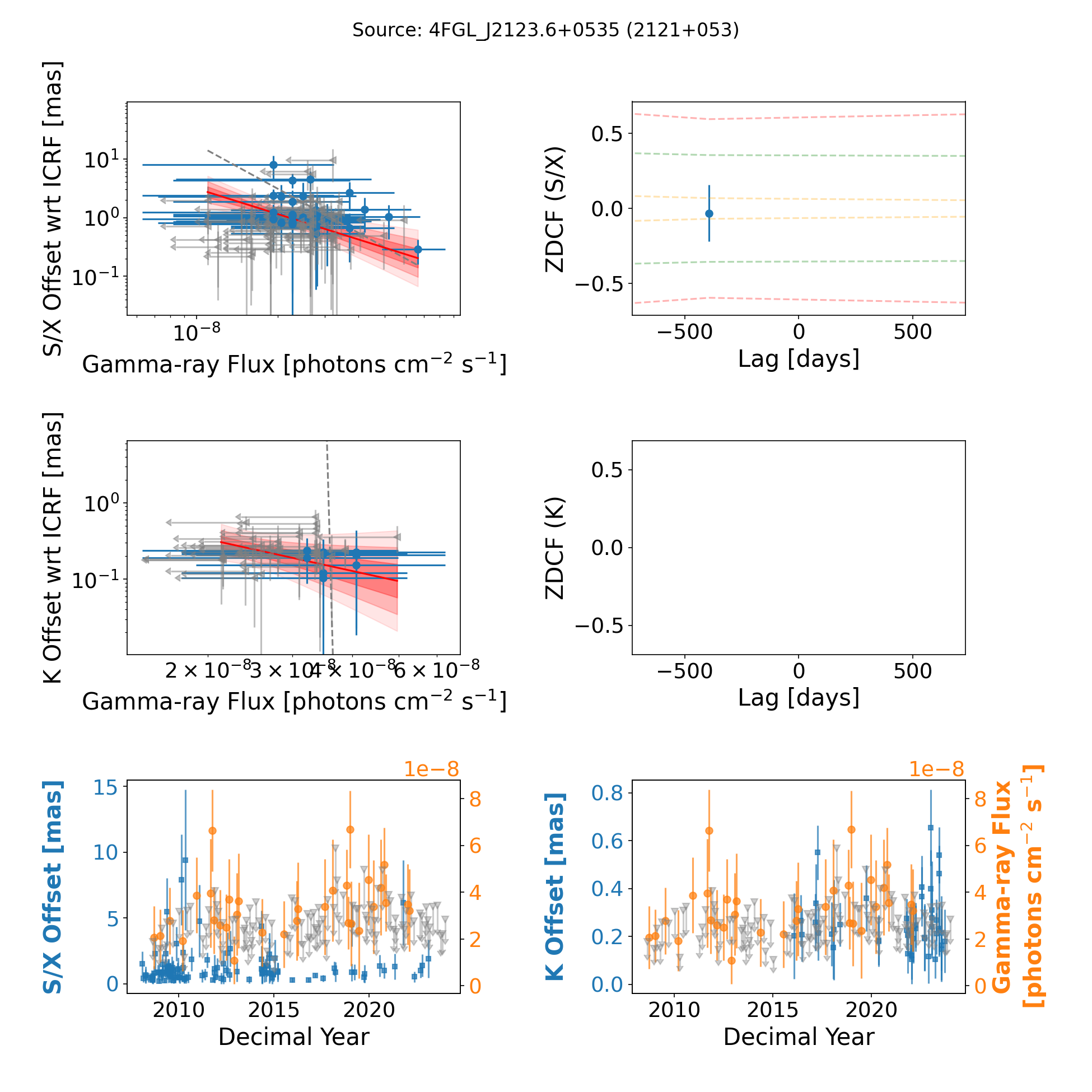}
    \caption{Same as Figure \ref{fig:correlations1}, but for the source 2121$+$053.}
    \label{fig:correlationsJ2123}
\end{figure}

\clearpage

\begin{figure}[ht!]
    \centering
    \includegraphics[width=\textwidth]{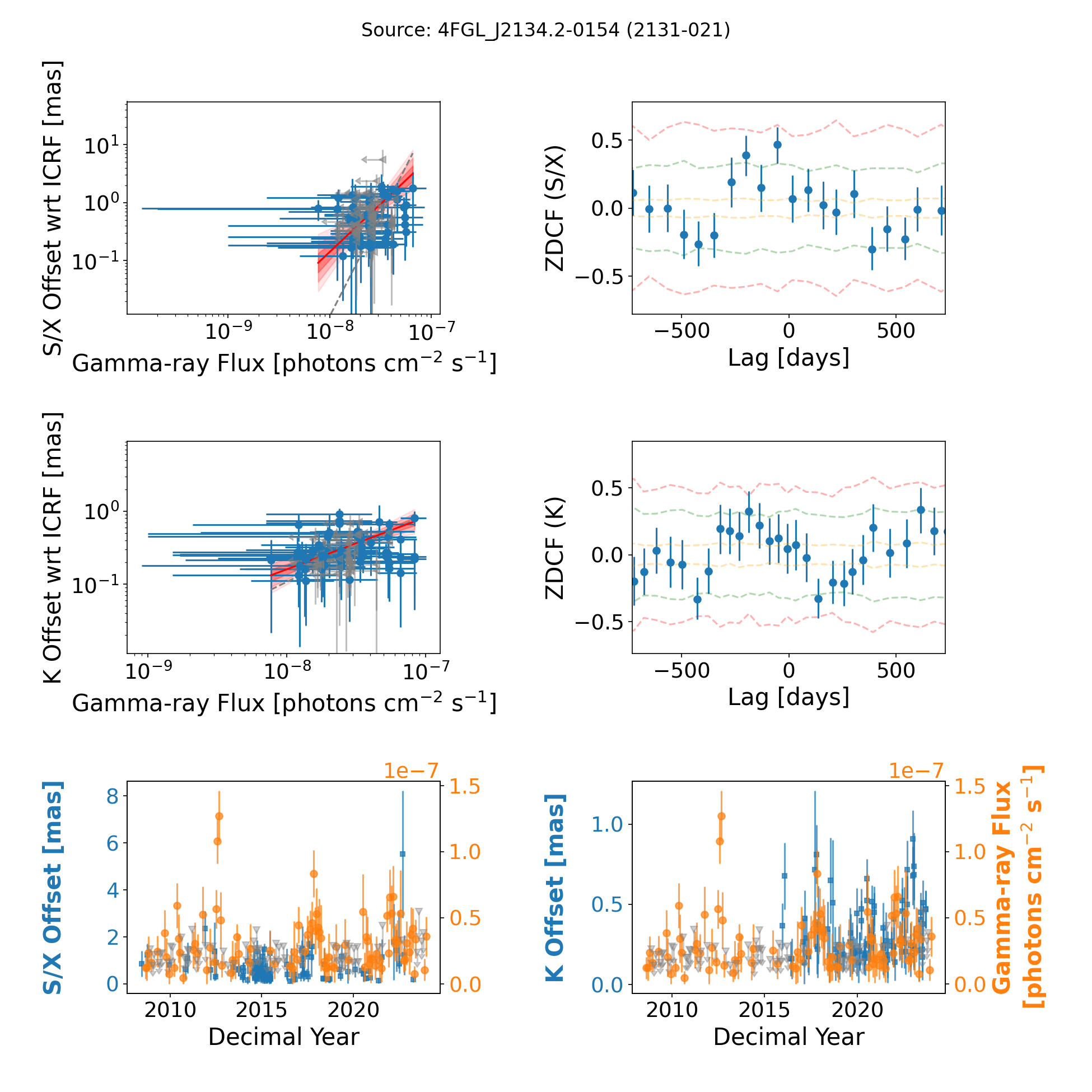}
    \caption{Same as Figure \ref{fig:correlations1}, but for the source 2131$-$021.}
    \label{fig:correlationsJ2134}
\end{figure}

\clearpage

\begin{figure}[ht!]
    \centering
    \includegraphics[width=\textwidth]{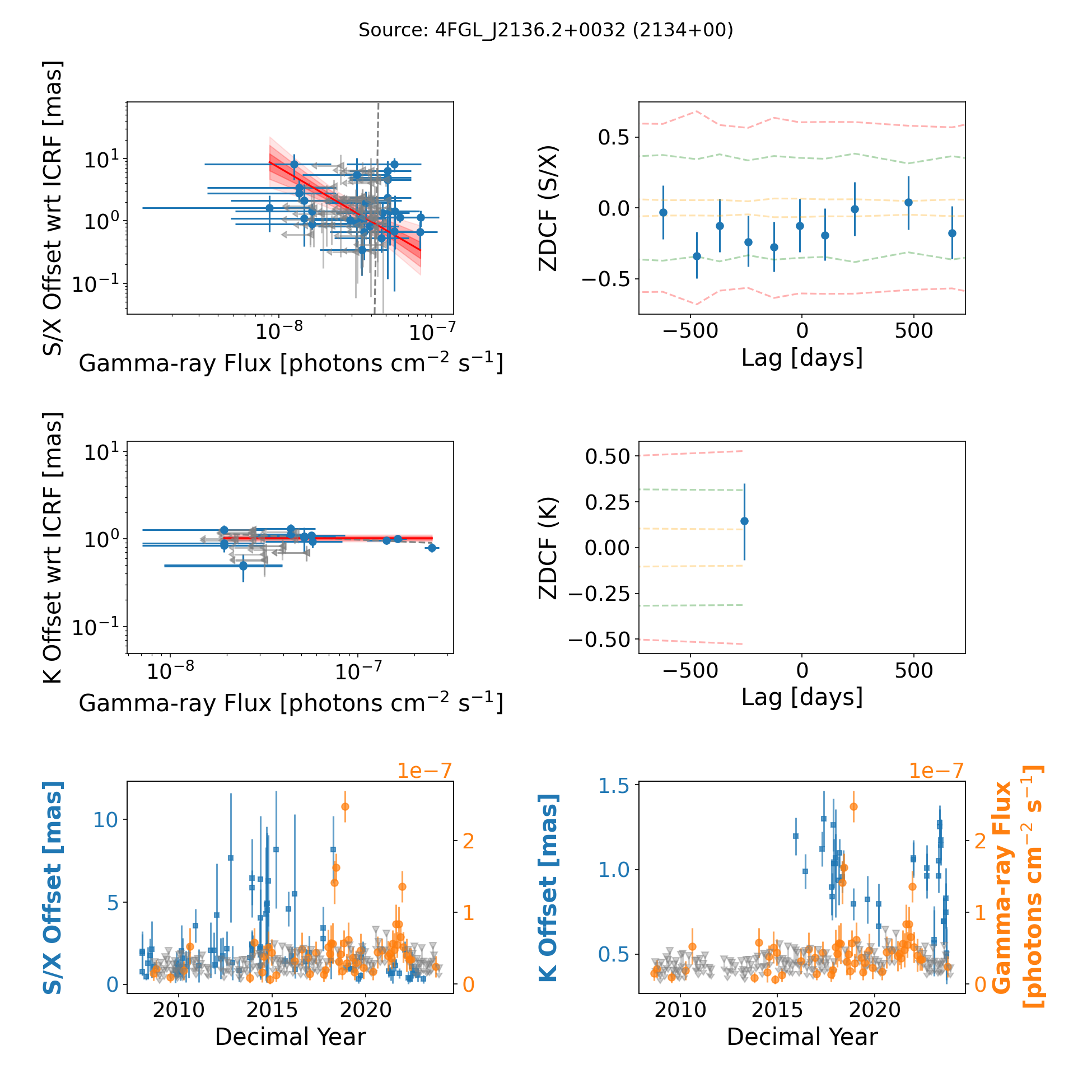}
    \caption{Same as Figure \ref{fig:correlations1}, but for the source 2134$+$00.}
    \label{fig:correlationsJ2136}
\end{figure}

\clearpage

\begin{figure}[ht!]
    \centering
    \includegraphics[width=\textwidth]{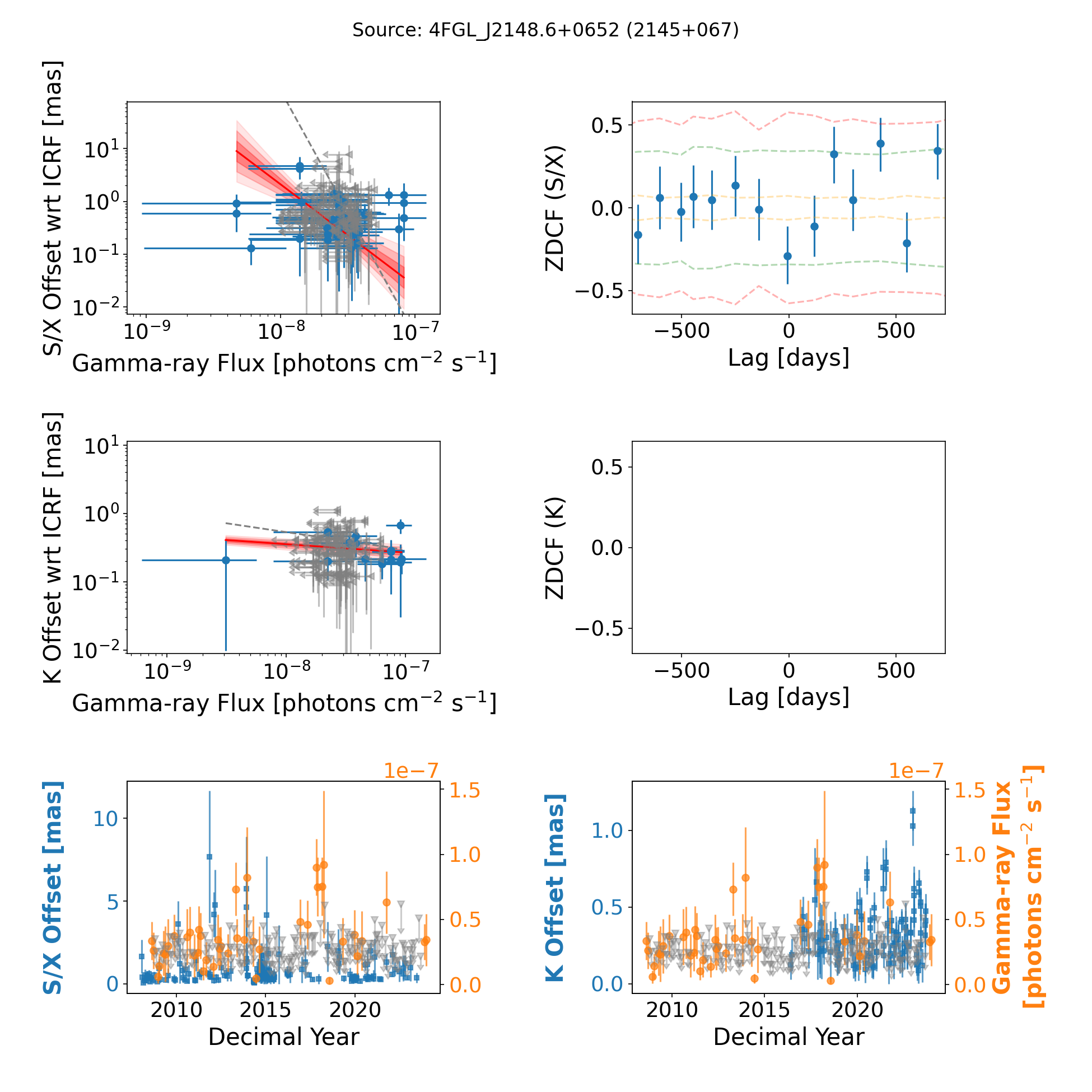}
    \caption{Same as Figure \ref{fig:correlations1}, but for the source 2145$+$067.}
    \label{fig:correlationsJ2148}
\end{figure}

\clearpage

\begin{figure}[ht!]
    \centering
    \includegraphics[width=\textwidth]{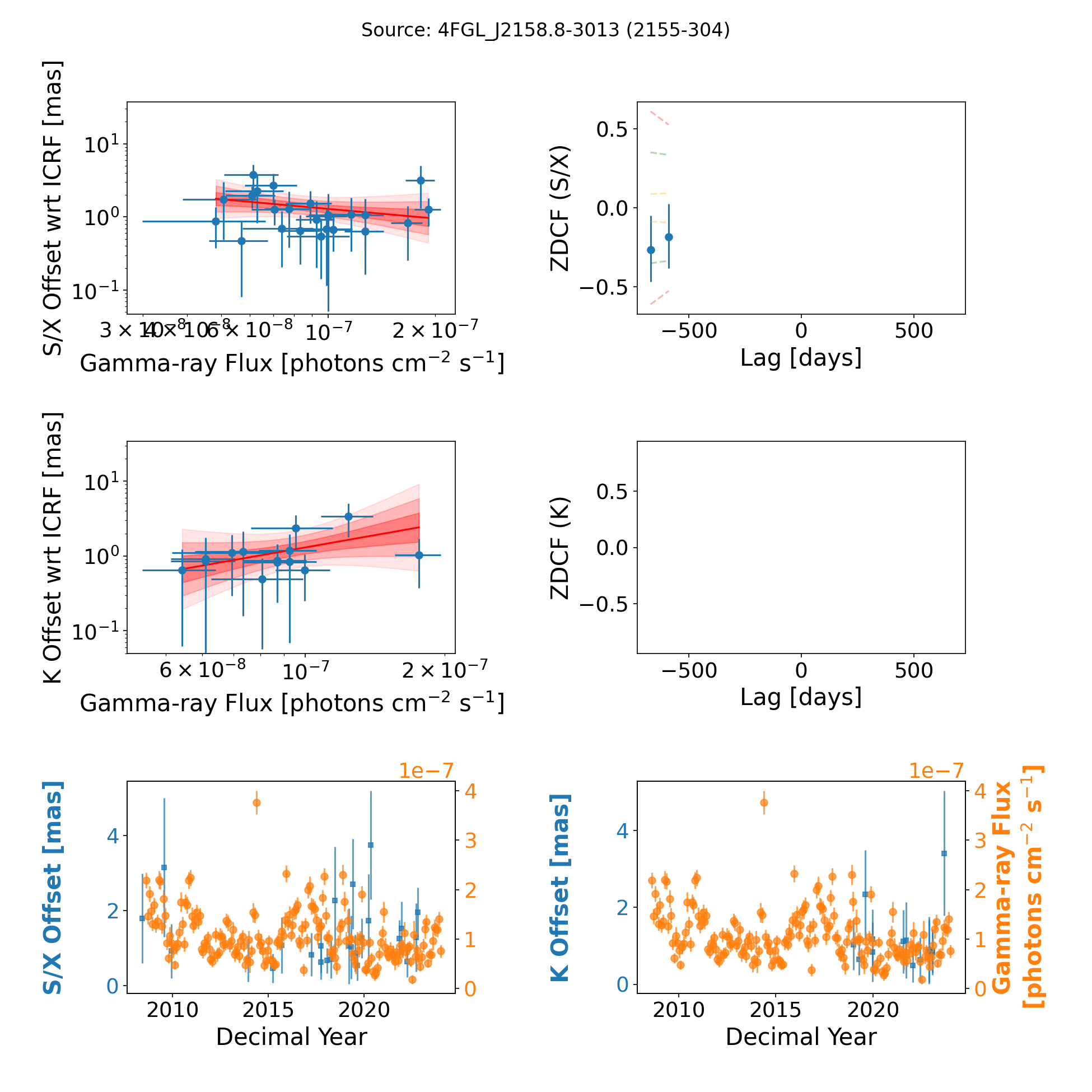}
    \caption{Same as Figure \ref{fig:correlations1}, but for the source 2155$-$304.}
    \label{fig:correlationsJ2158}
\end{figure}

\clearpage

\begin{figure}[ht!]
    \centering
    \includegraphics[width=\textwidth]{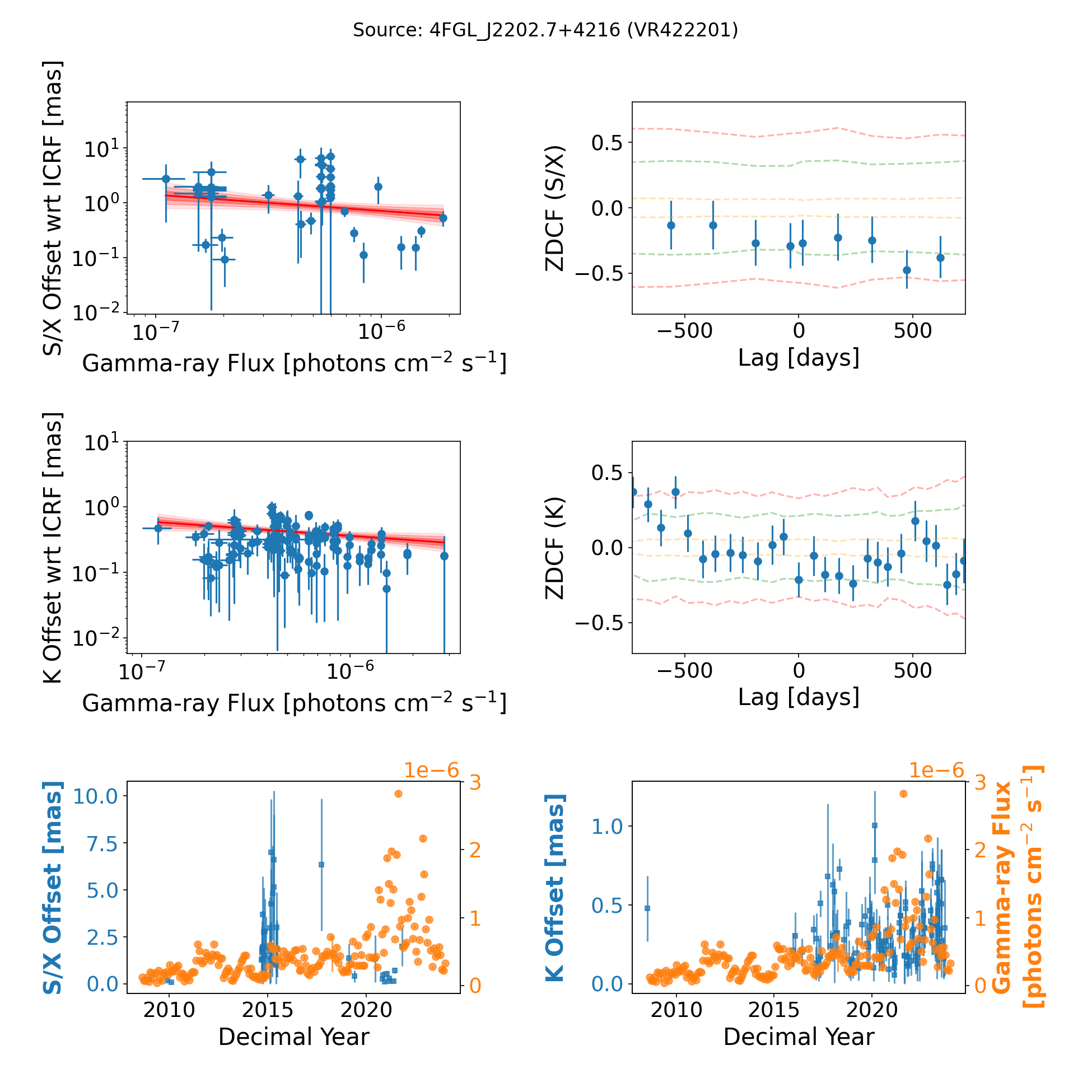}
    \caption{Same as Figure \ref{fig:correlations1}, but for the source BLLAC.}
    \label{fig:correlationsJ2202}
\end{figure}

\clearpage

\begin{figure}[ht!]
    \centering
    \includegraphics[width=\textwidth]{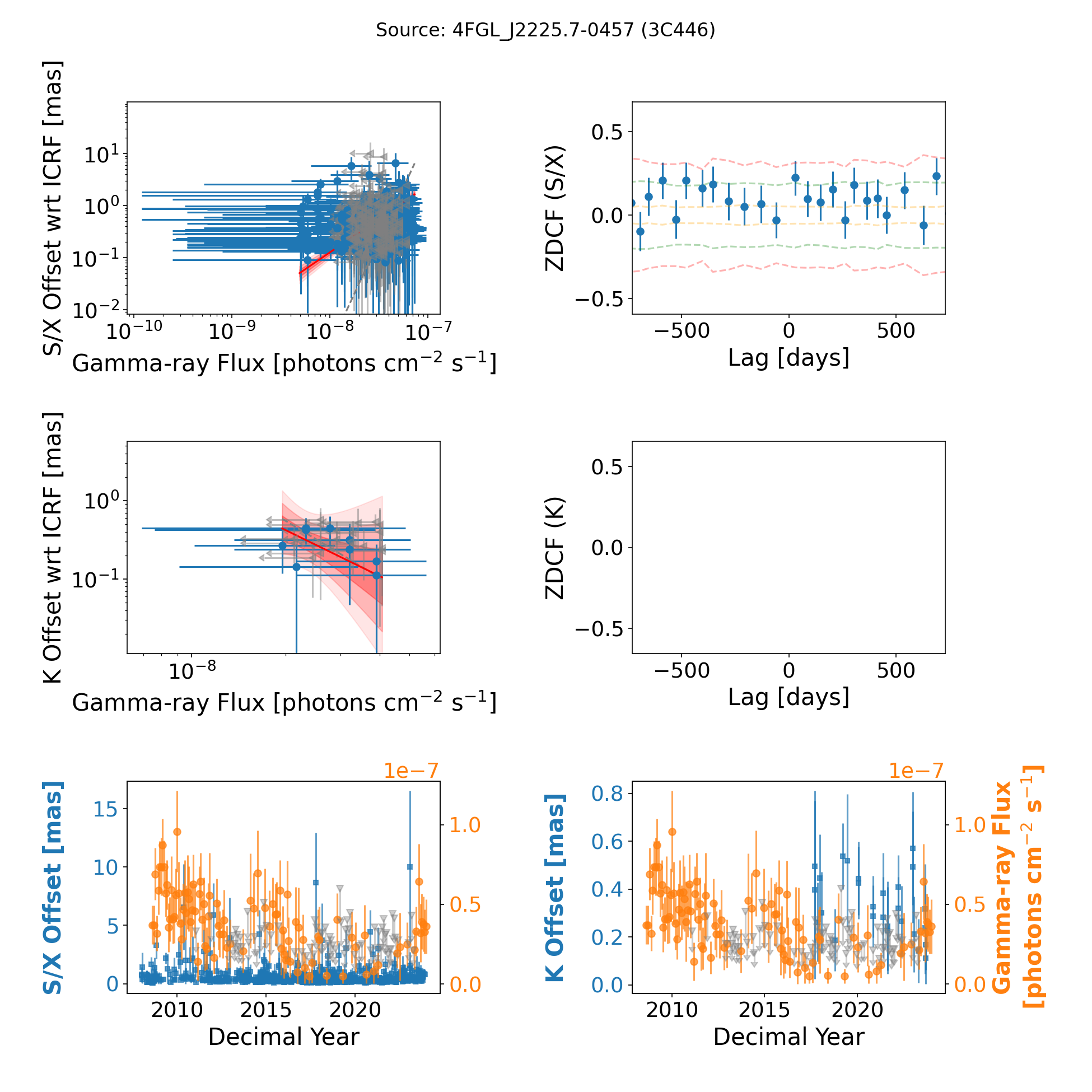}
    \caption{Same as Figure \ref{fig:correlations1}, but for the source 3C446.}
    \label{fig:correlationsJ2225}
\end{figure}

\clearpage

\begin{figure}[ht!]
    \centering
    \includegraphics[width=\textwidth]{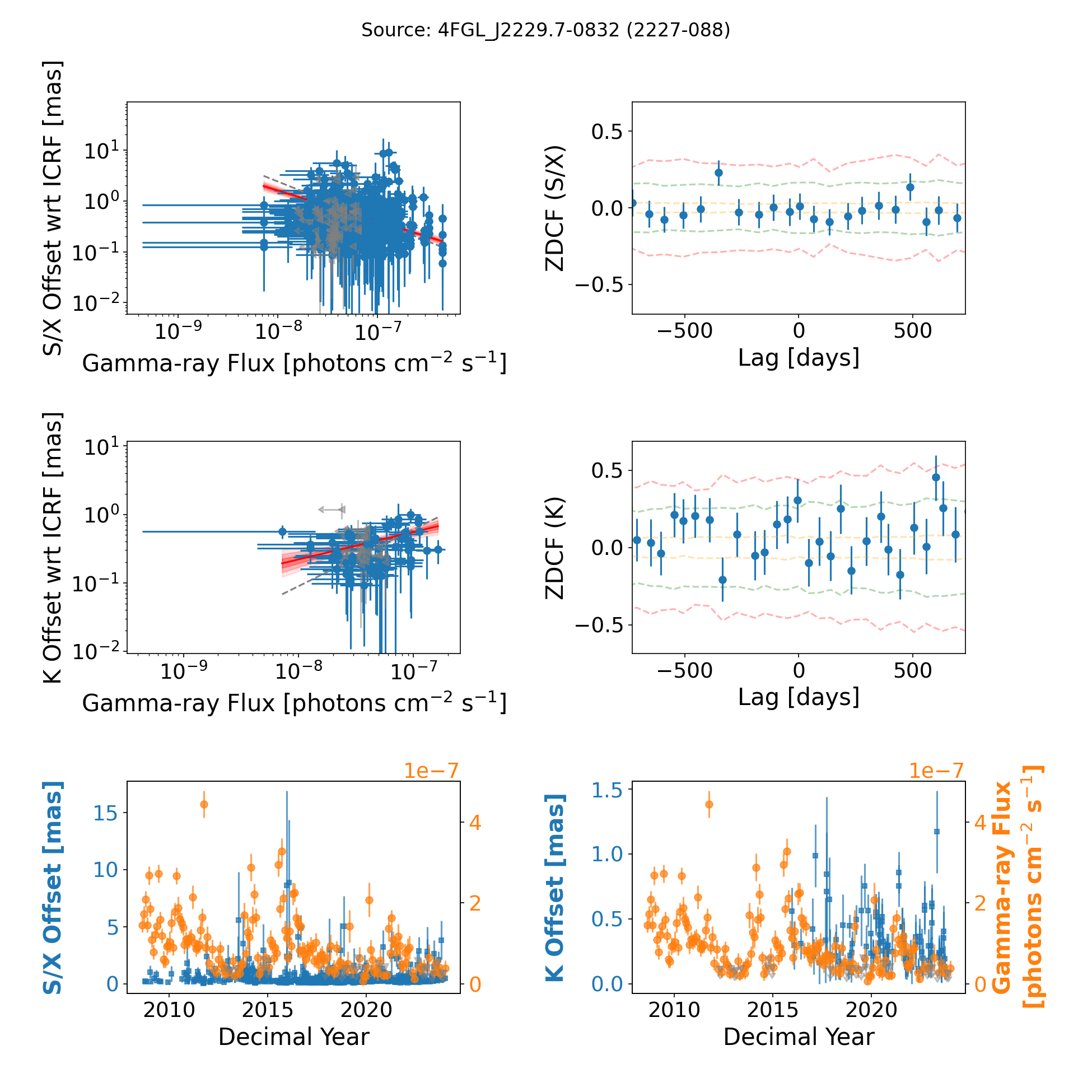}
    \caption{Same as Figure \ref{fig:correlations1}, but for the source 2227$-$088.}
    \label{fig:correlationsJ2229}
\end{figure}

\clearpage

\begin{figure}[ht!]
    \centering
    \includegraphics[width=\textwidth]{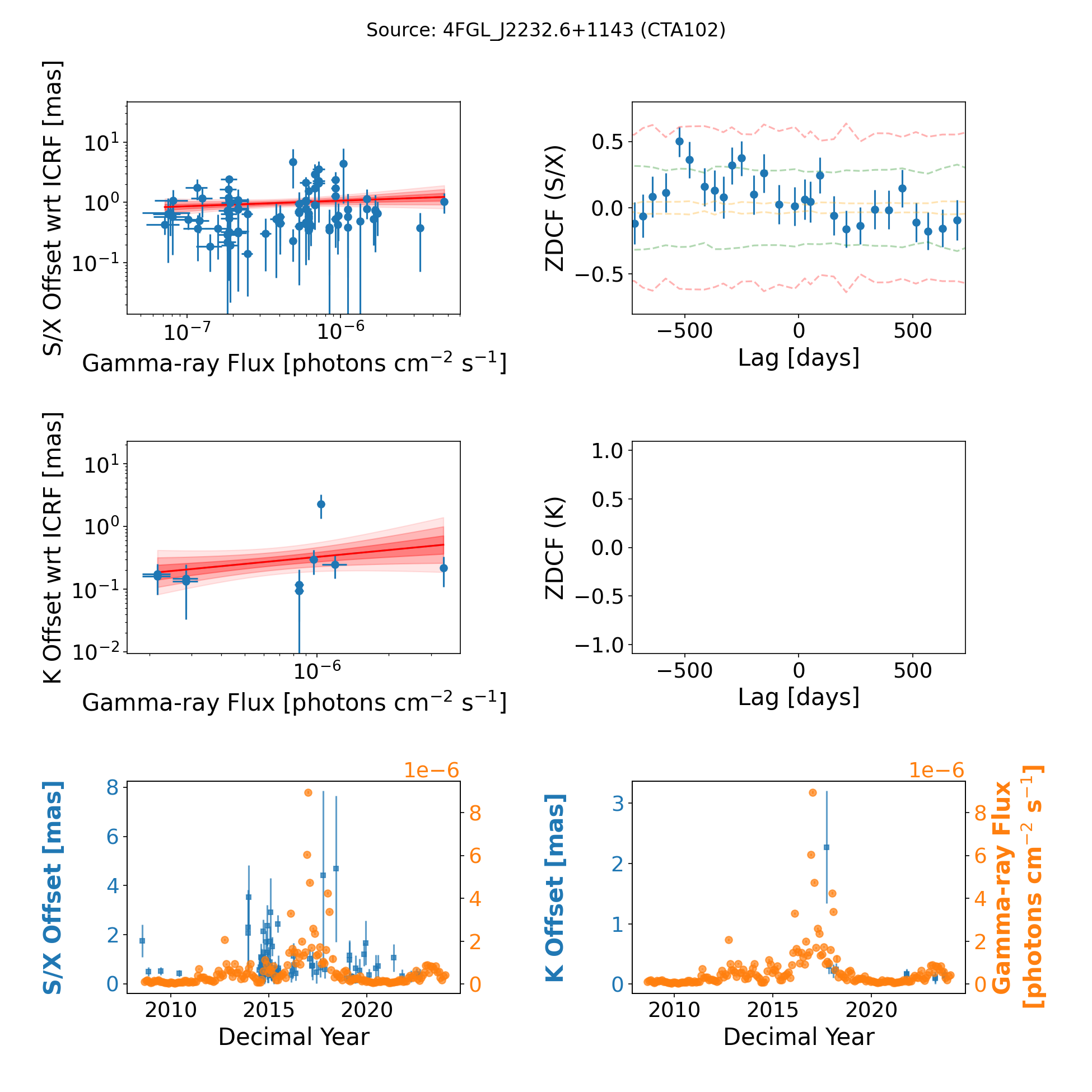}
    \caption{Same as Figure \ref{fig:correlations1}, but for the source CTA102.}
    \label{fig:correlationsJ2232}
\end{figure}

\clearpage

\begin{figure}[ht!]
    \centering
    \includegraphics[width=\textwidth]{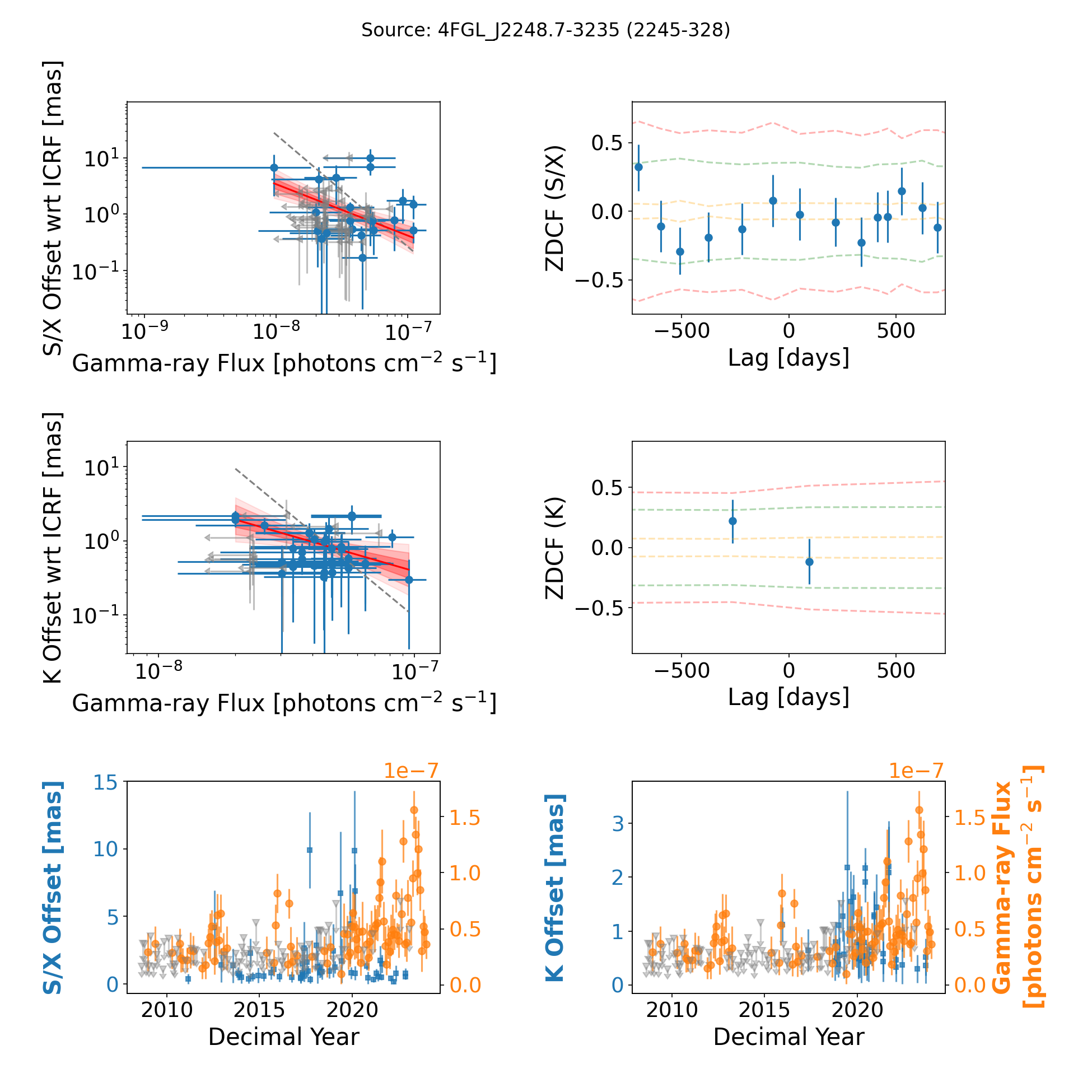}
    \caption{Same as Figure \ref{fig:correlations1}, but for the source 2245$-$328.}
    \label{fig:correlationsJ2248}
\end{figure}

\clearpage

\begin{figure}[ht!]
    \centering
    \includegraphics[width=\textwidth]{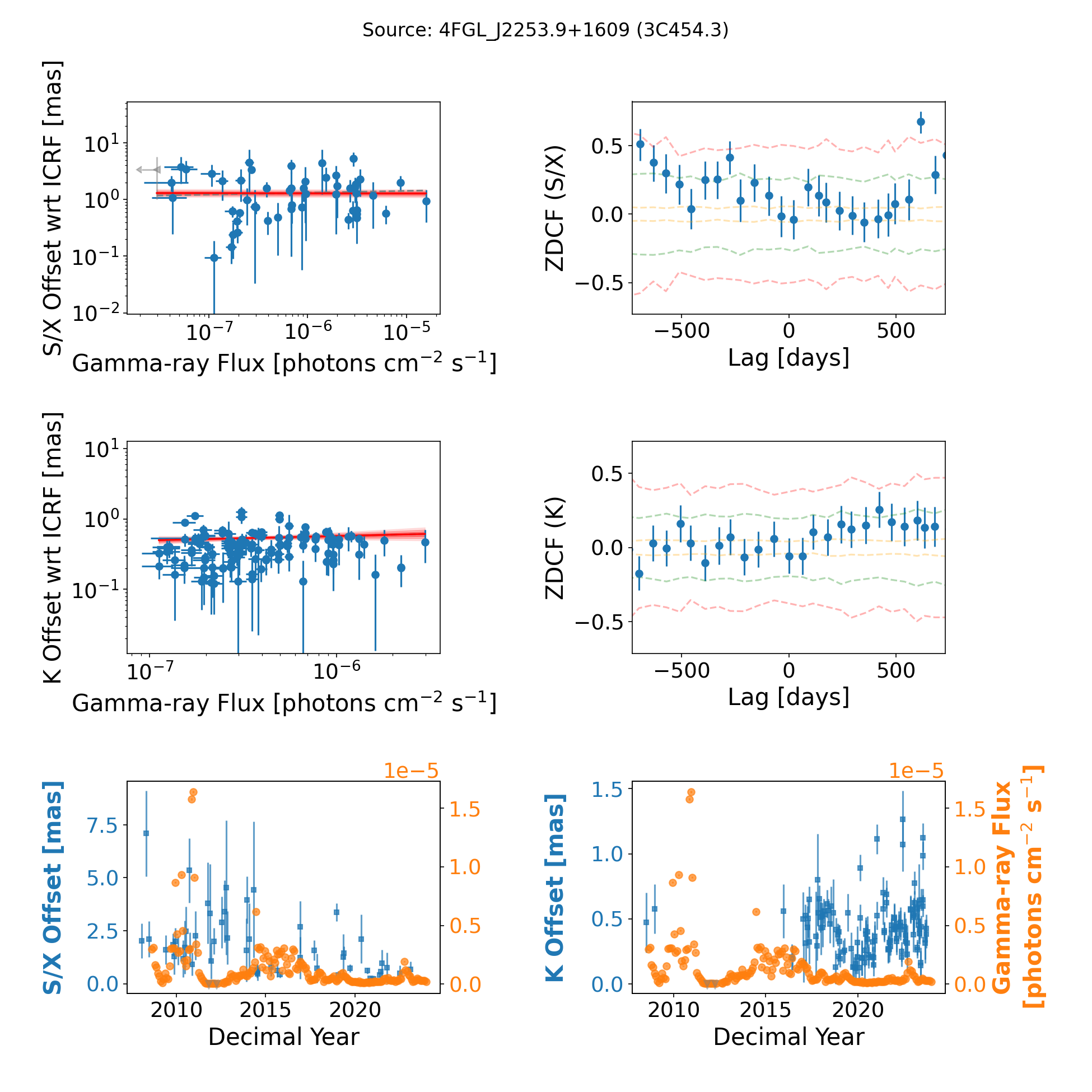}
    \caption{Same as Figure \ref{fig:correlations1}, but for the source 3C454.3.}
    \label{fig:correlationsJ2253}
\end{figure}

\clearpage

\begin{figure}[ht!]
    \centering
    \includegraphics[width=\textwidth]{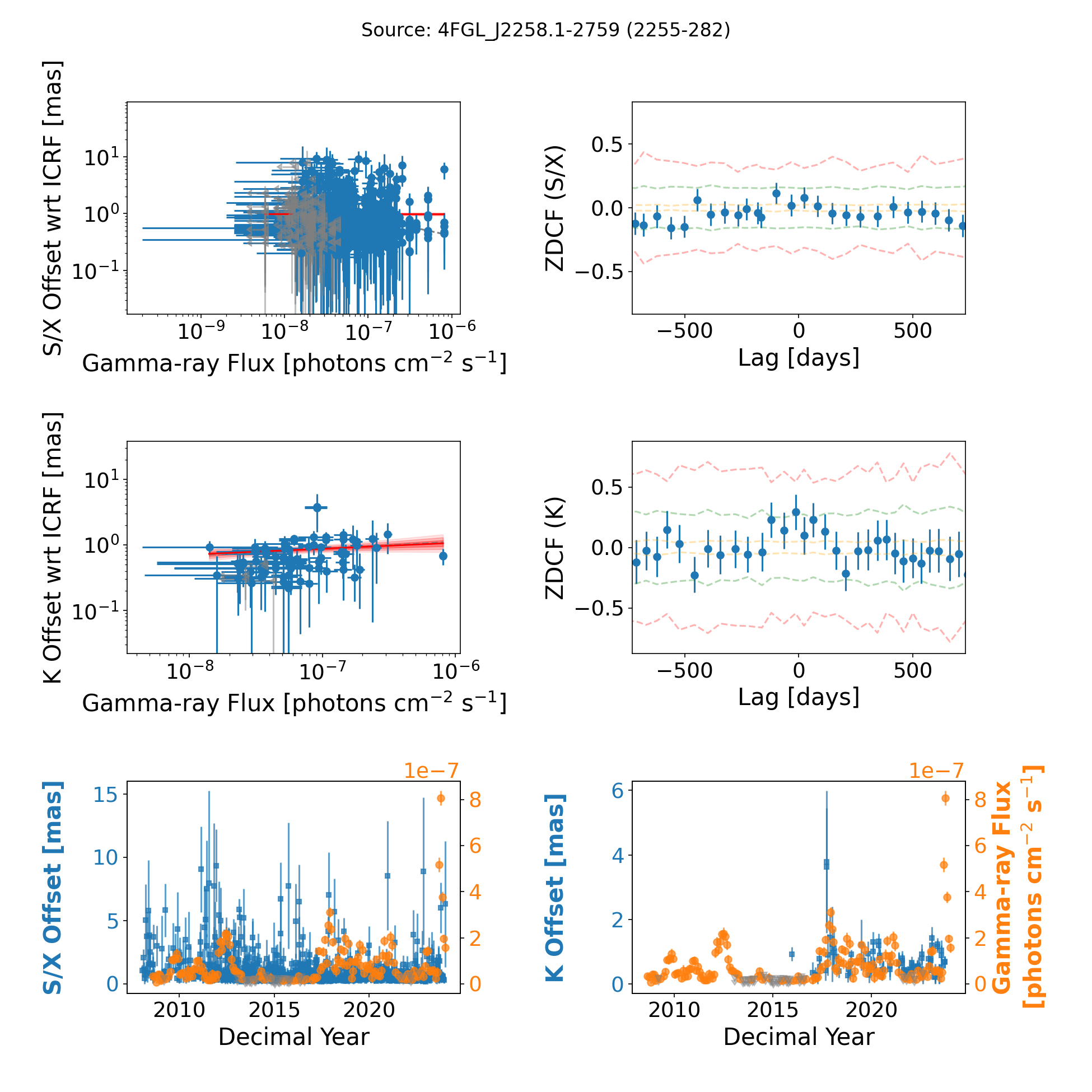}
    \caption{Same as Figure \ref{fig:correlations1}, but for the source 2255$-$282.}
    \label{fig:correlationsJ2285}
\end{figure}

\clearpage

\begin{figure}[ht!]
    \centering
    \includegraphics[width=\textwidth]{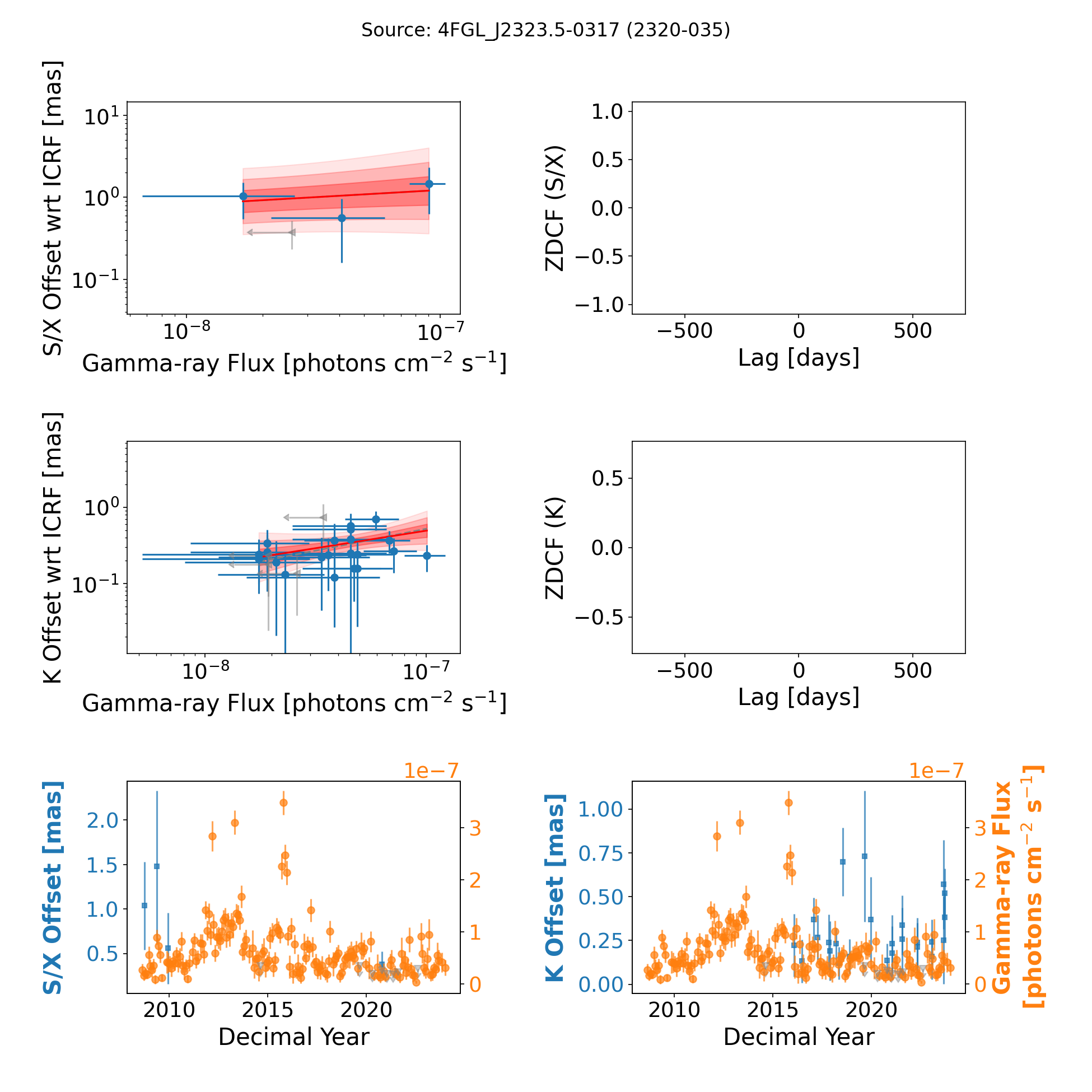}
    \caption{Same as Figure \ref{fig:correlations1}, but for the source 2320$-$035.}
    \label{fig:correlationsJ2323}
\end{figure}

\end{document}